\documentclass[10pt]{article}

\usepackage[a4paper, 
inner=1.5cm, top=1.5cm, outer=1.5cm, bottom=1.5cm, includefoot]{geometry}
\usepackage{bm}
\usepackage[max2]{authblk}
\usepackage{amssymb}
\usepackage{amsmath}
\usepackage{mathrsfs}
\usepackage{graphicx}
\usepackage{enumitem}
\usepackage[ruled,lined]{algorithm2e}
\usepackage[style=ieee,natbib=true,url=false,isbn=false]{biblatex}
\usepackage{amsthm}
\usepackage{appendix}
\usepackage{subdepth}
\usepackage{float}
\usepackage{hyperref}
\usepackage{mathtools}
\usepackage{xcolor}

\newcommand{\styy}[1]{\mathbb{#1}}
\newcommand{\ubar}[1]{\mkern 0.5mu\underline{\mkern-0.5mu#1\mkern-0.5mu}\mkern 0.5mu}
\newcommand{\uubar}[1]{\ubar{\ubar{#1}}}

\let\epsilon\varepsilon
\let\mtheta\theta
\let\theta\vartheta

\let\rho\varrho

\let\phi\varphi
\let\Gamma\varGamma
\let\Delta\varDelta
\let\Theta\varTheta
\let\Lambda\varLambda
\let\Xi\varXi
\let\Pi\varPi
\let\Sigma\varSigma
\let\Upsilon\varUpsilon
\let\Phi\varPhi
\let\Psi\varPsi

\let\Omega\varOmega

\newcommand{\grad}[3]{\nabla_{#1}^{#2} #3} 

\newcommand{\norm}[1]{\| #1 \|}
\newcommand{\abs}[1]{\vert #1 \vert}

\newcommand{\bbR}{\styy{R}}

\newcommand{\clE}{\mathcal{E}}

\newcommand{\clP}{\mathcal{P}}

\newcommand{\sT}{\mathsf{T}} 

\newcommand{\scH}{\mathscr{H}}

\newcommand{\scL}{\mathscr{L}}

\newcommand{\scT}{\mathscr{T}}

\newcommand{\rmi}{\mathrm{i}}

\newcommand{\ua}{\ubar{a}}
\newcommand{\ub}{\ubar{b}}
\newcommand{\uc}{\ubar{c}}

\newcommand{\uf}{\ubar{f}}

\newcommand{\ur}{\ubar{r}}

\newcommand{\uu}{\ubar{u}}
\newcommand{\uv}{\ubar{v}}
\newcommand{\uw}{\ubar{w}}

\newcommand{\uz}{\ubar{z}}
\newcommand{\uA}{\ubar{A}}

\newcommand{\uQ}{\ubar{Q}}

\newcommand{\uT}{\ubar{T}}

\newcommand{\uY}{\ubar{Y}}

\newcommand{\umtheta}{\ubar{\mtheta}}
\newcommand{\uNull}{\ubar{0}}
\newcommand{\uuA}{\uubar{A}}
\newcommand{\uuB}{\uubar{B}}
\newcommand{\uuC}{\uubar{C}}
\newcommand{\uuD}{\uubar{D}}

\newcommand{\uuS}{\uubar{S}}
\newcommand{\uuT}{\uubar{T}}
\newcommand{\uuU}{\uubar{U}}
\newcommand{\uuV}{\uubar{V}}

\newcommand{\<}{\langle}
\renewcommand{\>}{\rangle}

\newcommand{\uuubar}[1]{\ubar{\ubar{\ubar{#1}}}}
\newcommand{\uuuT}{\uuubar{T}}
\newcommand{\uuuA}{\uuubar{A}}
\newcommand{\uuuB}{\uuubar{B}}
\newcommand{\uuuC}{\uuubar{C}}
\newcommand{\uuuF}{\uuubar{F}}

\newtheorem{remark}{Remark}

\title{Equivariant Tensor Network Potentials}
\author[1]{M. Hodapp\thanks{corresponding author}\thanks{maxludwig.hodapp@mcl.at}}
\author[2]{A. Shapeev\thanks{a.shapeev@skoltech.ru}}
\affil[1]{Materials Center Leoben Forschung GmbH (MCL), Leoben (AT)}
\affil[2]{Skolkovo Institute of Science and Technology (Skoltech), Center for Artificial Intelligence Technology, Moscow (RU)}

\begin{document}

\maketitle

\begin{abstract}
    Machine-learning interatomic potentials (MLIPs) have made a significant contribution to the recent progress in the fields of computational materials and chemistry due to MLIPs' ability of accurately approximating energy landscapes of quantum-mechanical models while being orders of magnitude more computationally efficient.
    However, the computational cost and number of parameters of many state-of-the-art MLIPs increases exponentially with the number of atomic features.
    Tensor (non-neural) networks, based on low-rank representations of high-dimensional tensors, have been a way to reduce the number of parameters in approximating multidimensional functions, however, it is often not easy to encode the model symmetries into them.

    In this work we develop a formalism for rank-efficient equivariant tensor networks (ETNs), i.e., tensor networks that remain invariant under actions of SO(3) upon contraction.
    All the key algorithms of tensor networks like orthogonalization of cores and DMRG-based algorithms carry over to our equivariant case.
    Moreover, we show that many elements of modern neural network architectures like message passing, pulling, or attention mechanisms, can in some form be implemented into the ETNs.
    Based on ETNs, we develop a new class of polynomial-based MLIPs that demonstrate superior performance over existing MLIPs for multicomponent systems.
\end{abstract}

\tableofcontents

\section{Introduction}

\subsection{Motivation}

Machine learning has reached a level of maturity that has spawned technology in many fields of science now accessible to non-experts.
Prominent examples are DALL-E in computer vision, AlphaGo in board games, or ChatGPT in linguistics.
In atomistic modeling, machine-learning models, or, in the field-specific language, machine-learning interatomic potentials (MLIPs), have been developed to approximate the energy landscape of an underlying quantum-mechanical model, and reductions in computational cost of several orders of magnitude have been reported, while preserving the accuracy of quantum mechanics.
The three major classes of MLIPs are neural-network-based potentials \citep{behler_generalized_2007, schutt_schnet_2017, smith_ani1_2017, wang2018-deepmd, pun_physically_2019, takamoto_teanet_2022, batzner_equivariant_2022}, polynomial potentials \cite{thompson_spectral_2015, shapeev_moment_2016, drautz_atomic_2019}, and Gaussian process regression-based potentials \citep{bartok_gaussian_2010,jinnouchi2020-gaussian,vandermause2020-gaussian}.
It is believed, and the lack of successful counterexamples is a good proof of it, that rotational and permutational (i.e., with respect to permuting chemically equivalent atoms) must be incorporated in the MLIP's functional form.
For neural networks it has traditionally been achieved by using invariant input features \citep{behler_generalized_2007,schutt_schnet_2017,smith_ani1_2017,pun_physically_2019,takamoto_teanet_2022}, which was later generalized to covariant features and equivariant neural-network feature transformations \cite{batzner_equivariant_2022}.
The state-of-the-art polynomial models also work with covariant features and their algebraic transformations (addition and multiplication) in an equivariant manner  \cite{thompson_spectral_2015, shapeev_moment_2016, drautz_atomic_2019}.
Gaussian process-based models use invariant kernels \citep{bartok_gaussian_2010} or basis functions \citep{jinnouchi2020-gaussian,vandermause2020-gaussian}.
A very interesting case is \cite{kondor_clebsch_2018} which is both, a neural network model whose output depends polynomially on the input features.

A recent benchmark study \citep{zuo_performance_2020} has been conducted to compare the accuracy-vs-efficiency performance of different models on unary (i.e., with one chemical component) systems evaluated on a CPU.
This study, at least in its small-dataset, a polynomial model, MTP \cite{shapeev_moment_2016}, showed a comparatively favorable performance.
Despite this and other success stories, polynomial-based models suffer from exponentially growing complexity when the number of features is large.
In other words, neural-network potentials regain their advantage when applied to systems with complex quantum-mechanical phenomena (e.g., magnetism or electronic temperature) or a large chemical space (e.g., high-entropy alloys) since the latter models have a more controlled growth of the parameter space as the number of features increases \citep{artrith_efficient_2017,darby_compressing_2022,darby_tensorreduced_2023,lopanitsyna_modeling_2023}.

The problem of a large parameter space can be described as follows.
Polynomial MLIPs are potentials whose functional form is constructed by taking polynomials of atomic features (positions, atomic species, magnetic moments, etc.).
Polynomial MLIPs can be encoded with tensors, i.e., their functional form for a $d$-th order polynomial usually consists of (linear) combinations of scalar-valued tensor contractions of the following type (assuming the Einstein summation convention)
\begin{equation}\label{eq:poly-features}
    T_{i_1 \ldots i_d} \,
    v_{i_1} \ldots v_{i_d},
\end{equation}
where $T_{i_1 \ldots i_d}$ is the polynomial coefficient (parameter) tensor and $v_{i_1} \,\ldots\, v_{i_d}$ are the feature vectors.
The feature vectors themselves are multi-index vectors $v_i = v_{(j_1 \ldots j_n)} = F_{j_1 \ldots j_n}$, where $F$ is the feature tensor in which each dimension represents one atomic feature.
Evidently, the representation \eqref{eq:poly-features} becomes increasingly inefficient with an increasing number of features $n$ because the size of the feature vector is proportional to $\bar{m}^n$, where $\bar{m}$ is the average size over all dimensions of the feature tensor, and, so, contracting $T$ $d$-times with $v$ is proportional to $\bar{m}^{dn}$.
Even with only radial, angular, and species features, the amount of parameters can quickly increase to the order of one thousand for alloys with a large number of components, such as high-entropy alloys.

One approach to attempt reducing the amount of free parameters is to convert the coefficient tensor in \eqref{eq:poly-features} into a low-rank tensor format that approximates multivariate polynomials, e.g., the tensor train (TT) format \citep{oseledets_tensortrain_2011,cichocki_tensor_2017}.
Incorporation of symmetries into tensor networks if often not easy.
For example, in \citep{kostiuchenko2019-npj} a tensor train representation was used to approximate on-lattice atomic energies as functions of chemical degrees of freedom, and permutational symmetry was build in by explicitly symmetrizing the tensor over the corresponding point group of the lattice.
In the present setting, we require that group actions of SO(3) on atomic positions must leave \eqref{eq:poly-features} invariant; physically speaking, the (off-lattice) potential energy of a configuration of atoms must not change under pure rotations.
A related very recent development approaches this problem by utilizing a symmetric canonical tensor decomposition of the coefficient tensor \citep{jana_searching_2023}.

\subsection{Outline of the present work}

In this work we propose \textbf{equivariant tensor networks (ETNs)}, a formalism for constructing interatomic interaction models based on tensor networks that are invariant under the action of SO(3).
In particular, we develop a tensor network algebra (Section \ref{sec:tn-intro}) and use it to develop the following three building blocks that compose our ETNs (Section \ref{sec:SO(3)-ETNs}):
\begin{itemize}
    \item
    covariant vectors $v$, i.e., vectors that rotate correspondingly with a basis change under actions of SO(3),
    \item
    equivariant order-3 tensors $T$, i.e., tensors that describe equivariant maps of the covariant vectors, and
    \item
    feature contractions, i.e., parameterized contractions that map the feature tensors $F_{i_1 \ldots i_n}$ to some reduced covariant vector $v$.
\end{itemize}
Here and in what follows, the term ``SO(3)-covariant vector'' will be used in the context of a ``feature vector'' and refers to a collection, possibly large, of numbers that change (or remain constant) upon rotation, rather than specifically an element of $\mathbb{R}^3$.
We show that arbitrary ETNs expressing polynomials of arbitrary degree can be constructed entirely from these three building blocks.

We present the formalism in the language conventional to the field of numerical linear algebra, although we emphasize that many key ideas already exist in the field of quantum physics that have been used to, e.g., incorporate parity symmetries into tensor networks representing the wave functions of bosons and fermions (e.g., \citep{hutter_representing_2020}).
In particular, the recent developments \citep{orus_tensor_2019} concern SU(2)-equivariant tensor networks \citep{weichselbaum_nonabelian_2012} representing quantum states of qubit systems that recently receive a lot of attention in the context of quantum computing. 
In these tensor networks, SU(2) symmetries are incorporated using the Wigner-Eckhart Theorem that states that any spherical tensor $T$ with three multi-indices $\{ (\ell_i,m_{\ell_i},n_{\ell_i}) \}_{i=1,\ldots,3}$, where $\ell$ is the angular momentum, and $m$ is the phase, can be factorized to
\begin{equation}\label{eq:Wigner-Eckhart-Theorem}
    T_{(\ell_1,m_{\ell_1},n_{\ell_1})(\ell_2,m_{\ell_2},n_{\ell_2})(\ell_3,m_{\ell_3},n_{\ell_3})}
    =
    P_{(\ell_1,n_{\ell_1})(\ell_2,n_{\ell_2})(\ell_3,n_{\ell_3})}
    Q_{(\ell_1,m_{\ell_1})(\ell_2,m_{\ell_2})(\ell_3,m_{\ell_3})}
    ,
\end{equation}
where $P$ contains the degrees of freedom and is independent of the phase $m_{\ell_i}$,\footnote{in the quantum mechanics, $P$ is usually referred to as the ``reduced matrix element''} and $Q$ is the \emph{Clebsch-Gordan coefficient}, a constant tensor that is defined by the symmetry group.
Key algorithms like those based on DMRG (density matrix renormalization group) are formulated for such tensor networks \citep{orus_tensor_2019}.

In our work, we make full use of the structure of SO(3)-equivariance, namely, the possibility of building the irreducible representation based on \emph{real-valued spherical harmonics}.
In the real-valued case, vector contraction and dot product are the same, which enables us to build them using the \emph{Wigner 3-j coefficients}.
The latter are symmetric with respect to permutations of $(\ell_i,m_{\ell_i})$ which results into \emph{undirected networks} (they are directed in the SU(2) case due to the lack of symmetry of the Clebsch-Gordan coefficients, which results into network nodes of different types based on the direction of the adjacent edges).
This allows us to formulate the usual operations of tensor network algebra, like reshaping an order-3 tensor core into an order-2 tensor, combining two cores together, or performing decompositions and splitting of the cores, while preserving the symmetric structure of the network.
Furthermore, in our framework it is easy to also separate the reflection symmetric/antisymmetric features, resulting into the full O(3)-equivariance of the tensor network.

We exemplify the construction of ETNs by developing an equivariant version of the TT format
\begin{equation}\label{eq:ETT-features}
    T_{i_1j_2} T_{j_2i_2j_3} \ldots T_{j_di_d} v_{i_1} \ldots v_{i_d}.
\end{equation}
We will show that the number of parameters of this representation is proportional to $(d + n)\bar{m}\bar{r}^2$, where $\bar{r}$ is the average rank over all $T$'s.
Hence, if there is enough similarity between the features, the ranks can be made small and the representation \eqref{eq:ETT-features} will be more efficient than the ``raw'' polynomial \eqref{eq:poly-features}.
We also show how algorithms developed for conventional tensor networks, e.g., for optimizing tensor parameters, tensor orthogonalization, etc., carry over to the equivariant case (Section \ref{sec:algebra}).

With our ETN formalism, we develop a new class of MLIPs that we henceforth refer to as the class of \textbf{ETN potentials} (Section \ref{sec:potentials}).
In particular, we explicitly develop and implement an ETN potential that has atomic position and species features, and show that this ETN potential requires significantly fewer parameters than MTPs to reach the same level of accuracy for QM7, QM9, and several datasets for metallic alloys (Section \ref{sec:ETNP-performance}).

Further, we outline how our formalism can be used to construct more general ETN potentials (Section \ref{sec:generalizations}).
Examples include extensions to arbitrary ETN topologies, more general feature tensors, nonlocal atomic energies, and other symmetry groups.

The key advantage of our new formalism is its modularity due to the rather small set of required operations; in essence, tensor operations on up-to-order-3 tensors.
This heavily simplifies generic implementations and the automation of operations other than contractions to be performed on the ETNs, such as automatic differentiation, factorization, etc.
Our formalism for constructing ETN potentials can thus be, in some sense, considered as an interatomic potential modular building system (``interatomic potential Lego'') that allows for a rapid construction of efficient representations of polynomial MLIPs with arbitrary order and physical complexity.

\section{A Short Introduction to Tensor Networks}
\label{sec:tn-intro}

Here we introduce the notions and notations needed to present equivariant tensor networks.
We do not do it with maximal generality for the sake of conciseness of the presentation.
Nevertheless, many features present in the conventional tensor networks are also present in the equivariant tensor networks; we remark on this in Section \ref{sec:generalizations}.

A \emph{tensor} is a multidimensional array $T_{\< d\>} \in\bbR^{N_1\times N_2\times \ldots\times N_d}$, where $d$ will be called the \emph{number of dimensions} or simply the \emph{order} of the tensor (we reserve the term \emph{rank} to speak about, e.g., the rank of a matrix, or its generalization to tensors).
Since we will mostly work with tensors up to the order of three, we use a notation using underscores for those tensors, namely, $v_{\< 1\>}  =\uv$ for an order-1 tensor (or vector), $U_{\< 2\>} = \uuU$ for an order-2 tensor (or matrix), and $T_{\<3\>} = \uuuT$ for an order-3 tensor.
We will also use the index notation where, e.g., $\uv$, $\uuU$, and $\uuuT$, are written as $v_i$, $U_{ij}$, and $T_{ijk}$.
For tensor contractions we will generally adopt the Einstein notation.
In case we do not sum over all duplicate indices we will specify the contracted indices explicitly.

\subsection{Three is the magic tensor order}

A vector $\uu$ can be used to model a dependence $u_i$ on one discrete variable $i$, or one continuous variable upon choosing a (finite) basis $v_i$ and expanding the one-dimensional function in this basis with coefficients $u_i$.
The latter motivates us to consider a contraction $\alpha_{\uu}(\uv) = \uu \cdot \uv$ corresponding to $\uu$ and interpret it as a linear form over $\uv$.
This way, a tensor of order $d$, $T_{\< d\>}$, can be thought of describing a quantity dependening on $d$ variables and can be associated with a multilinear form
\begin{equation}\label{eq:tensor-contraction}
    \alpha = \alpha(\uv^1, \ldots, \uv^d) = \big( \ldots \big( T_{\< d\>} \uv^d \big) \uv^{d-1} \ldots \big) \uv^1.
\end{equation}

Of course, the simplest tensor is a scalar (zero-order tensor), but it is hardly sufficient to model anything of nontrivial complexity. A vector $\uv$ can be used to model a dependence on one variable, while an outer product of two vectors is an order-two tensor $\uu\otimes \uv$, however, it can only describe a two-dimensional function whose two variables are essentially independent of each other---again, we can conclude that a vector is too simple to describe complex dependencies.
Of course, it is known that any dependence can be represented as a sum of $\uu_k\otimes \uv_k$, however, this brings us to contractions of higher-order tensors.

Namely, the sum $\uu_k\otimes \uv_k$ is more conveniently interpreted as a product of two matrices $\uuU\uuV$---which is a matrix (order-two tensor) itself.
Here, by writing $AB$ we mean that the last dimension(s) of $A$ is contracted with the first dimension(s) of $B$.
By taking contractions of matrices one can only obtain tensors of order two or less which is, again, not sufficient to describe nontrivial multivariate dependencies.

Order-three tensors make a difference.
With two order-three tensors $\uuuA$ and $\uuuB$ by contracting one dimension in them one can obtain a nontrivial (in the sense defined above) order-four tensor $\uuuA\,\uuuB$; with three tensors one can obtain a nontrivial order-five tensor $\uuuA\,\uuuB\, \uuuC$, etc.
Thus, three is the minimal tensor order which allows representing nontrivial dependencies on any number of variables---in this sense we say that \emph{three} is the magic number for the tensor order.

\subsection{Tensor Diagrams}

\begin{figure}[htb]
    \centering
    \includegraphics[scale=0.8]{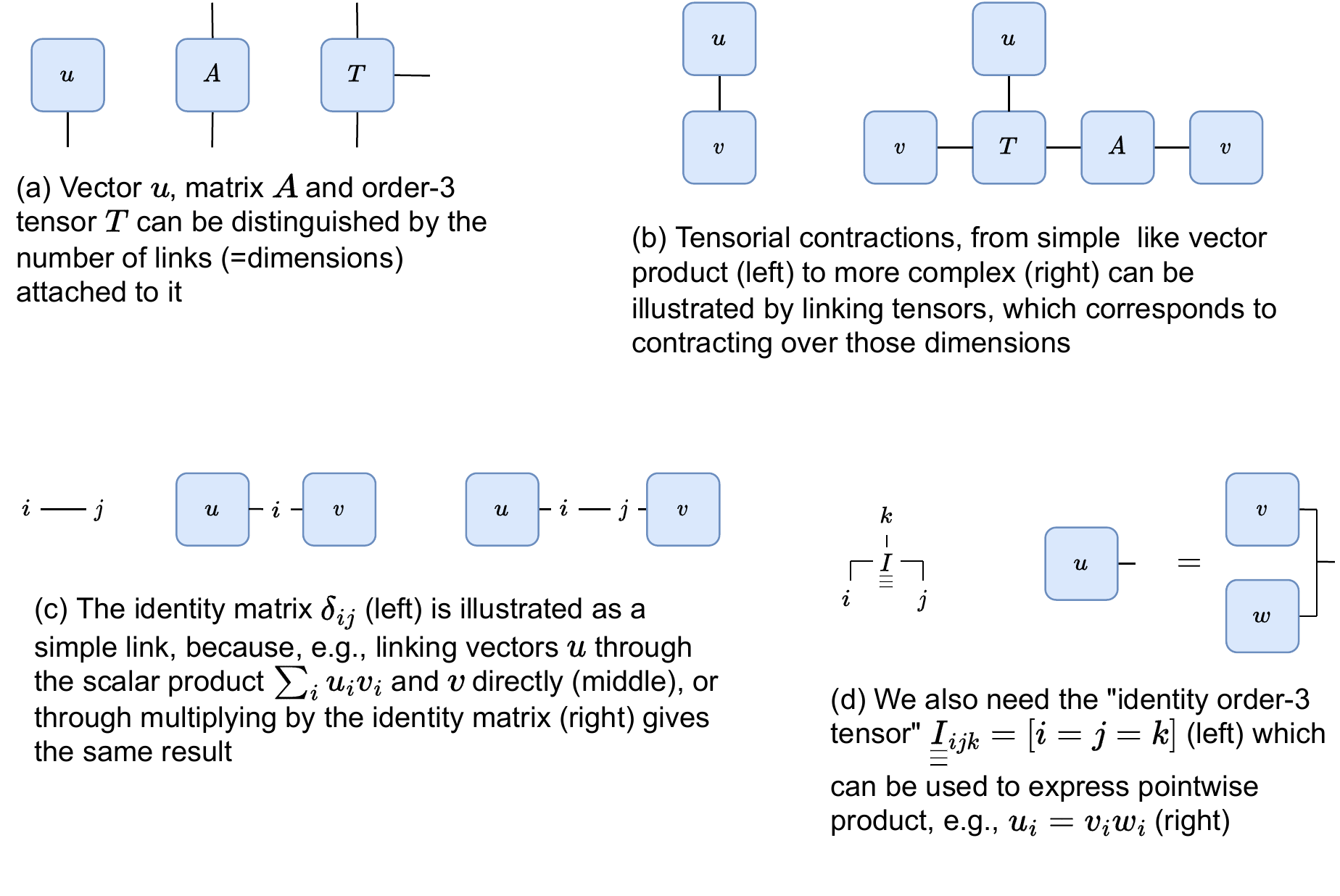}
    \caption{Example of diagrams of basic tensors and operations on them.
}
    \label{fig:diagrams-basic}
\end{figure}

Tensorial operations on vectors and matrices can easily be represented with formulas (as a single string of products of multiple vectors and matrices), however, when the tensor order goes above two, it is much easier to represent the operations with diagrams.
In Figure \ref{fig:diagrams-basic} we show the basic tensors and operations.
The tensors are represented with boxes, their dimensions are represented with links, and the diagram represents the result of contractions of tensors with themselves.
If two tensor dimensions are connected, it means that they are contracted in the result.
If there are no ``hanging'' links then the result is scalar.

As follows from the above, a multilinear form can be realized by contractions with multiple order-two and order-three tensors:
\begin{equation}\label{eq:tensor-train}
    \alpha(\uv^1,\ldots,\uv^d) = \uuT^1 \Big( \,\ldots\, \uuuT^{d-1} \Big( \uuT^d \uv^d \Big) \uv^{d-1} \,\ldots\, \Big) \uv^1
    ,
\end{equation}
where $\uuT^1$ and $\uuT^d$ are order-2 tensors and $\uuuT^2$, \ldots, $\uuuT^{d-1}$ are order-3 tensors. Thus, the right-hand side of \eqref{eq:tensor-train} is a product of two vectors and $d-2$ matrices yielding a scalar.
The multilinear form \eqref{eq:tensor-train} is nothing but the tensor train (TT) representation \citep{oseledets_tensortrain_2011} of the associated order-$d$ tensor which has been known in physics under the term matrix product state (MPS, see, e.g., \citep{fannes_finitely_1992,perez-garcia_matrix_2007}).
The graphical representation of \eqref{eq:tensor-train} in terms of tensor network diagrams is
\begin{equation}\label{eq:tensor-diagram}
\alpha(\uv^1,\ldots,\uv^d) = 
\begin{aligned}
    \includegraphics[width=0.43\textwidth]{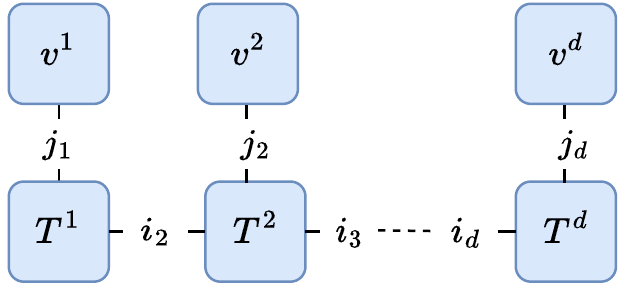}
\end{aligned}
.
\end{equation}
In these diagrams, a tensor is represented with a square block, with connections between blocks being contractions over the tensors' dimensions.
Our notation of tensor network diagrams is inspired by tensor network diagrams used in quantum physics (e.g., \citep{bridgeman_handwaving_2017}) and appears to us very convenient for representing tensors (and tensor operations) in our setting, in particular, in view of comparing different variants of equivariant tensor networks later on.
To show the power of this notation, we remark that the order-$d$ tensor $T_{\<d\>}$ corresponding to \eqref{eq:tensor-train} is not easy to write down in mathematical formulas, but very easy to represent as a visual diagram:
\begin{equation}\label{eq:tensor-diagram_no_v}
T_{\<d\>} = 
\begin{aligned}
    \includegraphics[width=0.43\textwidth]{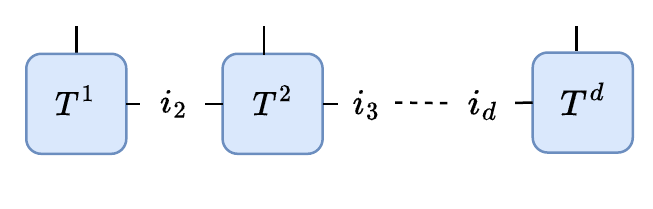}
\end{aligned}
.
\end{equation}
From this diagram one can see that an order-three tensor (like $T^2$) is a fundamental building block which allows to construct tensors of any order, which reiterates the point made in the previous subsection.

\section{SO(3)-Equivariant Tensor Networks}
\label{sec:SO(3)-ETNs}

Our motivation is to use \eqref{eq:tensor-contraction} to approximate O(3)-invariant functionals (e.g., interatomic interaction energies) of local atomic environments (or, more generally, clouds of points).
O(3) invariance means that, if the atomic environments are rotated or reflected in the three-dimensional space, then the functional $\alpha$ does not change.
We separate the problem of O(3) invariance into SO(3) (rotation) invariance and reflection invariance.
Mathematically, invariance of \eqref{eq:tensor-contraction} with respect to a group $G$ (for example, $G={\rm SO(3)})$, means that when $\uv$ changes under a group action (in our example, rotation) $\uv \mapsto g\cdot\uv$, then $\alpha$ remains constant.
Note that the model parameters, $T_{\<d\>}$, are not allowed to change with the group action (because the interatomic interaction model is independent of the frame of reference).
Often, together with the invariance comes the notion of \emph{equivariance}.
For instance, for \eqref{eq:tensor-train} to be invariant, the matrix $\uuT$ does not only have to be SO(3)-invariant component-wise, but also \emph{SO(3)-equivariant}, i.e., $g^{-1} \cdot \uuT (g\cdot \uv) = {\rm const}$ as $\uv$ rotates.
We note that the equivariance of the operator $\uuT$ is equivalent to the invariance of the quadratic form associated with $\uuT$, i.e., $(\uuT (g\cdot \uv)) \cdot (g\cdot \uv') = {\rm const}$.

We are hence interested in two kinds of objects: vectors (and sometimes tensors) rotating under the action of $G$, and tensors that remain constant but otherwise describe an equivariant transformation of the former vectors.
The first kind of objects are typically the input features of a model (with an arbitrarily chosen frame of reference) or transformed features occurring in the middle of a calculation, while the second kind are model coefficients that ``do not know'' which frame of reference was chosen for the input.
The first kind of objects are \emph{covariant vectors}, i.e., vectors that change under roations as $\uv' = \uuD^\sT \uv$, where $\uuD$ is a Wigner D-matrix, while the second kind of objects will be called \emph{equivariant} tensors.
In what follows, by ``contraction'' of tensors with anything else we mean invariant or equivariant contractions, depending on the context, unless it is explicitly mentioned otherwise.

Following the results of representation theory, for a covariant vector $\uv$, we consider its decomposition into an irreducible representation.
In the case of SO(3), this corresponds to spherical harmonics components with different angular momenta.
Hence, we consider $\uv$ as a \emph{multi-index} vector $v_{(\ell mn)}$, $\ell=0,\ldots,L$, where $L$ is the maximal angular momentum, $m\in\{-\ell,-\ell+1,\ldots,\ell\}$ is the phase, and $n=1,\ldots,N(\ell)$ is the number of the components corresponding to $\ell$.
We remark that we deviate intentionally from the notation $v_{(n\ell m)}$ commonly used in quantum physics that puts the index $n$, usually referred to as ``principal quantum number'', in front of $\ell m$ (cf., e.g., \citep{sakurai_modern_2020}).
Our reasoning in doing so is due to the fact that we will later on work with higher-dimensional tensors, where each multi-index may depend on a different $\ell$.
Since $m$ and $n$ are always assumed to depend on $\ell$, using the order $\ell mn$ appears more convenient to us for the sake of clarity.
In the spirit of machine learning, we will call $N(\ell)$ as the number of \emph{angular momentum channels} corresponding to each $\ell$.
We assume that the action of SO(3) on $v_{(\ell mn)}$ follows the action of SO(3) on the \emph{real} spherical harmonic (RSH) function $\dot Y_{\ell m}$ defined as follows
\begin{equation}\label{eq:real-sph}
    \dot Y_{\ell m} = \left\{
    \begin{aligned}
        \;& \frac{\rmi}{\sqrt{2}} \left( Y_{\ell -\abs{m}} - (-1)^m Y_{\ell\abs{m}} \right) &\quad& m<0, \\
        \;& Y_{\ell 0} && m=0, \\
        \;& \frac{1}{\sqrt{2}} \left( Y_{\ell-\abs{m}} + (-1)^m Y_{\ell\abs{m}} \right) && m>0,
    \end{aligned}
    \right.
\end{equation}
where the $Y_{\ell m}$'s are the (complex) spherical harmonics.

Using \eqref{eq:real-sph}, we introduce the mapping $U_{\ell mm'} : Y_{\ell m'} \rightarrow \dot Y_{\ell m}$ \citep{blanco_evaluation_1997}, i.e.,
\begin{equation}\label{eq:U_l}
    \uuU_\ell =
    \frac{1}{\sqrt{2}}
    \begin{pmatrix}
        \rmi &&&&&& -(-1)^\ell\,\rmi \\
        & \rmi &&&& -(-1)^{\ell-1}\,\rmi & \\
        && \ddots && \reflectbox{$\ddots$} && \\
        &&& \sqrt{2} &&& \\
        && \reflectbox{$\ddots$} && \ddots && \\
        & 1 &&&& (-1)^{\ell-1} & \\
        1 &&&&&& (-1)^\ell
    \end{pmatrix}.
\end{equation}
Since $\uuU_\ell$ is a unitary matrix, its inverse is its conjugate transpose $\uuU_\ell^{-1} = \uuU_\ell^{\sT\ast}$.
We then define the Wigner 3-j symbols (or just 3-j symbols) for RSHs as
\begin{equation}\label{eq:3-j-symbol-real-sph}
    \begin{Bmatrix}
        \ell_1 & \ell_2 & \ell_3 \\ m_1 & m_2 & m_3
    \end{Bmatrix}
    =
    \sum_{m_1',m_2',m_3'}
    \begin{pmatrix}
        \ell_1 & \ell_2 & \ell_3 \\ m_1' & m_2' & m_3'
    \end{pmatrix}
    U^{\sT\ast}_{\ell_1 m'_1m_1} U^{\sT\ast}_{\ell_2 m_2'm_2} U^{\sT\ast}_{\ell_3 m_3'm_3}.
\end{equation}
The 3-j symbols are nonzero only when the three $\ell$'s and $m$'s satisfy the usual two selection rules
\begin{align}
    \label{eq:triangular-inequality}
    \hypertarget{SR1}{\textbf{[SR1]}} & \quad {\rm TR}(\ell_1,\ell_2,\ell_3) \quad\text{if, by definition,}\quad \max_k(\ell_k) \leq \min_{i\ne j} (\ell_i+ \ell_j), \\
    \hypertarget{SR2}{\textbf{[SR2]}} & \quad m_1 + m_2 + m_3 = 0,
\end{align}
where ${\rm TR}(\ell_1,\ell_2,\ell_3)$ is the triangular inequality.
The 3-j symbols for RSHs can also be nonzero for other linear combinations of $m$'s.
More precisely, for the 3-j symbols for RSHs, the selection rule \hyperlink{SR2}{\textbf{[SR2]}} is replaced by the following two selection rules
\begin{align}\label{eq:SR3}
    \hypertarget{SR3}{\textbf{[SR3]}} & \quad \operatorname{sgn}(m_1)\operatorname{sgn}(m_2)\operatorname{sgn}(m_3) = \pm 1,
\end{align}
where $\operatorname{sgn}(\bullet)$ is the sign function, and the $+$ and $-$ hold for even or odd $\ell_1 + \ell_2 + \ell_3$, respectively, and
\begin{equation}\label{eq:SR4}
    \hypertarget{SR4}{\textbf{[SR4]}}
    \,\left\{
    \begin{aligned}
        \,\hypertarget{SR4.1}{\textbf{[SR4.1]}}
        && m_1 + m_2 + m_3 &= 0, \\
        \,\hypertarget{SR4.2}{\textbf{[SR4.2]}}
        && m_1 + m_2 - m_3 &= 0, \\
        \,\hypertarget{SR4.3}{\textbf{[SR4.3]}}
        && m_1 - m_2 + m_3 &= 0, \\
        \,\hypertarget{SR4.4}{\textbf{[SR4.4]}}
        && -m_1 + m_2 + m_3 &= 0.
    \end{aligned}
    \right.
\end{equation}
The selection rule \hyperlink{SR4}{\textbf{[SR4]}} holds true if (at least) one of the (sub-)selection rules \hyperlink{SR4.1}{\textbf{[SR4.1]}}--\hyperlink{SR4.4}{\textbf{[SR4.4]}} holds true.

The selection rules \hyperlink{SR3}{\textbf{[SR3]}} and \hyperlink{SR4}{\textbf{[SR4]}} follow from the properties of $\uuU_\ell^{\sT\ast}$ (see Appendix \ref{appdx:w3j_rsph_selection_rules}).
Although the additional selection rules suggest that the 3-j symbol for RSHs is much less sparse than the usual 3-j symbol, it is, in fact almost as sparse as the usual 3-j symbol and does not sacrifice computational efficiency (of tensor contractions) for convenience (undirected tensor networks) as shown in Appendix \ref{appdx:w3j_rsph_sparsity}.

Moreover, we refer to Appendix \ref{appdx:w3j_rsph_contraction_properties} for the equivalence of the 3-j symbol for RSHs and the Clebsch-Gordan coefficients in adding angular momenta.
However, we point out that the 3-j symbols for RSHs obey \emph{almost the same symmetries} as the usual 3-j symbol (cf., Appendix \ref{appdx:w3j_symmetries}).

In addition, we remark that the 3-j symbol for RSHs \eqref{eq:3-j-symbol-real-sph} is \emph{real} whenever $\ell_1 + \ell_2 + \ell_3$ is even, and \emph{imaginary} otherwise (cf., Appendix \ref{appdx:w3j_rsph_selection_rules} for further details).
This feature of the 3-j symbol for RSHs will be important for the full O(3) (reflection + rotation) invariance.

With our definition of the 3-j symbol for RSHs, we now define the basic operations to be performed on covariant vectors that we require to build equivariant tensor networks.

\begin{remark}
    At this point, we remark that the choice of the basis is not unique.
    One could likewise choose invariant polynomials (invariant with respect to rotation and reflection) as the basis set \citep{shapeev_moment_2016}.
    However, within the tensor network formalism we propose here, the operators that generate those invariant polynomials are not available off-the-shelf (such as for spherical harmonics).
    It could be nevertheless intersting to explore other basis sets that would lead to potentially more sparse representations.
\end{remark}

\subsection{Order-1 Tensors (Vectors)}

The basic operations we require are the contraction of two covariant vectors
\begin{equation}
    \uu \cdot \uv = \sum_{\ell,m,n} u_{(\ell mn)} v_{(\ell mn)},
\end{equation}
and \emph{equivariant} contractions of a vector $\uc$ with a covariant vector. The latter can only be performed with respect to the 0-th angular momenta such that
\begin{equation}\label{eq:vector-contraction}
    \alpha = \alpha(\uv) = \<\uc, \uv\> = \sum_n c_n v_{00n}.
\end{equation}
In other words, \eqref{eq:vector-contraction} is the most general equivariant contraction.

\subsection{Order-2 Tensors (Matrices)}

Instead of defining the outer product of two covariant vectors as simply $u_{(\ell mn)} v_{(\ell'm'n')}$,
we instead reinterpret it as a covariant vector
\begin{equation}\label{eq:outer-g-vectors}
    (\uu \otimes \uv)_{(\ell m \,(\ell' \ell'' n'n'')\,)}
    =
    \sum_{m',m''}
    \begin{Bmatrix}
        \ell' & \ell'' & \ell \\
        m' & m'' & m
    \end{Bmatrix}
    u_{(\ell'm'n')} v_{(\ell''m''n'')}
    .
\end{equation}
Here, the index on the left-hand side should be understood as follows: for each $n'$, $n''$ and a pair of $\ell'$ and $\ell''$ satisfying ${\rm TR}(\ell',\ell'',\ell)$ we have one angular momentum channel $n=(\ell' \ell'' n'n'')$.
Altogether, for each $\ell$, the outer product has
$
\sum_{\ell',\ell'':{\rm TR}(\ell,\ell',\ell'')} N'(\ell') N''(\ell'')
$
angular momenta channels.
We chose the 3-j symbols rather than the Clebsch-Gordan coefficients to emphasize their symmetry with respect to permuting the tensorial dimensions.

We remark that we use the notation \eqref{eq:outer-g-vectors} instead of, e.g.,
the one used in quantum physics
(cf., bipolar spherical harmonics \citep{varshalovich_quantum_1988}), \begin{equation}
    \big(\uu_{\ell'} \otimes \uv_{\ell''}\big)_{(\ell mn)}
    =
    \sum_{m',m''}
    \begin{Bmatrix}
        \ell' & \ell'' & \ell \\
        m' & m'' & m
    \end{Bmatrix}
    u_{(\ell'm'n')} v_{(\ell''m''n'')}
    ,
\end{equation}
implying that this is nonzero only when ${\rm TR}(\ell',\ell'',\ell)$, because we would like that the number of channels is explicitly featured in the notation.

An equivariant contraction of an arbitrary order-2 tensor $\uuC$ and two covariant vectors is then given by
\begin{equation}\label{eq:order-2-tensor-contraction}
    \alpha
    =
    \alpha(\uu,\uv)
    =
    \sum_{\ell,m,n,n'} C_{\ell nn'} \, u_{(\ell mn)} v_{(\ell mn')}.
\end{equation}
Note that we write $C_{\ell nn'}$ with three indices, and, so, it appears to be an order-3 tensor---but we use it in the following as a shorthand notation for a block-diagonal order-2 tensor $C_{(\ell n)(\ell n')}$, with blocks $\uuC_\ell$ corresponding to different $\ell$ being arbitrary, regular $N(\ell) \times N'(\ell)$ matrices.
Hence, many operations that can be done on matrices, such as the QR- or SVD-decomposition, can be done in a block-wise fashion, as shown in Section \ref{sec:algebra}.

The shortest (but not necessarily simplest) way to prove \eqref{eq:order-2-tensor-contraction} is to consider the outer product of $\uu$ and $\uv$ \eqref{eq:outer-g-vectors} (only the $\ell=0$ component matter%
\footnote{
We hence use the fact that \citep{varshalovich_quantum_1988}
\begin{equation}
    \begin{pmatrix}
        \ell & \ell & 0 \\
	    m & -m & 0
    \end{pmatrix}
    =
    \frac{(-1)^{\ell - m}}{\sqrt{2\ell + 1}}.
\end{equation}
})
\begin{align}
    (\uu \otimes \uv)_{(00nn')}
    =\,&
    \sum_{m,m'}
    \begin{Bmatrix}
        \ell & \ell' & 0 \\
	    m & m' & 0
    \end{Bmatrix}
    u_{(\ell mn)} v_{(\ell'm'n')} \\
    \overset{\mathclap{\eqref{eq:w3j-rsph-expr2.1}}}{=}\,&
    \sum_{m}
    (-1)^{m}
    \begin{pmatrix}
        \ell & \ell & 0 \\
	    m & -m & 0
    \end{pmatrix}
    u_{(\ell mn)} v_{(\ell mn')} \\
    =\,&
    \sum_{m}
    \frac{(-1)^{\ell}}{\sqrt{2\ell + 1}} \,
    u_{(\ell mn)} v_{(\ell mn')}
    ,
\end{align}
and then using the arbitrary vectorial contraction \eqref{eq:vector-contraction} noting that the scaling $\frac{(-1)^{\ell}}{\sqrt{2\ell + 1}}$ can be adsorbed into $C_{\ell n n'}$.

\subsection{Order-3 Tensors (where the Magic Happens)}

In the previous subsection we have already encountered a special case of an order-3 equivariant tensor---the 3-j symbol for RSHs.
If we consider an arbitrary order-3 tensor $\uuuT$ given by a trilinear form
\begin{equation}\label{eq:order-3-tensor-contraction}
    \alpha = \alpha(\uu,\uv,\uw) =
    T_{(\ell_1 m_1 n_1) (\ell_2 m_2 n_2) (\ell_3 m_3 n_3)} \,
    u_{(\ell_1 m_1 n_1)} \,
    v_{(\ell_2 m_2 n_2)} \,
    w_{(\ell_3 m_3 n_3)}.
\end{equation}
and require that $\alpha$ stays invariant as SO(3) acts simultaneously on $\uu$, $\uv$, and $\uw$, we will arrive at the following general representation of $\uuuT$,
\begin{equation}
    T_{(\ell_1 m_1 n_1) (\ell_2 m_2 n_2) (\ell_3 m_3 n_3)}
    = C_{(\ell_1 n_1) (\ell_2 n_2) (\ell_3 n_3)} \,
    \begin{Bmatrix}
        \ell_1 & \ell_2 & \ell_3 \\
        m_1 & m_2 & m_3
    \end{Bmatrix},
\end{equation}
where the coefficient tensor $\uuuC$ can be arbitrary.
Due to the properties of the 3-j symbols for RSHs, $C_{(\ell_1 n_1) (\ell_2 n_2) (\ell_3 n_3)}$ can be defined as nonzero only when the triangular inequality \eqref{eq:triangular-inequality} is satisfied.

The bilinear and linear forms can be obtained by simply substituting $(\ell_i m_i n_i) = (0,0,0)$ for one or two indices of the above equations.
In particular, a bilinear form can be written with $\uuuT$ in the following manner
\begin{equation}
\begin{aligned}
    \alpha = \alpha(\uu,\uv) = \< \uu, \uuT\uv \>
    &=
    T_{(\ell_1 m_1 n_1) (\ell_2 m_2 n_2) (0 0 0)} \,
    u_{(\ell_1 m_1 n_1)} \,
    v_{(\ell_2 m_2 n_2)} \\
    &=
    T_{(\ell m n_1) (\ell m n_2)} \,
    u_{(\ell m n_1)} \,
    v_{(\ell m n_2)}, \\
\end{aligned}
\end{equation}
or
\begin{equation}
\begin{aligned}
    \alpha = \alpha(\uv,\uw) = \< \uv, \uuT\uw \>
    &=
    T_{(0 0 0) (\ell_2 m_2 n_2) (\ell_3 m_3 n_3)} \,
    v_{(\ell_2 m_2 n_2)} \,
    w_{(\ell_3 m_3 n_3)} \\
    &=
    T_{(\ell m n_2) (\ell m n_3)} \,
    v_{(\ell m n_2)} \,
    w_{(\ell m n_3)}.
\end{aligned}
\end{equation}
For completeness, linear forms can be written with $\uuuT$ as follows
\begin{equation}
\begin{aligned}
    \alpha = \alpha(\uu) = \< \uT, \uv \>
    &=
    T_{(0 0 0) (\ell_2 m_2 n_2) (0 0 0)} \,
    v_{(\ell_2 m_2 n_2)} \\
    &=
    T_{(0 0 n_2)} \,
    v_{(0 0 n_2)}.
\end{aligned}
\end{equation}

In the following, we tacitly refer to $T_{(\ell_1 m_1 n_1) (\ell_2 m_2 n_2) (\ell_3 m_3 n_3)}$ (and also $\uuuT$) as \emph{equivariant tensors} without explicitly mentioning the dimension.

It is known that equivariant contractions of higher-order (higher than three) covariant vectors can be expressed through products of the 3-j (or Clebsch-Gordan) coefficients, so we do not have to explicitly consider higher-order equivariant tensors.
This turns into a very interesting coincidence: similarly to how order-3 tensors were sufficient to generate a tensor of any order by repeatedly contracting the order-3 tensors, a third-order equivariant tensor contain the needed information to encode equivariance in any higher-order tensors.
\emph{The algorithmic combination of these two facts is the core of this work}; everything else can be thought of as a corollary of this.

\subsection{Construction of Equivariant Tensor Networks}

Using our definition of equivariant tensors, we now exemplify the construction of equivariant tensor networks (ETNs) using the tensor train format \citep{oseledets_tensortrain_2011}.
We emphasize that, in principle, \emph{any topology of tensor networks tensors can be realized also with the equivariant tensors,} including the already mentioned hierarchical Tucker decomposition \citep{grasedyck_hierarchical_2010}, the ring tensor format \citep{zhao_learning_2019}, or more advanced networks like PEPs \citep{verstraete_renormalization_2004}.

\subsubsection{Equivariant Tensor Trains}

In the tensor train format, a multilinear form is a product of equivariant tensors with covariant vectors, leading to the representation
\begin{equation}\label{eq:ETT}
\begin{aligned}
    \alpha(\uv^1, \ldots, \uv^d)
    =&\,\, \uuT^1 \Big( \,\ldots\, \uuuT^{d-1} \Big( \uuT^d \uv^d \Big) \uv^{d-1} \,\ldots\, \Big) \uv^1 \\
    =&\,\, \left(T^1_{(\ell_1' m_1' n_1') (\ell_1 m_1 n_1)} v^1_{(\ell_1' m_1' n_1')}\right)
           \left(T^2_{(\ell_1 m_1 n_1) (\ell_2' m_2' n_2') (\ell_2 m_2 n_2)} v^2_{(\ell_2' m_2' n_2')}\right) \\
     &\,\, \ldots
           \left(T^d_{(\ell_{d-1} m_{d-1} n_{d-1}) (\ell_d' m_d' n_d')} v^d_{(\ell_d' m_d' n_d')}\right).
\end{aligned}
\end{equation}
We denote this representation as the \emph{equivariant tensor train (ETT)} representation of a contraction of a general ($d$-dimensional) equivariant tensor with $d$ covariant input feature vectors.
Within the ETT representation, every coefficient tensor $\uuuC^i$ corresponding to an equivariant tensor $\uuuT^i$ has
$
\sum_{\ell,\ell',\ell'':{\rm TR}(\ell,\ell',\ell'')} N_{\rm ett}^{\rm rank}(i,\ell) N(i,\ell') N_{\rm ett}^{\rm rank}(i,\ell'')
$
angular momenta channels.
As a boundary condition for $\alpha$ being a scalar, we require that $N_{\rm ett}^{\rm rank}(0,0) = N_{\rm ett}^{\rm rank}(d+1,0) = 1$.
The invariance of $\alpha$ with respect to SO(3) follows from the properties of the equivariant tensors as discussed in the previous sections.

\subsubsection{Invariance under O(3)}
\label{sec:O(3)-invariance}

For our application of interatomic potentials, we also require reflection invariance of the multilinear forms $\alpha$ \eqref{eq:ETT}.
To that end, we recall that, under space inversions $\clP v_{(\ell mn)}(\widehat\ur) = v_{(\ell mn)}(-\widehat\ur)$, we have
\begin{equation}
    v_{(\ell mn)}(\widehat\ur) = \clP^{-1} v_{(\ell mn)}(\widehat\ur) = (-1)^\ell v_{(\ell mn)}(-\widehat\ur).
\end{equation}
To ensure the invariance of $\alpha$ with respect to O(3), we require, according to the previous section, that $\ell'_1 + \ell'_2 + \ldots + \ell'_d$ is even.
Recalling the definition of the 3-j symbol for RSHs, this immediately implies that $\alpha$ is always a \emph{real} quantity since, in this case, the sum over all $\ell$'s
\begin{equation}
    (\ell'_1 + \ell_1) + (\ell_1 + \ell_2' + \ell_2) + \ldots + (\ell_{d-1} + \ell_d') = \sum_{i=1}^d \ell_i' + \sum_{i=1}^{d-1} 2\ell_i
\end{equation}
is obviously also even.
This also implies that we can neglect imaginary parts for ETTs with $<$3 cores since we only need to consider even $\ell_1 + \ell_2' + \ell_2$.
For $>$3 cores, we need to keep track of both real and imaginary parts.
Algorithms \ref{algo:ett-contraction_up-to-3-cores} and \ref{algo:ett-contraction_>3-cores} are possible implementations of these two cases (therein, a tensor with superscripted ``${\rm re}$'' denotes its real part, whereas a tensor with superscripted ``${\rm im}$'' denotes its imaginary part).

\begin{algorithm}[hbt]
    \SetAlgoSkip{bigskip}
    \LinesNumbered
    \SetKwInput{Input}{Input}
    \SetKwInput{Output}{Output}
    \caption{ETT contraction with three or less cores}
    \label{algo:ett-contraction_up-to-3-cores}
    \Input{Input feature vectors $\uv^i$}
    $\uu^{d-1,{\rm re}} \,\leftarrow\, \uuT^{d,{\rm re}} \uv^d$; \\
    \For{$i = d-1, \ldots, 1$}{
        $\uu^{i-1,{\rm re}} \,\leftarrow\, \left(\uuuT^{i,{\rm re}} \uu^{i,{\rm re}}\right) \uv^i$; \\
    }
    \Output{$u^{0,{\rm re}}_0$}
\end{algorithm}

\begin{algorithm}[hbt]
    \SetAlgoSkip{bigskip}
    \LinesNumbered
    \SetKwInput{Input}{Input}
    \SetKwInput{Output}{Output}
    \caption{ETT contraction with four or more cores}
    \label{algo:ett-contraction_>3-cores}
    \Input{Input feature vectors $\uv^i$}
    $\uu^{d-1,{\rm re}} \,\leftarrow\, \uuT^{d,{\rm re}} \uv^d$; \\
    $\uu^{d-2,{\rm re}} \,\leftarrow\, \left(\uuuT^{d-1,{\rm re}} \uu^{d-1,{\rm re}}\right) \uv^{d-1}$; \\
    $\uu^{d-2,{\rm im}} \,\leftarrow\, \left(\uuuT^{d-1,{\rm im}} \uu^{d-1,{\rm re}}\right) \uv^{d-1}$; \\
    \For{$i = d-2, \ldots, 3$}{
        $\uu^{i-1,{\rm re}} \,\leftarrow\, \left( \uuuT^{i,{\rm re}} \uu^{i,{\rm re}} + \uuuT^{i,{\rm im}} \uu^{i,{\rm im}} \right) \uv^i$; \\
        $\uu^{i-1,{\rm im}} \,\leftarrow\, \left( \uuuT^{i,{\rm re}} \uu^{i,{\rm im}} + \uuuT^{i,{\rm im}} \uu^{i,{\rm re}} \right) \uv^i$; \\
    }
    $\uu^{1,{\rm re}} \,\leftarrow\, \left( \uuuT^{2,{\rm re}} \uu^{2,{\rm re}} + \uuuT^{2,{\rm im}} \uu^{2,{\rm im}} \right) \uv^{2}$; \\
    $u_0^{0,{\rm re}} \,\leftarrow\, \left(\uuT^{1,{\rm re}} \uu^{1,{\rm re}}\right) \uv^1$; \\
    \Output{$u_0^{0,{\rm re}}$}
\end{algorithm}

There is another way to understand the real/imaginary splitting of reflection symmetric{\slash}antisymmetric components by simply noting that for the conventional complex-valued harmonics, written in the x,y,z coordinates (such that $x^2+y^2+z^2=1$), it holds that $Y_{\ell m}(x,y,z) = (Y_{\ell m}(x,-y,z))^\ast$, tracking which through our definitions of RSHs and the 3-j symbols yields an alternative proof.

\section{Equivariant Tensor Network Numerical Algebra}\label{sec:algebra}

The power of conventional tensor networks lies in the rich family of algorithms working with those.
In this section we show that all the key algorithms manipulating with conventional tensor networks can be adapted to work with the equivariant tensor networks as well.

\subsection{Trivial algorithms: gradient descent and alternating least squares}

Although having a nontrivial structure, the evaluation of tensor networks reduces to sequential multiplication and addition of numbers, which can be effectively differentiated with respect to tensor core parameters using back propagation---hence, the gradient descent algorithm of optimizing tensor parameters (weights) can trivially (from the mathematical point of view) be formulated.
A nontrivial generalization would be Riemannian gradient descent (i.e., the one with Riemannian gradient), but we leave this to future work.

Another trivially transferable algorithm is alternating least squares (ALS)---the tensor depends linearly on each of the cores $C_{(\ell_1 n_1)(\ell_2 n_2)(\ell_3 n_3)}$ and they can be independently optimized, each time solving a simple quadratic optimization problem.

We next come to more advanced algorithms, generalization of which to the equivariant case is less trivial.

\subsection{QR or SVD normalization of cores}

In the course of optimization, it is beneficial to orthogonalize the cores to avoid, e.g., that the weights unnecessarily grow during the optimization.

\subsubsection{Order-2 cores}\label{sec:order-2-decompositions}
\begin{figure}[htb]
    \centering
    \includegraphics[width=0.9\textwidth]{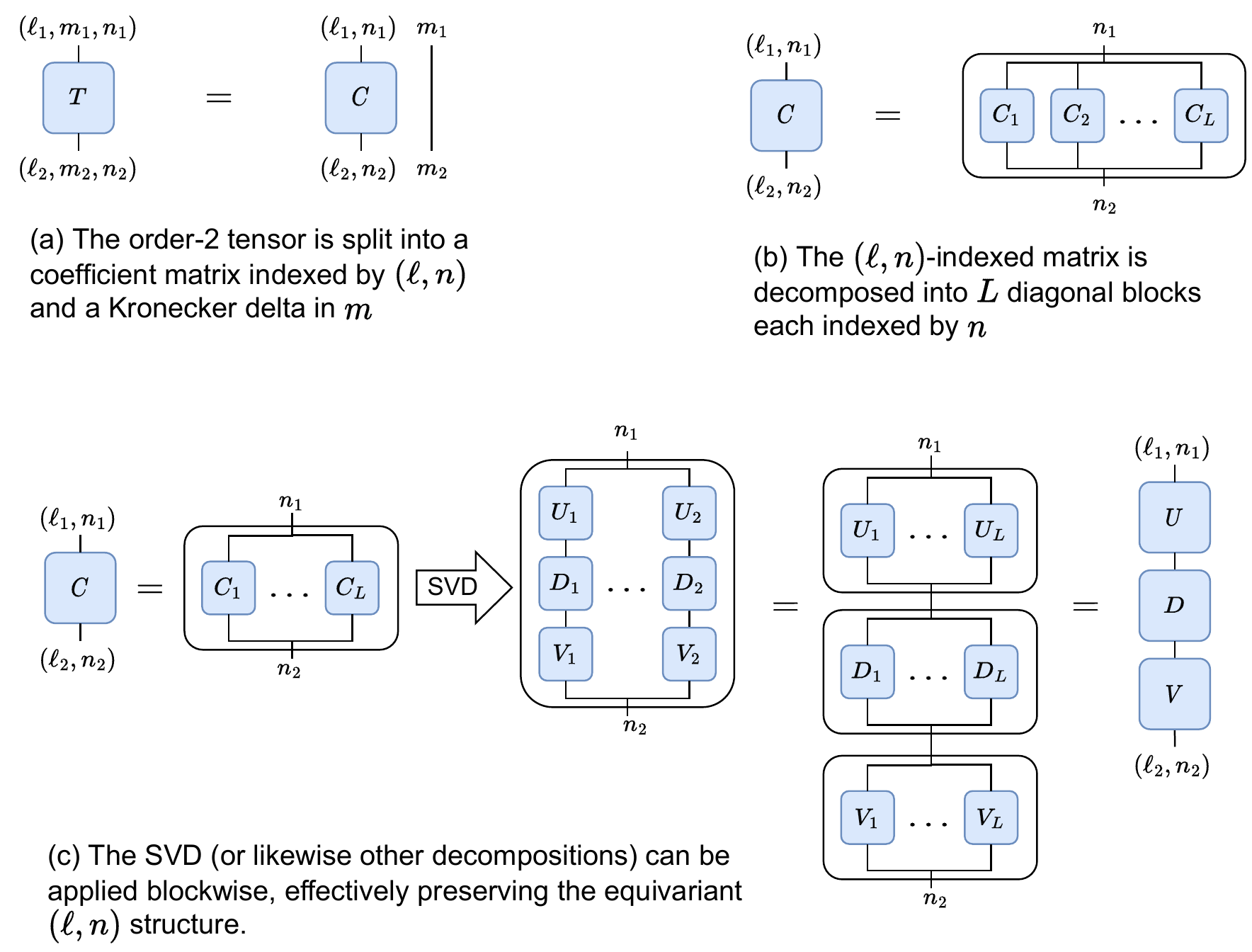}
    \caption{Diagrammatic illustration of concepts in and operations on order-2 tensors.
}
    \label{fig:order-2-algebra}
\end{figure}

To that end, we consider two order-2 cores (cf., Figure \ref{fig:order-2-algebra})
\begin{align*}
\tilde{A}_{(\ell_1 m_1 n_1)(\ell_2 n_2 m_2)}
&= A_{(\ell_1 n_1)(\ell_1 n_2)} \delta_{\ell_1-\ell_2} \delta_{m_1-m_2},\quad\text{and}
\\
\tilde{B}_{(\ell_1 m_1 n_1)(\ell_2 n_2 m_2)}
&= B_{(\ell_1 n_1)(\ell_1 n_2)} \delta_{\ell_1-\ell_2} \delta_{m_1-m_2},
\end{align*}
where $\delta$ is the Kronecker delta symbol.
Notice that we have explicitly kept all three indices $(\ell m n)$ for each tensor dimension and used the Kronecker delta to indicate which indices are effectively equal.
Assuming these tensors are adjacent in the tensor network, they form another order-2 product
\begin{align*}
\tilde{C}_{(\ell_1 m_1 n_1)(\ell_3 n_3 m_3)}
&=
\sum_{\ell_2, m_2, n_2}
A_{(\ell_1 n_1)(\ell_2 n_2)} \delta_{\ell_1-\ell_2} \delta_{m_1-m_2}
B_{(\ell_2 n_2)(\ell_3 n_3)} \delta_{\ell_2-\ell_3} \delta_{m_2-m_3}
\\&=
\delta_{\ell_1-\ell_3} 
\delta_{m_1-m_3} 
\sum_{\ell_2, m_2, n_2}
\delta_{\ell_1-\ell_2} 
\delta_{m_1-m_2} 
A_{(\ell_1 n_1)(\ell_2 n_2)} 
B_{(\ell_2 n_2)(\ell_3 n_3)}.
\\&=
\delta_{\ell_1-\ell_3} 
\delta_{m_1-m_3} 
\sum_{\ell_2, n_2}
\delta_{\ell_1-\ell_2} 
A_{(\ell_1 n_1)(\ell_2 n_2)} 
B_{(\ell_2 n_2)(\ell_3 n_3)}.
\end{align*}
What we see inside the sum is, effectively, a product of two block-diagonal matrices where different values of $\ell$ comprise independent blocks.
We therefore have that
\[
C_{(\ell n_1) (\ell n_3)} =
\sum_{n_2} A_{(\ell n_1) (\ell n_2)} B_{(\ell n_2) (\ell n_3)}
\]
for each $\ell$.
Orthogonalization is thus simple: we can, e.g., get a $QR$ decomposition for $A_{(\ell \bullet) (\ell \bullet)} = QR$ and reassign the cores $A'_{(\ell \bullet) (\ell \bullet)} := Q$ and $B'_{(\ell \bullet) (\ell \bullet)} = R B_{(\ell \bullet) (\ell \bullet)}$ for each $\ell$.
The new $A'$ and $B'$ make up the same product $C$ as the old ones, but now the rows of $A$ are orthogonalized.
Likewise, for each $\ell$ we can apply the SVD (singular value decomposition) $C_{(\ell \bullet) (\ell \bullet)} = U D V^\sT$ and assign
$A'_{(\ell \bullet) (\ell \bullet)} = U D^{1/2}$ and 
$B'_{(\ell \bullet) (\ell \bullet)} = D^{1/2} V^\sT$.
In this case both, rows of $A'$ and columns of $B'$, will be orthogonal.

\subsubsection{Order-3 cores}\label{sec:order-3-decompositions}
We now consider two adjacent order-3 cores.
This is a harder case and requires some preparation.

Consider an order-3 tensor
\[
\tilde{A}_{(\ell_1 m_1 n_1) (\ell_2 m_2 n_2) (\ell_3 m_3 n_3)}
=
A_{(\ell_1 n_1) (\ell_2 n_2) (\ell_3 n_3)}
\begin{Bmatrix}
    \ell_1 & \ell_2 & \ell_3 \\
    m_1 & m_2 & m_3
\end{Bmatrix}
\]
and the corresponding trilinear form
\[
\alpha
=
A_{(\ell_1 n_1) (\ell_2 n_2) (\ell_3 n_3)}
\begin{Bmatrix}
    \ell_1 & \ell_2 & \ell_3 \\
    m_1 & m_2 & m_3
\end{Bmatrix}
u_{(\ell_1 m_1 n_1)} v_{(\ell_2 m_2 n_2)} w_{(\ell_3 m_3 n_3)}.
\]
We later contract this tensor along the third dimension while the first two will remain uncontracted.
This motivates us to reexpand $u_{(\ell_1 m_1 n_1)} v_{(\ell_2 m_2 n_2)}$ as
\[
U_{(\ell m (\ell_1 \ell_2 n_1 n_2))} = 
\sum_{m_1,m_2}
\begin{Bmatrix}
    \ell_1 & \ell_2 & \ell \\
    m_1 & m_2 & m
\end{Bmatrix}
u_{(\ell_1 m_1 n_1)} v_{(\ell_2 m_2 n_2)}.
\]
We fix $\ell_1$ and $\ell_2$, and let $\ell$ go from $|\ell_1-\ell_2|$ to $\ell_1 + \ell_2$.
We then introduce an auxiliary $\ell_1,\ell_2$-indexed family of matricies
\[
(J_{\ell_1,\ell_2})_{(\ell m)(m_1 m_2)}
:= 
\begin{Bmatrix}
    \ell_1 & \ell_2 & \ell \\
    m_1 & m_2 & m
\end{Bmatrix}.
\]
The first dimension of each of these matrices is $\ell m$ and the second one is $m_1 m_2$.
The size of each matrix is exactly $(2\ell_1+1) (2\ell_2+1) \,\times\, (2\ell_1+1) (2\ell_2+1)$.
We denote its inverse as $(J_{\ell_1,\ell_2})^{-1}$ (it exists since the 3-j coefficients form a one-to-one mapping) and hence express
\[
u_{(\ell_1 m_1 n_1)} v_{(\ell_2 m_2 n_2)}
=
(J_{\ell_1,\ell_2})^{-1}_{(m_1 m_2)(\ell m)} U_{(\ell m (\ell_1 \ell_2 n_1 n_2))}.
\]
Our trilinear form $\alpha$, after substitution of $U$ instead of $u$ and $v$ becomes a bilinear form
\begin{align*}
\alpha
&=
A_{(\ell_1 n_1) (\ell_2 n_2) (\ell_3 n_3)}
\begin{Bmatrix}
    \ell_1 & \ell_2 & \ell_3 \\
    m_1 & m_2 & m_3
\end{Bmatrix}
(J_{\ell_1,\ell_2})^{-1}_{(m_1 m_2)(\ell m)} U_{(\ell m (\ell_1 \ell_2 n_1 n_2))} w_{(\ell_3 m_3 n_3)}
\\&=
A_{(\ell_1 n_1) (\ell_2 n_2) (\ell_3 n_3)}
(J_{\ell_1,\ell_2})_{(\ell_3 m_3)(m_1 m_2)}(J_{\ell_1,\ell_2})^{-1}_{(m_1 m_2)(\ell m)} U_{(\ell m (\ell_1 \ell_2 n_1 n_2))} w_{(\ell_3 m_3 n_3)}
\\&=
A_{(\ell_1 n_1) (\ell_2 n_2) (\ell_3 n_3)}
\delta_{\ell-\ell_3} \delta_{m-m_3}
U_{(\ell m (\ell_1 \ell_2 n_1 n_2))} w_{(\ell_3 m_3 n_3)}.
\end{align*}
The last expression should remind us of the order-2 tensor that we have considered in the last subsection.
In other words, if we denote
\[
\hat{A}_{(\ell_3 (\ell_1 \ell_2 n_1 n_2))\,(\ell_3 n_3)} := A_{(\ell_1 n_1) (\ell_2 n_2) (\ell_3 n_3)},
\]
it is the coefficient matrix of the order-2 tensor.

This motivates us to consider the following definition of reshaping an order-3 tensor to an order-2 tensor:
\[
\big({\rm reshape}_{((1,2),3)} A\big)_{(\ell (\ell_1 \ell_2 n_1 n_2))\,(\ell n)}
:= A_{(\ell_1 n_1) (\ell_2 n_2) (\ell n)},
\]
where the index $((1,2),3)$ indicates that we are merging dimension 1 and 2 together.

The rest is relatively straightforward.
If the tensor $\tilde{A}_{(\ell_1 m_1 n_1) (\ell_2 m_2 n_2) (\ell_3 m_3 n_3)}$ is contracted with another tensor $\tilde{B}_{(\ell_3 m_3 n_3) (\ell_4 m_4 n_4) (\ell_5 m_5 n_5)}$, we reshape the latter's coefficients to $
\hat{B}_{(\ell_3 n_3) \, (\ell_3 (\ell_4 \ell_5 n_4 n_5))}
$
and proceed as in Section \ref{sec:order-2-decompositions} for order-2 tensors.
That is, for the QR decomposition we let, independently for each $\ell_3$,
$
\hat{A}
= QR,
$
and assign
$\hat{A}' := Q$ and
$\hat{B}' := R \hat{B}$.
Likewise, for SVD we decompose
$\hat{A} \hat{B} = U D V^\sT$ and assign
$\hat{A}' := U D^{1/2}$ and $\hat{B}' := D^{1/2} V^\sT$.
After setting the new cores $\hat{A}'$ and $\hat{B}'$ we trivially reshape them back to the sought order-3 cores $A'$ and $B'$.

\begin{figure}[htb]
    \centering
    \includegraphics[width=0.9\textwidth]{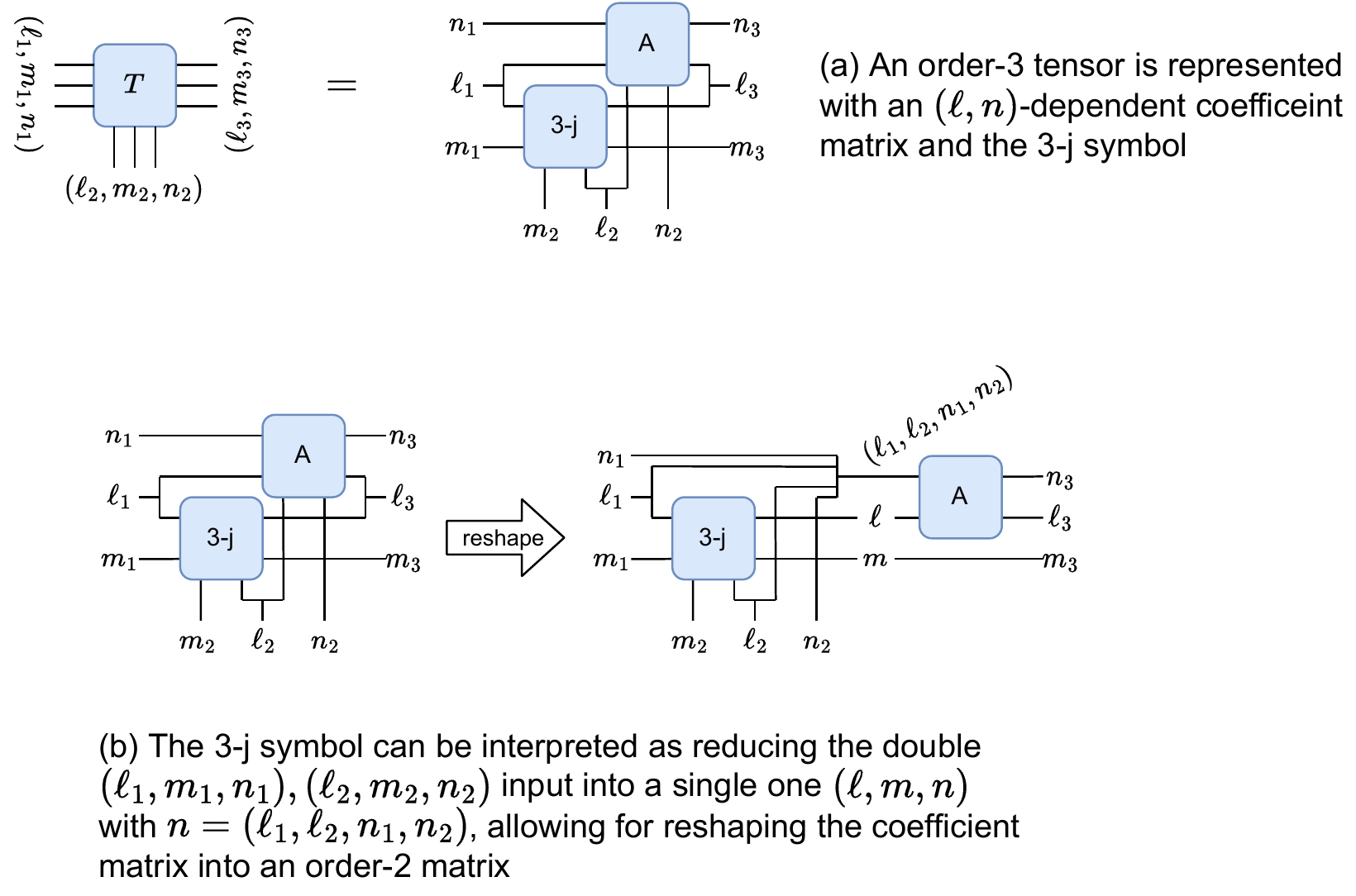}
    \caption{Diagrammatic illustration of concepts in and operations on order-3 tensors.
}
    \label{fig:order-3-algebra}
\end{figure}

The ideas presented above are illustration in Figure \ref{fig:order-3-algebra}.
The case of contracting an order-3 tensor with an order-2 tensor is, in the view of the above, is straightforward: we need to reshape the order-3 tensor and keep it as an order-2 tensor.

\begin{remark}
    Assuming a QR decomposition of $\hat{A}\hat{B}$ such that $\hat{A} = Q$, an interesting property that follows directly from the orthogonality of $\hat{A}$ is the unitarity (more precisely, left-unitarity) of the associated equivariant tensor $A$.
    To see this, let
    \begin{equation}\label{eq:T-unitarity}
    \begin{aligned}
        A^{\sT\ast} A
        :=\;\,&
        A_{(\ell(\ell_1\ell_2 m_1m_2n_1n_2))(\ell mn)}
        A_{(\ell(\ell_1\ell_2 m_1m_2n_1n_2))(\ell' m'n')} \\
        =\;\,&
        \sum_{\substack{\ell_1,m_1,n_1, \\ \ell_2,m_2,n_2}}
        \hat{A}_{(\ell(\ell_1\ell_2 n_1n_2))(\ell n)}
        \begin{Bmatrix}
            \ell_1 & \ell_2 & \ell \\ m_1 & m_2 & m
        \end{Bmatrix}^\ast
        \hat{A}_{(\ell(\ell_1\ell_2 n_1n_2))(\ell' n')}
        \begin{Bmatrix}
            \ell_1 & \ell_2 & \ell' \\ m_1 & m_2 & m'
        \end{Bmatrix} \\
        =\;\,&
        \sum_{\substack{\ell_1,n_1, \\ \ell_2,n_2}}
        \hat{A}_{(\ell(\ell_1\ell_2 n_1n_2))(\ell n)}
        \hat{A}_{(\ell(\ell_1\ell_2 n_1n_2))(\ell' n')}
        \sum_{m_1,m_2}
        \begin{Bmatrix}
            \ell_1 & \ell_2 & \ell \\ m_1 & m_2 & m
        \end{Bmatrix}^\ast
        \begin{Bmatrix}
            \ell_1 & \ell_2 & \ell' \\ m_1 & m_2 & m'
        \end{Bmatrix} \\
        \overset{\mathclap{\eqref{eq:w3j-rsph_orth}}}{=}\;\,&
        \delta_{\ell\ell'} \delta_{mm'}
        \sum_{\substack{\ell_1,n_1, \\ \ell_2,n_2}}
        \hat{A}_{(\ell(\ell_1\ell_2 n_1n_2))(\ell n)}
        \hat{A}_{(\ell(\ell_1\ell_2 n_1n_2))(\ell' n')} \\
        =\;\,&
        \delta_{\ell\ell'} \delta_{mm'} \delta_{nn'}
    \end{aligned}
    ,
    \end{equation}
    where we have adsorbed the prefactor $(2\ell + 1)$ from \eqref{eq:w3j-rsph_orth} into $\hat{A}$.
    Hence, there appears to us no obstacle in extending convergence results for the usual TT-ALS algorithm to the equivariant case; the unitarity property \eqref{eq:T-unitarity} hints that an ETT-ALS algorithm that is stabilized using QR decompositions should be guaranteed to converge (cf., Theorem 1 in \citep{holtz_alternating_2012}).
\end{remark}

\begin{remark}[a potentially more efficient version of SVD-based orthogonalization]
The dimension $\ell_3 (\ell_1 \ell_2 n_1 n_2)$ of the reshaped matrix may be rather large. 
Instead of applying SVD to the product $\hat{A} \hat{B}$ which can be rather large, one can consider an SVD applied separately to $\hat{A} = U_1 D_1 V_1^\sT$, $\hat{B} = U_2 D_2 V_2^\sT$, in which case $\hat{A} \hat{B} = U_1 (D_1 V_1^\sT U_2 D_2) V_2^\sT$.
The expression in the parenthesis is a square matrix with reduced dimension $(\ell_3 n_3)$ which itself can be decomposed as $(D_1 V_1^\sT U_2 D_2) = U_3 D_3 V_3^\sT$ so that the final SVD can be obtained as $\hat{A} \hat{B} = U_1 U_3 D_3 V_3^\sT V_2^\sT$.
\end{remark}

\subsection{Tensor optimization algorithms}

Reshaping equivariant tensors opens the way to applying a number of algorithms that are standard for the conventional tensor networks.
For instance, algorithms based on the Density Matrix Renormalization Group (DMRG) algorithm \citep{white_density_1992,perez-garcia_matrix_2007}, like TT-DMRG-cross \citep{savostyanov_fast_2011} that is applied to the problem of minimizing a quadratic functional of the multidimensional tensor, can now be easily formulated.

Indeed, if we have two adjacent cores $A$ and $B$ (like in Section \ref{sec:order-3-decompositions}), we can reshape and combine these cores into a large matrix $\hat{A} \hat{B}$, with respect to which, with all other cores fixed, the optimization problem is quadratic.
We can hence solve this quadratic optimization problem, apply the SVD algorithm to decompose, reshape back, and this forms the new cores $A'$ and $B'$.

\begin{figure}[htb]
    \centering
    \includegraphics[scale=0.8]{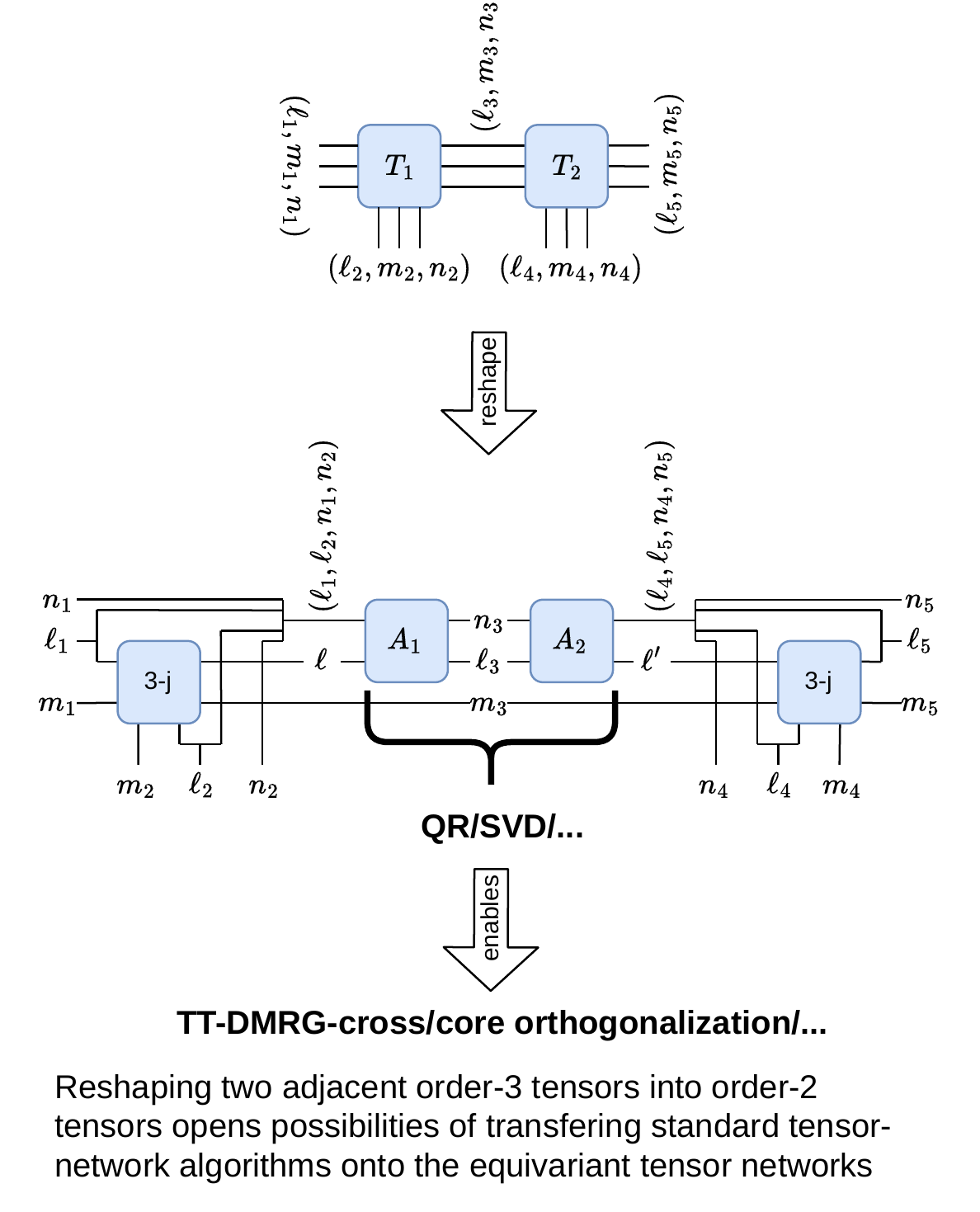}
    \caption{Diagrammatic illustration of how one can conduct QR, SVD, TT-DMRG-cross, or other algorithms on two adjacent order-3 equivariant tensors.
    They are reduced (reshaped) to the product of two order-2 tensors, operations on which (e.g., matrix multiplication) can be done block-wise with respect to $\ell$.
}
    \label{fig:dmrg}
\end{figure}

For example, for a tensor-train representation of $T_{\< d\>}$ given by
\begin{align*}
T_{(\ell_1' m_1' n_1')\ldots(\ell_d' m_d' n_d')} &=
\sum_{(\ell_1 m_1 n_1)\ldots}
\ldots \tilde{T}^{i}_{(\ell_{i-1} m_{i-1} n_{i-1}) (\ell_i' m_i' n_i') (\ell_i m_i n_i)} 
\\&\mathstrut\phantom{=\sum_{(\ell_1 m_1 n_1)\ldots}
\ldots}
\tilde{T}^{i+1}_{(\ell_i m_i n_i) (\ell_{i+1}' m_{i+1}' n_{i+1}') (\ell_{i+1} m_{i+1} n_{i+1})}
\ldots
,
\end{align*}
(compared to \eqref{eq:ETT} our $(\ell m n)$-dependent cores have tildes above them to be consistent with the rest of the notations in this section) the optimization problem ${\mathcal F}(T) \to \min$ when all cores except for $\tilde{T}^{(i-1)}$ and $\tilde{T}^{(i)}$ are fixed, reduces to the problem schematically written as
\[
{\mathcal F}\big(
\ldots \tilde{T}^{i}_{(\ell_{i-1} m_{i-1} n_{i-1}) (\ell_i' m_i' n_i') (\ell_i m_i n_i)} 
\tilde{T}^{i+1}_{(\ell_i m_i n_i) (\ell_{i+1}' m_{i+1}' n_{i+1}') (\ell_{i+1} m_{i+1} n_{i+1})}
\ldots
\big)
.
\]
The coefficients of these two tensors are reshaped to the matrices
$\hat{T}^i_{(\ell' (\ell_{i-1} \ell_i n_{i-1} n_i))\, (\ell' n')}$ and $\hat{T}^{i+1}_{(\ell' n') \, (\ell' (\ell_i \ell_{i+1} n_i n_{i+1}))}$
and then ${\mathcal F}$ is quadratic in terms of $C := \hat{T}^i \hat{T}^{i+1}$.
It is hence solved, decomposed into $U D V^\sT$, and then the updated cores are set as $\hat{T}^i = U D^{1/2}$ and $\hat{T}^{i+1} = D^{1/2} V^\sT$.

The idea of how the TT-DMRG-cross algorithm can be applied to the two adjacent tensor network cores is diagrammatically illustrated in Figure \ref{fig:dmrg}.

\section{Application to Interatomic Potentials}
\label{sec:potentials}

We now use ETNs to construct interatomic potentials.
We present our ETN potentials with the classical way using formulas defining the features, then the functional form, etc., as well as the tensor network diagram notation that we have developed in the previous sections.
This allows us to easier understand the proposed functional forms, but also to compare those of different potentials.
In particular, we present tensor network diagrams for several state-of-the-art MLIPs in Appendix \ref{appdx:TN-MLIPs}, including MTP, SNAP, and ACE.

\subsection{Equivariant Tensor Train Representation of Per-Atom Energies}

Before applying our ETN formalism, we first fix some additional notation: in the following, we denote a configuration of atoms $\ur_{i}$ by $\{ \ur_{i} \}$, and a neighborhood of an atom $\ur_{i}$ by $\{ \ur_{ij} \}$.
We then define the input feature vectors more explicitly as functions of atomic neighborhoods $\{ \ur_{ij} \}$
\begin{equation}
    v_{(\ell mn)} = v_{(\ell mn)}(\{ \ur_{ij} \}) = \sum_j \dot Y_{lm}(\widehat\ur_{ij}) f_n,
\end{equation}
where $\widehat\ur_{ij}$ is the angular contribution to $\ur_{ij}$, and $f_n$ is a vector of pair-wise non-angular features that we will define momentarily.
We can then write the per-atom energy as follows
\begin{equation}
    \clE = \clE(\{ \ur_{ij} \}) =
    \uuT^1 \Big( \,\ldots\, \uuuT^{d-1} \Big( \uuT^d \uv \Big) \uv \,\ldots\, \Big) \uv.
\end{equation}
The total energy of a configuration $\{ \ur_i \}$ is then given as follows
\begin{equation}
 \Pi = \Pi(\{ \ur_i \}) = \sum_i \clE(\{ \ur_{ij} \}).
\end{equation}

\begin{remark}
    In fact, we consider the vectors $\uv' = \big( 1, \uv^\sT \big)^\sT$ as input vectors to the ETT in order to generate complete polynomials.
    We omit this detail as it would require some tedious notation, which is not essential to explain the basic construction of the potentials.
\end{remark}

\subsection{Non-Angular Features}

In the following, we consider multicomponent systems with $N_{\rm spec}$ atomic species.
That is, the feature vectors do not only depend on the atomic positions in the neighborhood of an atom $i$ but also on the species of the $i$-th atom $\uz^i$ and the species of all atoms $\uz^j$ in its neighborhood.
We assume a lexicographical order of the atomic species $s = \{0, 1, 2, \ldots, N_{\rm spec}-1 \}$ and define
\begin{align}\label{eq:type_features}
    z^i_\beta = \delta_{\beta s_i}, &&& z^j_\gamma = \delta_{\beta s_j},
\end{align}
where $s_i$ and $s_j$ are the species of the $i$-th and $j$-th atoms, respectively.
We may then write the non-angular feature vectors as
\begin{equation}
    f_n = f_{(\mu\beta\gamma)}(\norm{\ur_{ij}}, \uz^i, \{\uz^j\}) = Q_\mu(\norm{\ur_{ij}}) z^i_\beta z^j_\gamma,
\end{equation}
where $Q_\mu(\norm{\ur_{ij}})$ are radial basis functions.
The corresponding feature vectors then read
\begin{equation}
    v_{(\ell m\mu\beta\gamma)}(\{ \ur_{ij} \}, \uz^i, \{ \uz^j \})
    = \sum_j \dot Y_{lm}(\widehat\ur_{ij}) Q_\mu(\norm{\ur_{ij}}) z^i_\beta z^j_\gamma.
\end{equation}
A trilinear form of type \eqref{eq:order-3-tensor-contraction} with such $\uv$'s is then given by
\begin{equation}
    \alpha = \alpha(\uu,\uv,\uw) =
    T_{(\ell_1 m_1 n_1) (\ell_2 m_2 \mu\beta\gamma) (\ell_3 m_3 n_3)} \,
    u_{(\ell_1 m_1 n_1)} \,
    v_{(\ell_2 m_2 \mu\beta\gamma)} \,
    w_{(\ell_3 m_3 n_3)}.
\end{equation}

\subsection{Equivariant Tensor Network Potentials}
\label{sec:ETN-potentials}

One problem that arises when defining a potential energy as done in the previous sections is the growth of the parameter space when the number of features becomes large.
Even considering only radial and species features can already lead to a very large parameter space, e.g., for systems with many components, such as high-entropy alloys.
To address this problem, we propose in the following a contraction of features before entering the ETT, leading to \emph{equivariant tensor network (ETN) potentials}.

In order to illustrate this idea, we consider a contraction of only non-angular features and leave the spherical harmonic basis untouched.
To that end, we reshape the input feature vectors $v_{(\ell m \mu\beta\gamma)}$ into three-dimensional covariant tensors
\begin{equation}
    F_{\ell m\mu(\beta\gamma)}
    = \sum_j \dot Y_{\ell m}(\widehat\ur_{ij}) Q_\mu(\norm{\ur_{ij}}) z^i_\beta z^j_\gamma.
\end{equation}
We then contract $\mu$ and $(\beta\gamma)$ before entering the ETT, that is, we define the \emph{contracted feature vectors} as
\begin{equation}\label{eq:reduced_ett_input_vector}
\begin{aligned}
    v_{(\ell m n)}
    &= B_{\ell n\alpha\lambda} A_{\ell\lambda(\beta\gamma)} F_{\ell m\alpha(\beta\gamma)} \\
    &= \sum_j \dot Y_{\ell m}(\widehat\ur_{ij}) \left( B_{\ell n\alpha\lambda} Q_\alpha(\norm{\ur_{ij}}) \left( A_{\ell\lambda(\beta\gamma)} z^i_\beta z^j_\gamma \right) \right),
\end{aligned}
\end{equation}
with the $\ell$-dependent coefficient tensors
\begin{align}
    \uuA_\ell \in \bbR^{N_{\rm spec}^{\rm rank}(\ell) \times N_{\rm spec}^2}
    &&\text{and}&&
    \uuuB_\ell \in \bbR^{N_{\rm rad}^{\rm rank}(\ell) \times N_{\rm rad}(\ell) \times N_{\rm spec}^{\rm rank}(\ell)}.
\end{align}
We remark here that we never explicitly compute $A_{\lambda(\beta\gamma)} z^i_\beta z^j_\gamma$ but rather take the $(z^i_{s_i} z^j_{s_j})$-th column of the matrix $\uuA$, since this is the only nonzero entry in $\uz^i \otimes \uz^j$ according to \eqref{eq:type_features}.
Algorithm \ref{algo:ett-input-vectors} summarizes our implementation of computing $\uv$.

\begin{algorithm}[hbt]
    \SetAlgoSkip{bigskip}
    \LinesNumbered
    \SetKwInput{Input}{Input}
    \SetKwInput{Output}{Output}
    \caption{Computation of the contracted feature vectors for the ETN potential \eqref{eq:tnp}}
    \label{algo:ett-input-vectors}
    \Input{Neighborhood $\{ \ur_{ij} \}$ of an atom $\ur_i$, species features $\uz^i$, $\{\uz^j\}$}
    $\uv \,\leftarrow\, \uNull$; \\
    \For{\textnormal{\bf all} $\ur_{ij} \in \{ \ur_{ij} \}$}{
        \For{$\ell = 0, \ldots, L$}{
            $\ua \,\leftarrow\, \uA_{\ell (z^i_{s_i} z^j_{s_j})}$; \\
            $\ub \,\leftarrow\, (\uuuB_\ell \, \ua) \, \uQ(\norm{\ur_{ij}})$; \\
            $\uv_\ell \,\leftarrow\, \uv_\ell + \operatorname{reshape}_{(1,2)}\left( \uY_\ell(\widehat\ur_{ij}) \otimes \ub \right)$; \\
        }
    }
    \Output{$\uv$}
\end{algorithm}

In the tensor network diagram notation, the per-atom energy then reads
\begin{equation}\label{eq:tnp}
\clE(\uuuF) = \;
\begin{aligned}
    \includegraphics[width=0.7\textwidth]{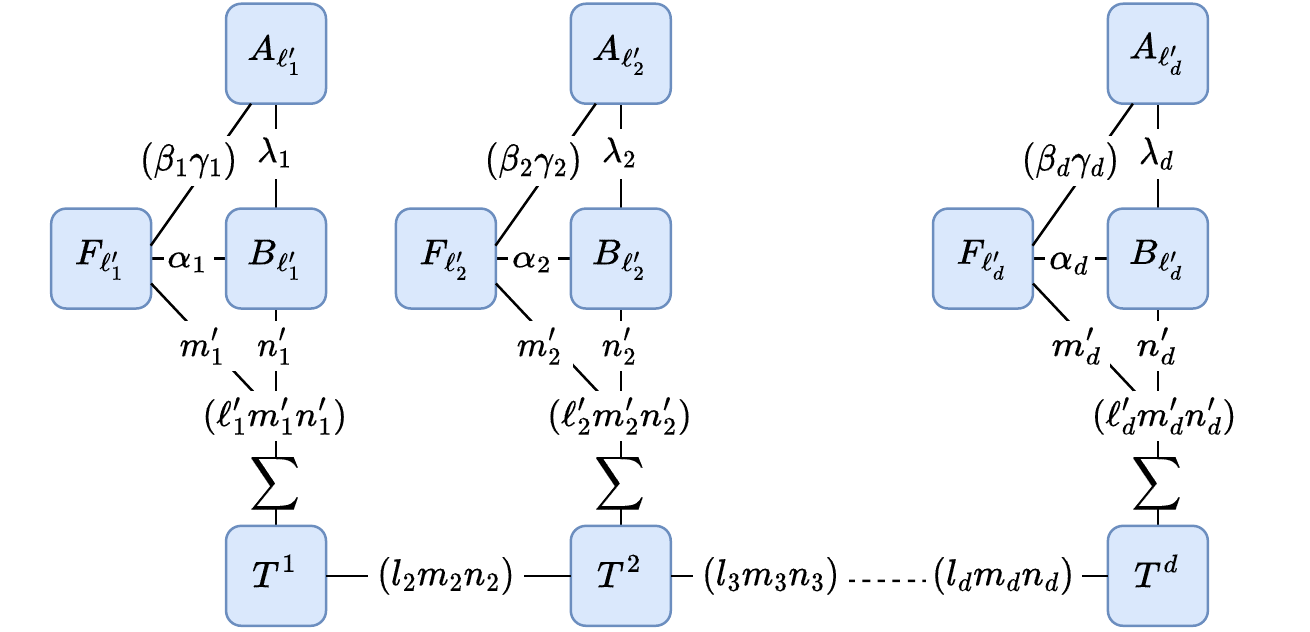}
\end{aligned}.
\end{equation}
This is, of course, not the only possibility to construct a tensor network potential.
For example, if the number of atomic species becomes large, one may envision a separation of $\uz^i$ and $\uz^j$ leading to
\begin{equation}\label{eq:tnp2}
\clE(F_{\< 4\>}) = \;
\begin{aligned}
    \includegraphics[width=0.7\textwidth]{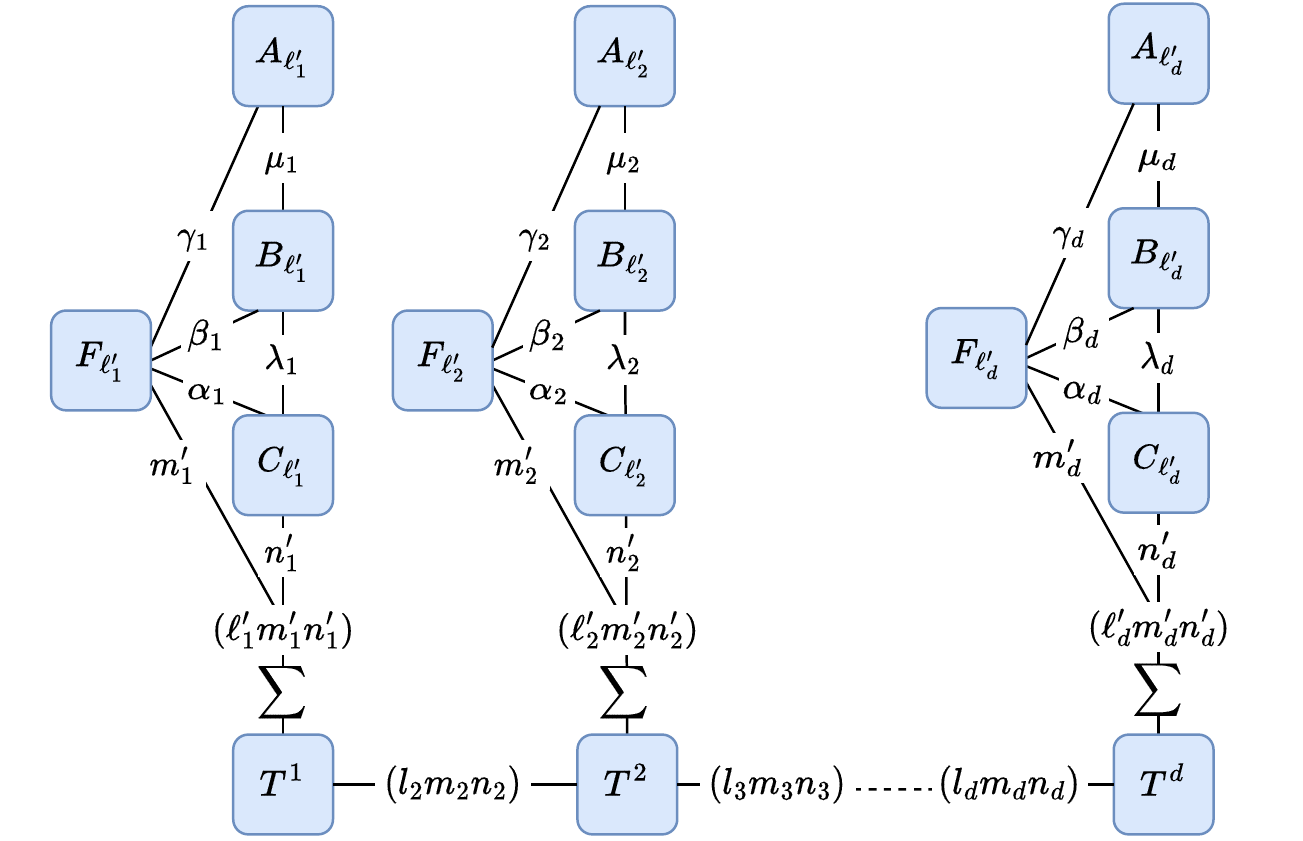}
\end{aligned}.
\end{equation}
We call the $T^i$ tensors in \eqref{eq:tnp} and \eqref{eq:tnp2} ETT cores and the entire lower part the ETT.

We emphasize that we have used weight sharing between ETT cores for the two ETN potentials defined above, i.e., the parameter tensors $\uuA$, $\uuuB$, and $\uuuC$, do not depend on $d$---as opposed to $\uuuT$.
We could have likewise used a core-dependent feature contraction, which could simplify the training since the ETNs then linearly depend on each parameter tensor (refer to our remarks in Section \ref{sec:generalizations}).

\subsection{Hyperparameters of Tensor Network Potentials}

MLIPs contain a number of hyperparameters, such as the cut-off radius, the choice of the radial basis, etc.
In the following, we list the hyperparameters that are specific to ETN potentials and do not occur in other MLIPs.
The ETN-specific hyperparameters are related to the ranks of the tensors $\uuA$, $\uuuB$, and $\uuuT$, as shown in Table \ref{tab:tensor-dimensions-TNP}.
All tensorial dimensions are defined per core $i$ and angular momentum $\ell$.
We remark that the ETT dimension $d$ and the total number of angular momenta for the feature vectors and ETT ranks are also hyperparameters that define the body-order and the polynomial degree of the ETN potential.

\begin{table}[hbt]
    \centering
    \begin{tabular}{|c|c|}
        \hline
        Tensor & Dimensions \\ \hline\hline
        $\uuA_{\ell'}$ & $N_{\rm spec}^{\rm rank}(\ell') \times N_{\rm spec} \times N_{\rm spec}$ \\ \hline
        $\uuuB_{\ell'}$ & $N_{\rm rad}^{\rm rank}(\ell') \times N_{\rm rad}(\ell') \times N_{\rm spec}^{\rm rank}(\ell')$ \\ \hline
        $\uuuT^i_{(\ell m)(\ell'm')(\ell''m'')}$ & $N_{\rm ett}^{\rm rank}(i,\ell) \times N_{\rm rad}^{\rm rank}(\ell') \times N_{\rm ett}^{\rm rank}(i,\ell'')$ \\ \hline
    \end{tabular}
    \caption{Tensorial dimensions for the ETN potential \eqref{eq:tnp}}
    \label{tab:tensor-dimensions-TNP}
\end{table}

\section{Performance of Equivariant Tensor Network Potentials}
\label{sec:ETNP-performance}

\subsection{Optimization of the Hyperparameters}

In order to identify the ``best'' hyperparameters for the ETN potentials, i.e., the hyperparameters that are close-to the Pareto front that minimizes some function with respect to the number of fitting parameters, we propose a \emph{stochastic gradient descent algorithm}.
The function to be minimized could be, e.g., the validation loss, the root-mean-square error of per-atom energies, or some other quantity of interest.
Since the total number of hyperparameters becomes large when the ETT dimension and the total number of angular momenta become large, we first place some assumptions based on our  ``educated guess'' for the selection of the hyperparameters:
\begin{itemize}
    \item
    The total number of angular momenta of the ETT ranks is constant over all cores and, moreover, equal to the total number of angular momenta $L$ of the feature vectors.
    \item
    The total number of angular momenta channels of the ETT ranks are constant over all cores, i.e., $N_{\rm ett}^{\rm rank}(i,\ell) = N_{\rm ett}^{\rm rank}(\ell)$.
    \item
    Any number of angular momentum channels $N(\ell)$ decays exponentially with increasing $\ell$ according to the function
    \begin{equation}\label{eq:decay-channels}
        N(\ell) = (N(0) - 1) \mathrm{e}^{-\ell/a} + 1
    \end{equation}
    rounded to the nearest integer, where $a$ specifies the decay.
    Specifically, we set $a = 1$ for $N_{\rm rad}(\ell)$, and $a = 5$ for $N_{\rm spec}^{\rm rank}(\ell)$, $N_{\rm ett}^{\rm inp}(\ell)$, and $N_{\rm ett}^{\rm rank}(\ell)$.
    So, the number of radial functions is assumed to decay slowly with $\ell$, while all the ranks decay fast with increasing $\ell$.
\end{itemize}
Based on the previous assumptions we are left with the following six hyperparameters as degrees of freedom:
\begin{itemize}
    \item
    the maximum number of radial functions $N_{\rm rad} = N_{\rm rad}(0)$,
    \item
    the order/dimension $d$ of the ETT,
    \item
    the maximum number of angular momenta $L$,
    \item
    the maximum species rank $N_{\rm spec}^{\rm rank} = N_{\rm spec}^{\rm rank}(0)$,
    \item
    the maximum radial rank $N_{\rm rad}^{\rm rank} = N_{\rm rad}^{\rm rank}(0)$,
    \item
    the maximum ETT rank $N_{\rm ett}^{\rm rank} = N_{\rm ett}^{\rm rank}(0)$.
\end{itemize}
We thus define the set of hyperparameters as
\begin{equation}
    \scH = \left\{ N_{\rm rad}, d, L, N_{\rm spec}^{\rm rank}, N_{\rm rad}^{\rm rank}, N_{\rm ett}^{\rm rank} \right\}.
\end{equation}

Given an ETN potential with hyperparameters $\scH$, we then define the loss function
\begin{equation}
    \scL(\scH,\umtheta,\scT) =
    \frac{1}{\#\scT}
    \sum_{\{ \ur_{i} \} \in \scT} \left(
	w_{\rm e} \Big( \Pi (\{ \ur_{i} \}) - \Pi^{\rm qm}(\{ \ur_{i} \}) \Big)^2
	+
	w_{\rm f} \sum_{\ur_i \in \{ \ur_{i} \}} \| \uf_{r_i} - \uf^{\rm qm}_{r_i} \|^2 
    \right)
    ,
\end{equation}
where $\#\scT$ is the size of the training set, $\umtheta$ is the vector of fitting parameters, $\scT$ is the dataset set which may contain quantum-mechanical energies $\Pi^{\rm qm}$ and forces $\uf^{\rm qm}$ of a configuration, and $w_{\rm e}$ and $w_{\rm f}$ are weighting parameters.
We assume in the following a training set $\scT^{\rm train}$ and a validation set $\scT^{\rm valid}$.
We exemplify the optimization of the hyperparameters by minimizing the mean loss of $\scT^{\rm valid}$ with respect to $\scH$.
Our algorithm is then as follows:
\begin{enumerate}[label={\bf [S\arabic*]}]
    \item
    We start with a pair potential with five radial basis functions and minimal ranks, i.e., $\scH = \{5,1,1,1,1,1\}$.
    \item
    We then minimize the loss of $\scT^{\rm train}$ $N$ times using different random initializations of the parameters $\umtheta$ and compute the mean of the validation loss
    \begin{equation}
        \bar{\scL}^\mathrm{valid}(\scH) = \frac{1}{N} \sum_{i=1}^N \scL(\scH,\umtheta_i,\scT^{\rm valid}).
    \end{equation}
    \item
    We then increase each of the hyperparameters by 1 individually and denote these updates by $\scH + \Delta_i\scH$ ($i=1,\ldots,6$).
    For each hyperparameter update we minimize the training loss $N$ times and compute the mean validation losses $\bar{\scL}^\mathrm{valid}(\scH + \Delta_i\scH)$.
    \item
    For each hyperparameter update, we compute the stochastic gradient of the mean validation loss as follows
    \begin{equation}\label{eq:parameter_gradient}
        \grad{i}{} \bar{\scL}^\mathrm{valid} = - \frac{\log\bar{\scL}^\mathrm{valid}(\scH + \Delta_i\scH) - \log\bar{\scL}^\mathrm{valid}(\scH)}{\log\#\umtheta(\scH + \Delta_i\scH) - \log\#\umtheta(\scH)},
    \end{equation}
    where $\#\umtheta(\scH)$ and $\#\umtheta(\scH + \Delta_i\scH)$ are the number of parameters before and after the $i$-th hyperparameter update.
    Note that we use logarithms to adjust the order of $\bar{\scL}^\mathrm{valid}$ and $\#\umtheta$. 
    \item
    We now increase $\scH$ by the $\Delta_i\scH$ that maximizes \eqref{eq:parameter_gradient} and return to {\bf [S3]}.
\end{enumerate}
We repeat the steps {\bf [S3]}--{\bf [S5]} until the mean validation loss is as small as we want it to be.

\subsection{Numerical Examples}\label{sec:numerical}

We test the algorithm on various datasets for multicomponent systems from the literature.
For the optimization of the loss function, we use python's plain BFGS solver.
During optimization, we monitor the validation loss. After 1000 iterations, we stop the optimization and select the minimal validation loss from all the iterations.
We repeat the optimization three times to compute the mean validation loss.
For the weights we choose
\begin{align}
    w_{\rm e} = 1 \, {\rm eV^2}, && w_{\rm f} = 0.01 \, {\rm (eV/\text{\AA})^2}.
\end{align}
Further, for all computational results, presented in the following, we use a cut-off radius of 5\,{\AA}, if not specified otherwise.

We test our algorithm on various datasets from the literature for multicomponent systems and compare the performance of the ETN potentials to moment tensor potentials (MTPs) in terms of the required number of parameters; we refer to Appendix \ref{appdx:MTPs} for the functional form of MTPs.
We will analyze the error in the validation loss and the root-mean-square error for the per-atom energies and forces, referred to as $RMSE$ and $RMSF$ in the following.

\subsubsection{QM7}

The QM7 dataset \citep{blum_970_2009,rupp_fast_2012} contains 7211 small molecules with up to 23 atoms and six atomic species (C, N, O, S, Cl, and H) in the their equilibrium position.
We select 6186 configurations for the training set, and 1025 configurations for the validation set.
We fit to total energies.

\begin{figure}[t!]
    \begin{minipage}{0.5\textwidth}
        \centering
        (a)
    \end{minipage}\hfill
    \begin{minipage}{0.5\textwidth}
        \centering
        (b)
    \end{minipage}\\[0.2em]
    \begin{minipage}{0.5\textwidth}
        \centering
        \includegraphics[width=0.8\textwidth]{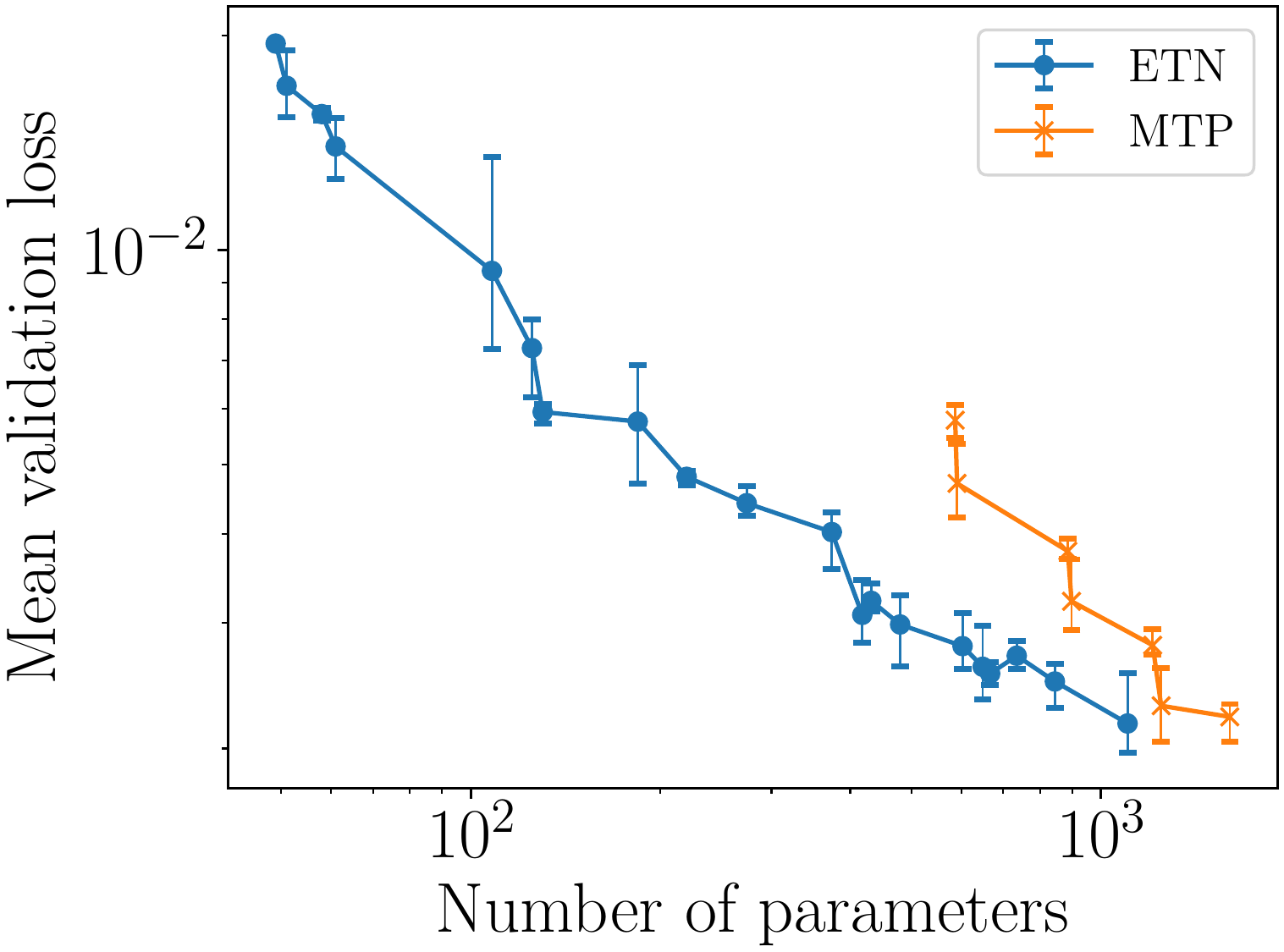}
    \end{minipage}\hfill
    \begin{minipage}{0.5\textwidth}
        \centering
        \includegraphics[width=0.8\textwidth]{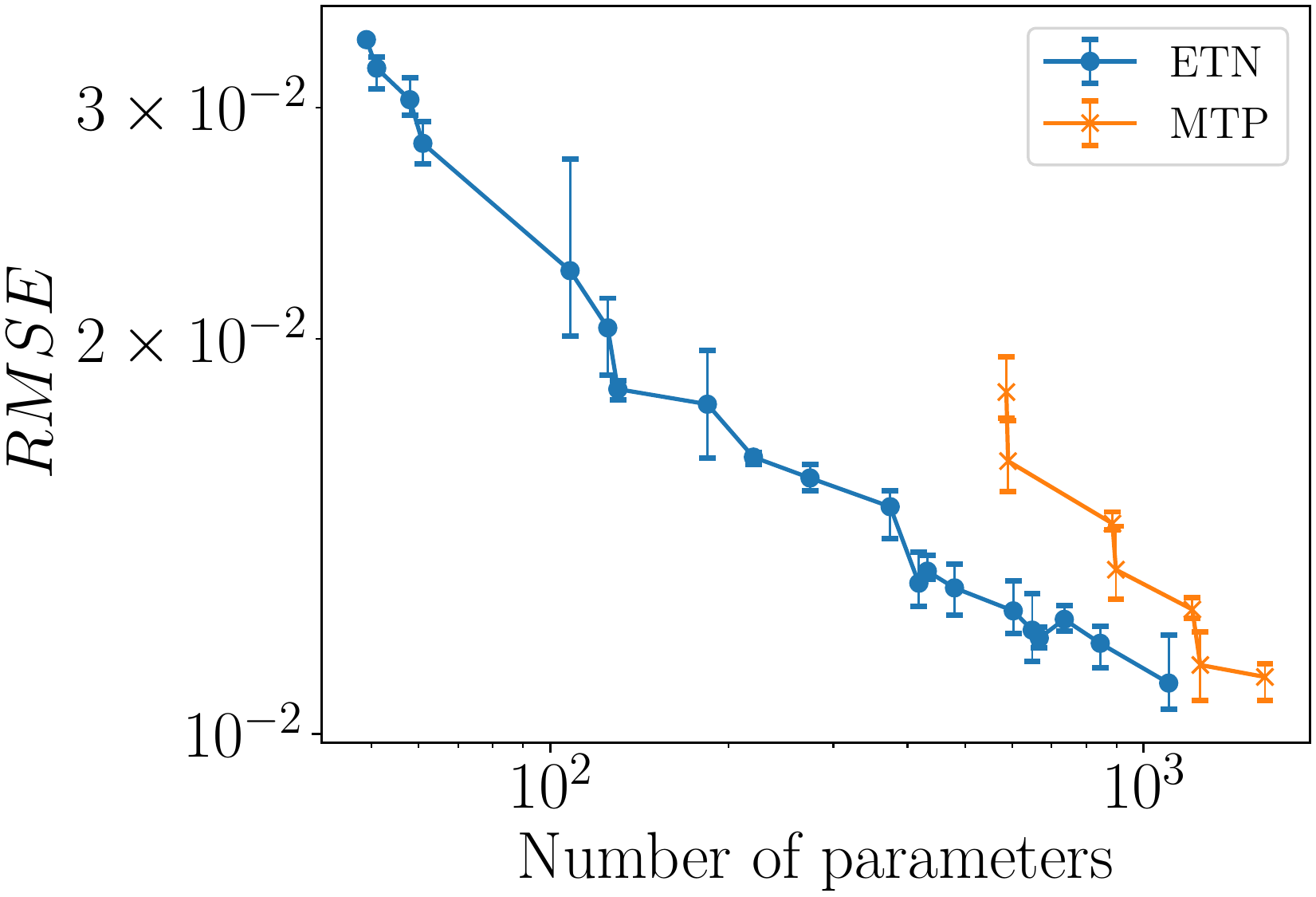}
    \end{minipage}
    \caption{(a) Mean validation loss, and (b) $RMSE$, as a function of the number of parameters for QM7.
    The horizontal bars correspond to the minimum and maximum values}
    \label{fig:results_QM7}
\end{figure}

The decay of the mean validation loss, computed using the stochastic gradient descent algorithm, is shown in Figure \ref{fig:results_QM7} (a).
In general, the ETN potentials require much fewer parameters than MTPs for a comparable level of accuracy, especially the ETN potentials with lower complexity.

\subsubsection{QM9}
\label{sec:QM9}

The QM9 dataset \citep{ruddigkeit_enumeration_2012,ramakrishnan_quantum_2014} contains 134k small organic molecules with up to five atomic species (C, H, O, N, F) in the their equilibrium position.
We select a random subset of 20k configurations from the full dataset.
From these 20k configurations, we use 17k configurations for the training set and 3k configurations for the validation set.
We fit to total energies.

\begin{figure}[t!]
    \begin{minipage}{0.5\textwidth}
        \centering
        (a)
    \end{minipage}\hfill
    \begin{minipage}{0.5\textwidth}
        \centering
        (b)
    \end{minipage}\\[0.2em]
    \begin{minipage}{0.5\textwidth}
        \centering
        \includegraphics[width=0.8\textwidth]{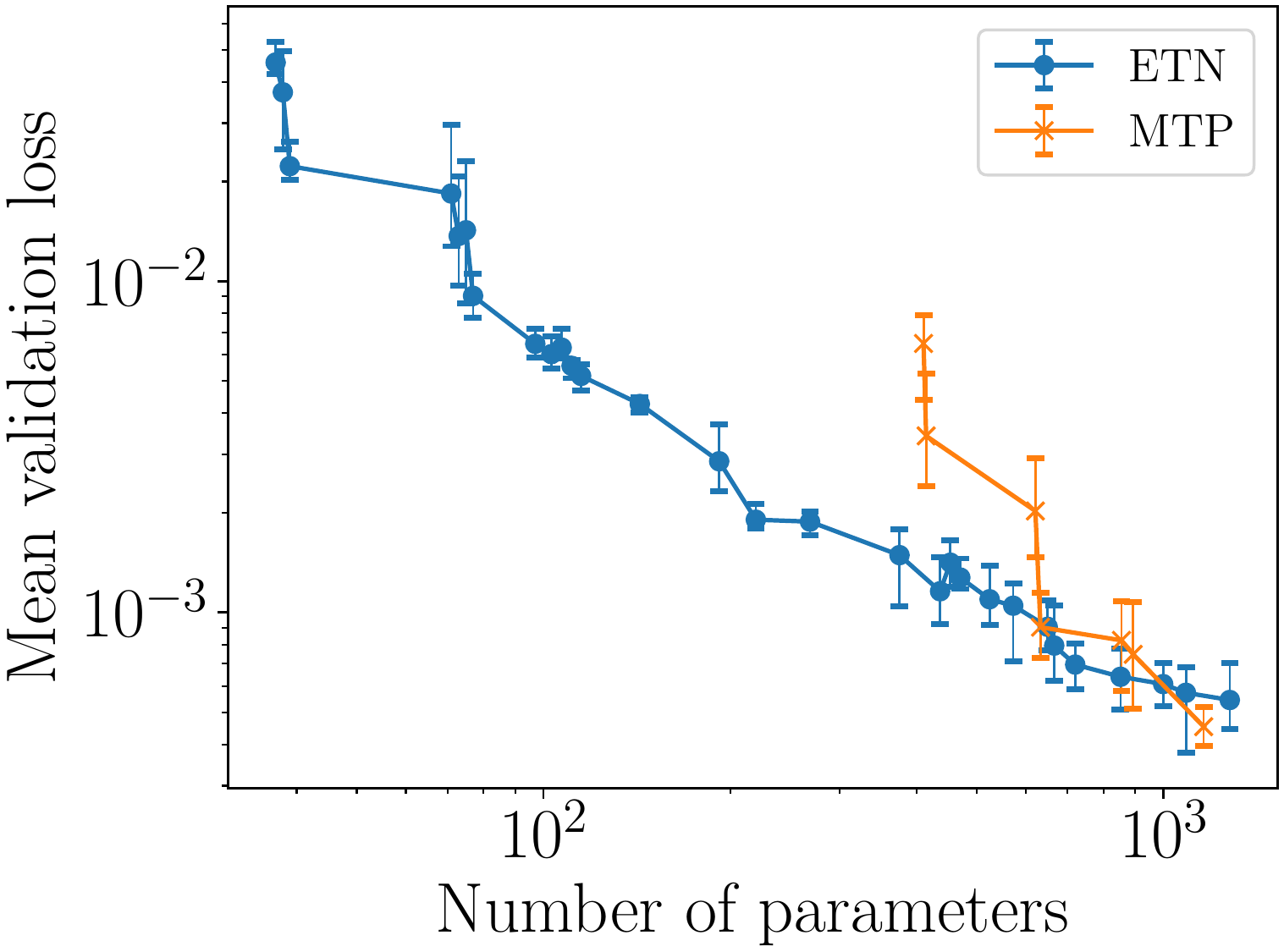}
    \end{minipage}\hfill
    \begin{minipage}{0.5\textwidth}
        \centering
        \includegraphics[width=0.8\textwidth]{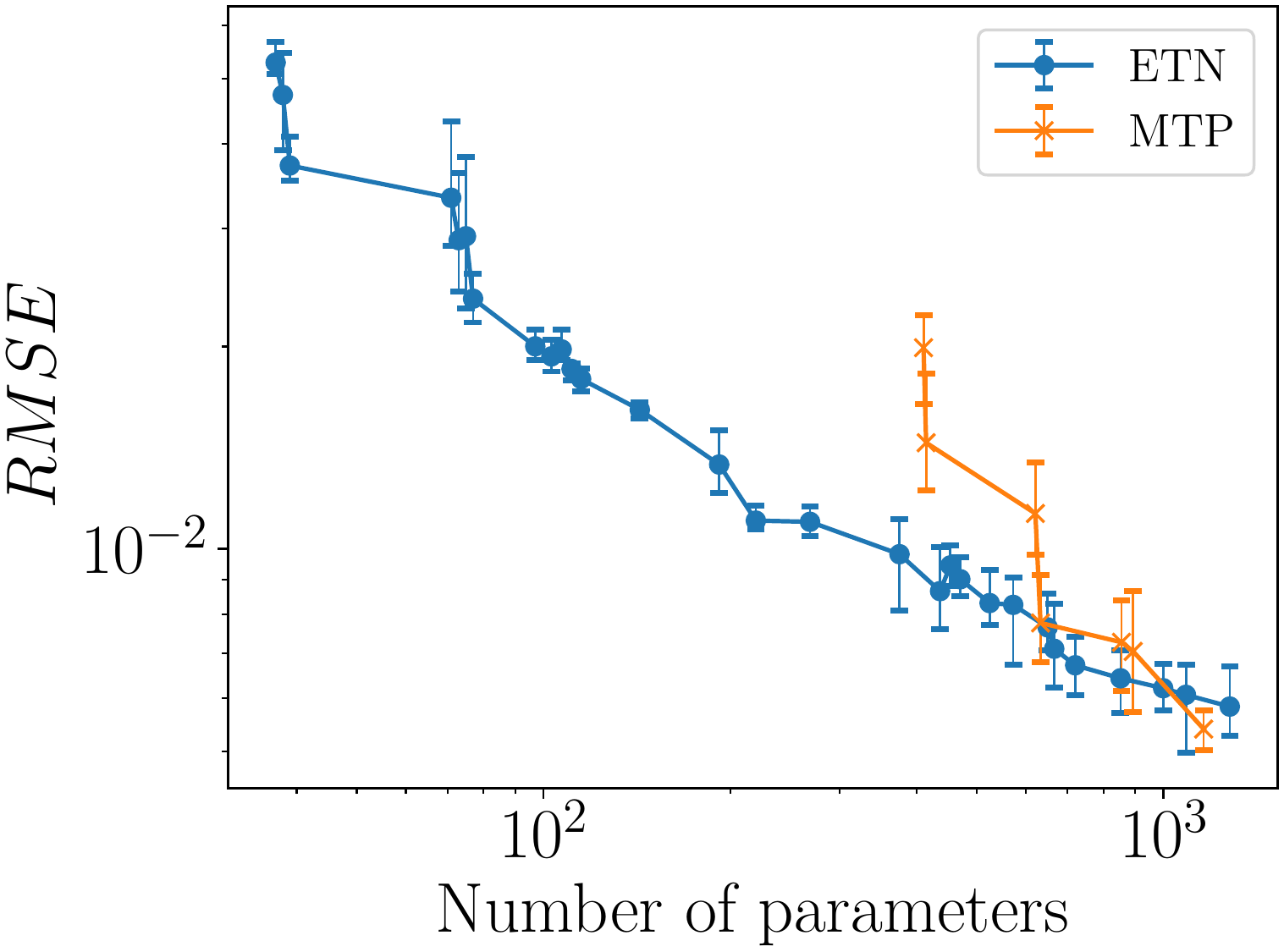}
    \end{minipage}
    \caption{(a) Mean validation loss, and (b) $RMSE$, as a function of the number of parameters for QM9.
    The horizontal bars correspond to the minimum and maximum values}
    \label{fig:results_QM9}
\end{figure}

As for QM7, ETN potentials with lower complexity require much fewer parameters than MTPs.
However, when approaching the realm of higher accuracy ($RMSE$ $<$\,10\,meV/atom, cf., Figure \ref{fig:results_QM9} (b)), there is no difference anymore between ETN potentials and MTPs.
Presently, we attribute this to python's BFGS solver not being the optimal choice for training ETN potentials.
Indeed, we have observed premature convergence (after $\sim$\,50 iterations) for approximately 50\,\% of the training runs per iteration when the ETN potentials reach a higher complexity (i.e., from $\sim$\,400--500 coefficients).
On the other hand, python's BFGS is also not optimal for MTPs, but MTPs already have a more mature functional form that appears to perform better on QM9.

\subsubsection{MoNbTaW Medium-Entropy Alloy}

This dataset has been developed to construct a spectral neighbor analysis potential (SNAP) for the MoNbTaW medium-entropy alloy \citep{li_complex_2020}.
Hence, we refer to this dataset as MoNbTaW-MEA in the following.
The dataset contains 5529 configurations containing various configuration types, e.g., configurations with free surfaces, snapshots from molecular dynamics simulations at finite temperature, etc.
We use 4983 configurations for the training set and 546 for the validation set (the validation set is constructed such that every configuration type is represented).
We fit to total energies and forces.

\begin{figure}[t!]
    \begin{minipage}{0.5\textwidth}
        \centering
        (a)
    \end{minipage}\hfill
    \begin{minipage}{0.5\textwidth}
        \centering
        (b)
    \end{minipage}\\[0.2em]
    \begin{minipage}{0.5\textwidth}
        \centering
        \includegraphics[width=0.8\textwidth]{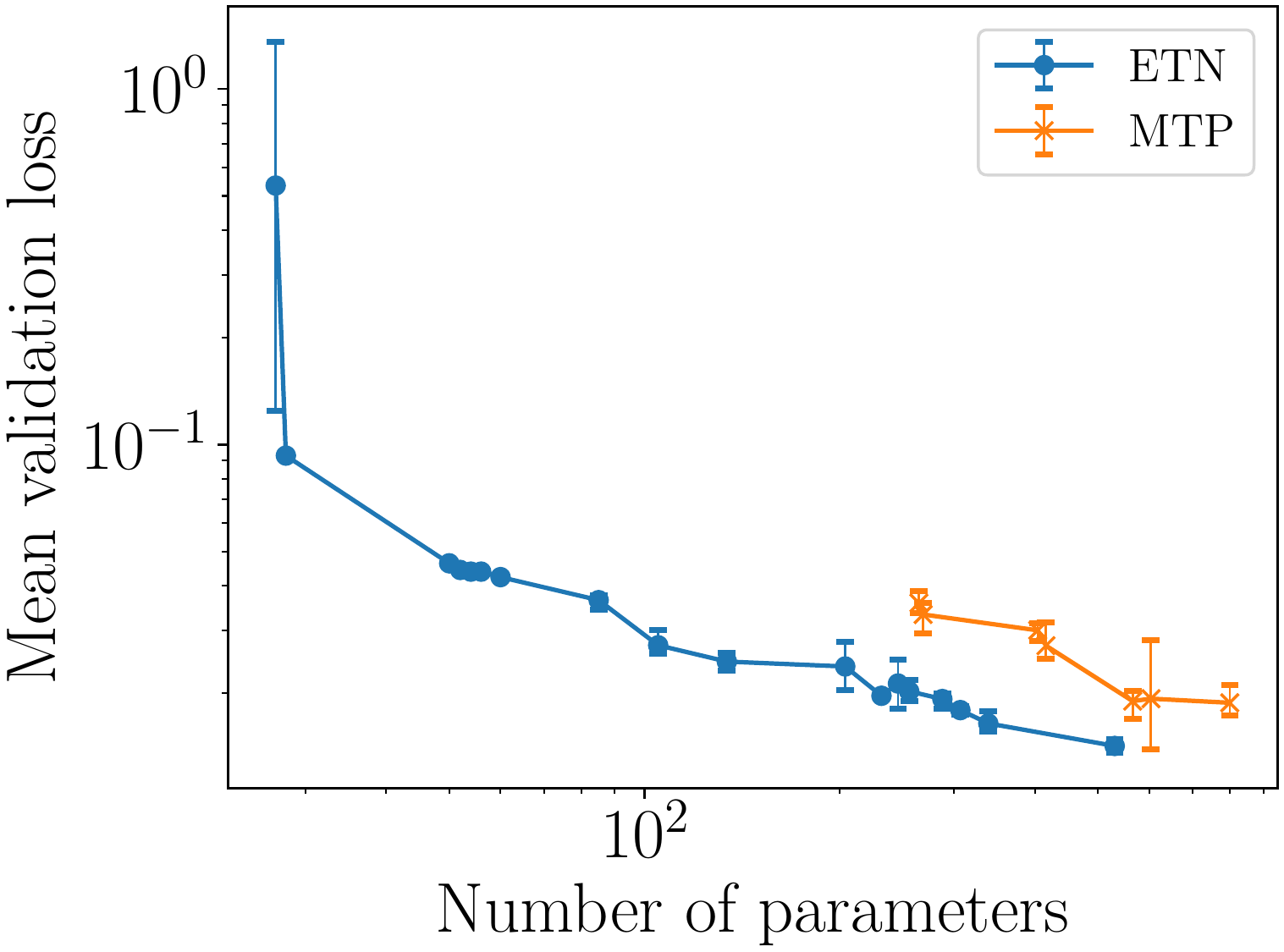}
    \end{minipage}\hfill
    \begin{minipage}{0.5\textwidth}
        \centering
        \includegraphics[width=0.8\textwidth]{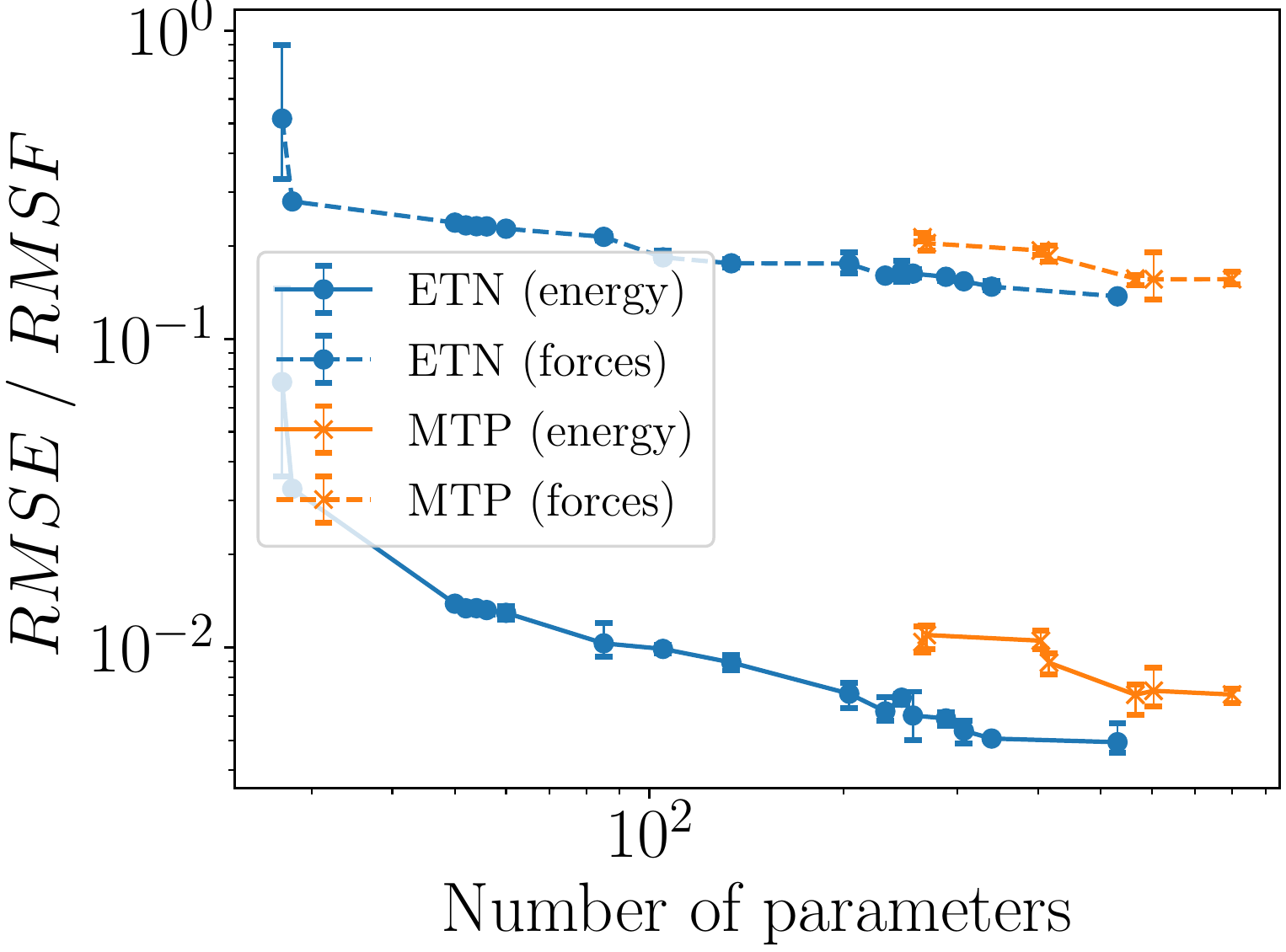}
    \end{minipage}
    \caption{(a) Mean validation loss, and (b) $RMSE$/$RMSF$, as a function of the number of parameters for MoNbTaW-MEA.
    The horizontal bars correspond to the minimum and maximum values}
    \label{fig:results_MoNbTaW-MEA}
\end{figure}

Over the entire range of potentials, the amount of coefficients is universally lower than the amount of coefficients for MTPs when both potentials give similar accuracy, as shown in Figure \ref{fig:results_MoNbTaW-MEA} (a) and (b).
Even when approaching the realm of high accuracy ($RMSE \sim \text{5--6}\,\text{meV/atom}$, $RMSF \sim \text{150--180}\,\text{meV/\AA}$), ETN potentials require 2--3 times less parameters than MTPs.
We attribute this excellent performance to the ability of ETN potentials to learn similarities between atomic species by varying the species rank $N_{\rm spec}^{\rm rank}$ since all atoms belong to the group of transition metals.

\subsubsection{MoNbTaVW High-Entropy Alloy}

This dataset, henceforth referred to as MoNbTaVW-HEA, appears to us the first dataset generated for a high-entropy alloy (five components or more) and is more general than the MoNbTaW-MEA in the sense that it contains a more diverse set of configurations, including various defects (vacancies, di- and tri-vacancies, self-interstitial atoms, stacking faults) and liquid states \citep{byggmastar_modeling_2021}.
The whole dataset contains 2859 configurations.
The dataset also contains many out-of-equilibrium configurations with very large forces (up to $\sim$\,150\,eV/{\AA}), such as dimers.
We have removed configurations that contain forces $>$\,5\,eV/{\AA} leading to a subset of the full dataset containing 1423 configurations.
From this subset we use 1210 configurations for training, and 213 for validation.
We fit to total energies and forces.

\begin{figure}[t!]
    \begin{minipage}{0.5\textwidth}
        \centering
        (a)
    \end{minipage}\hfill
    \begin{minipage}{0.5\textwidth}
        \centering
        (b)
    \end{minipage}\\[0.2em]
    \begin{minipage}{0.5\textwidth}
        \centering
        \includegraphics[width=0.8\textwidth]{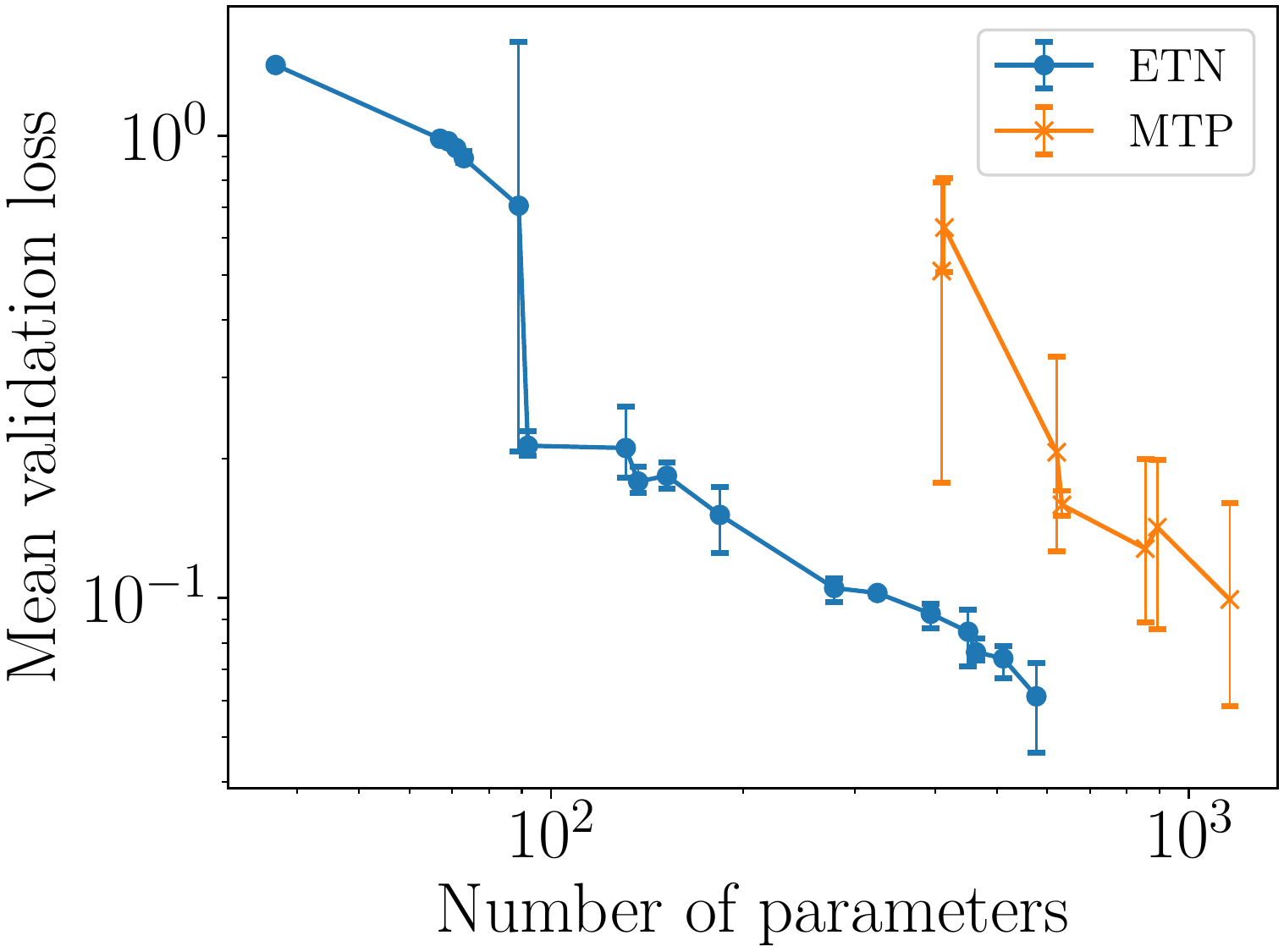}
    \end{minipage}\hfill
    \begin{minipage}{0.5\textwidth}
        \centering
        \includegraphics[width=0.8\textwidth]{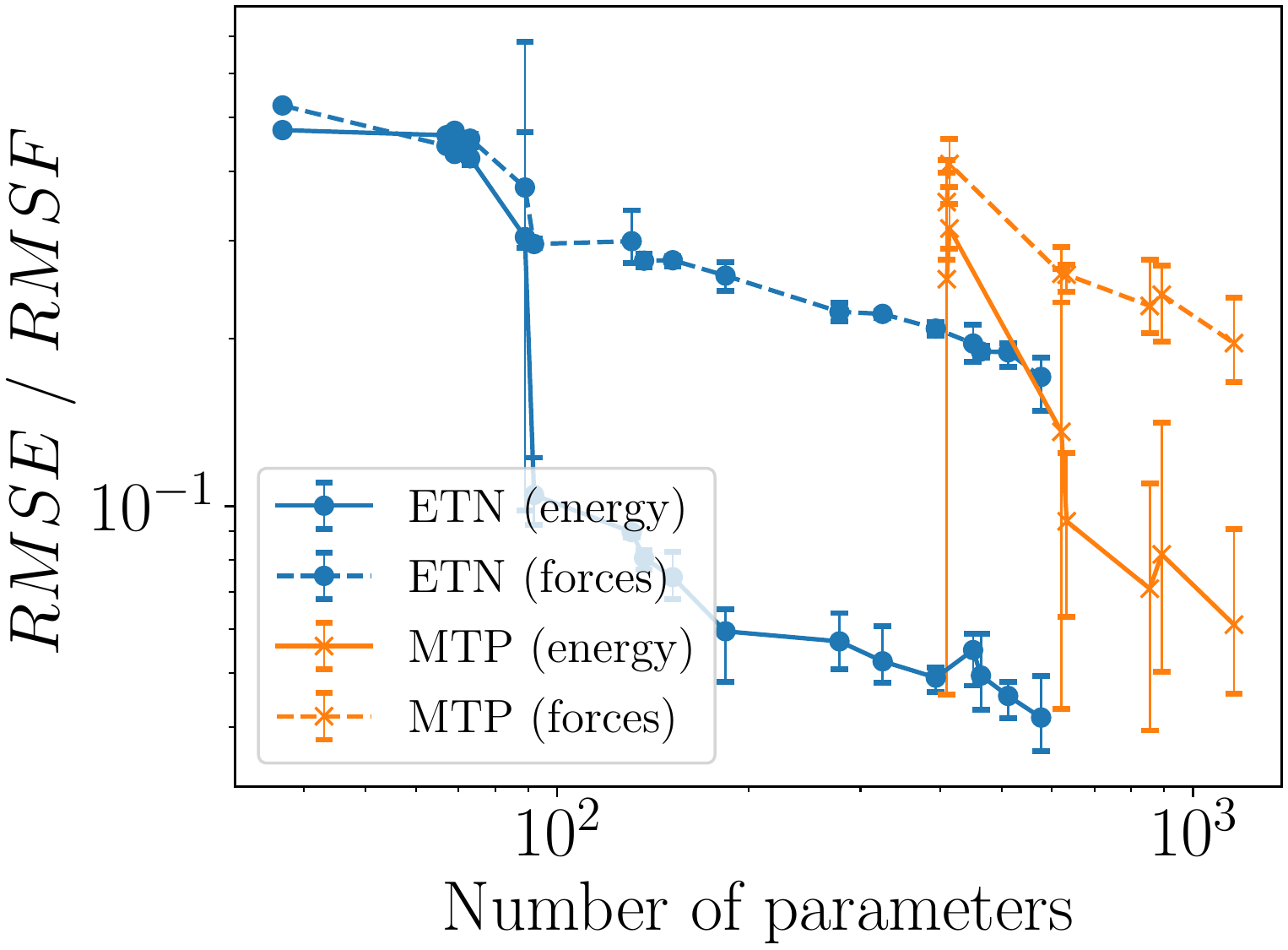}
    \end{minipage}
    \caption{(a) Mean validation loss, and (b) $RMSE$/$RMSF$, as a function of the number of parameters for MoNbTaVW-HEA.
    The horizontal bars correspond to the minimum and maximum values}
    \label{fig:results_MoNbTaVW-HEA}
\end{figure}

From Figure \ref{fig:results_MoNbTaVW-HEA}, we observe that the performance of ETN potentials is similar to MoNbTaW-MEA.
In comparison to MTPs, ETN potentials require many times less coeffients to reach a comparable mean validation loss. 
We yet remark that, for MoNbTaVW-HEA, it now appears to be more difficult to fit MTPs as the fluctuation in the loss function is much larger than for ETN potentials.
Nevertheless, even when comparing the \emph{mean} validation loss of ETN potentials and the \emph{minimum} validation loss of MTPs, ETN potentials still require 2--3 times less coefficients when both potentials give comparable accuracy.

\subsubsection{Binary Alloys with 10 different Species}

We now test the performance of ETN potentials on a dataset containing 10 binary alloys \citep{nyshadham_machinelearned_2019}, with 10 different atomic species in total (AgCu, AlFe, AlMg, AlNi, AlTi, CoNi, CuFe, CuNi, FeV, and NbNi). We refer to it as BA10 in the following.
BA10 contains all possible fcc, bcc, and hcp structures, with up to eight atoms in the unit cell.
In total, BA10 contains 15950 configurations.
We select 13558 configurations that serve as the training set, the remaining 2392 configurations serve as the validation set.
Since BA10 only contains equilibrium structures we only fit to total energies.
Further, we choose a cut-off radius of 7\,{\AA}.

\begin{figure}[t!]
    \begin{minipage}{0.5\textwidth}
        \centering
        (a)
    \end{minipage}\hfill
    \begin{minipage}{0.5\textwidth}
        \centering
        (b)
    \end{minipage}\\[0.2em]
    \begin{minipage}{0.5\textwidth}
        \centering
        \includegraphics[width=0.8\textwidth]{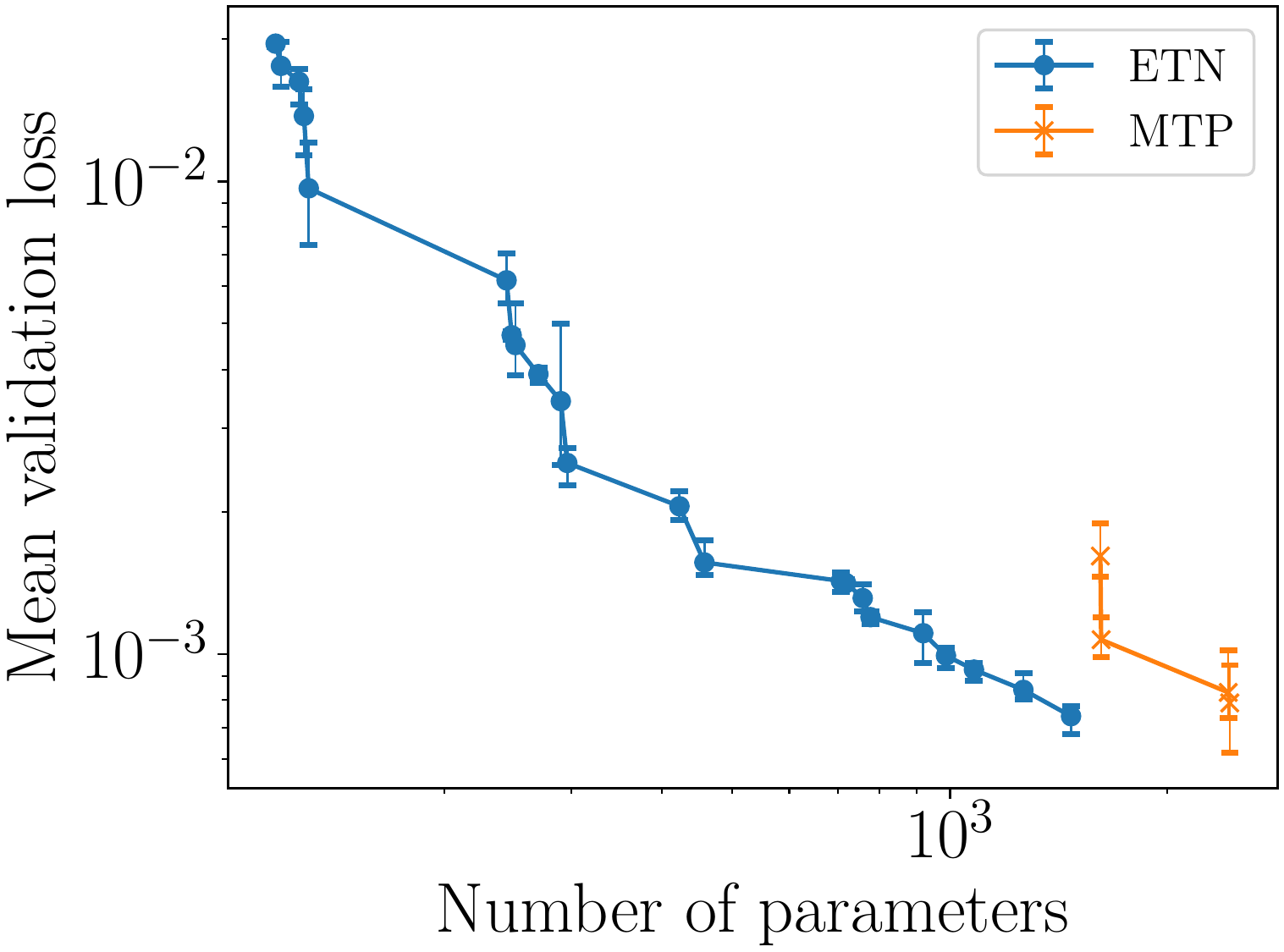}
    \end{minipage}\hfill
    \begin{minipage}{0.5\textwidth}
        \centering
        \includegraphics[width=0.8\textwidth]{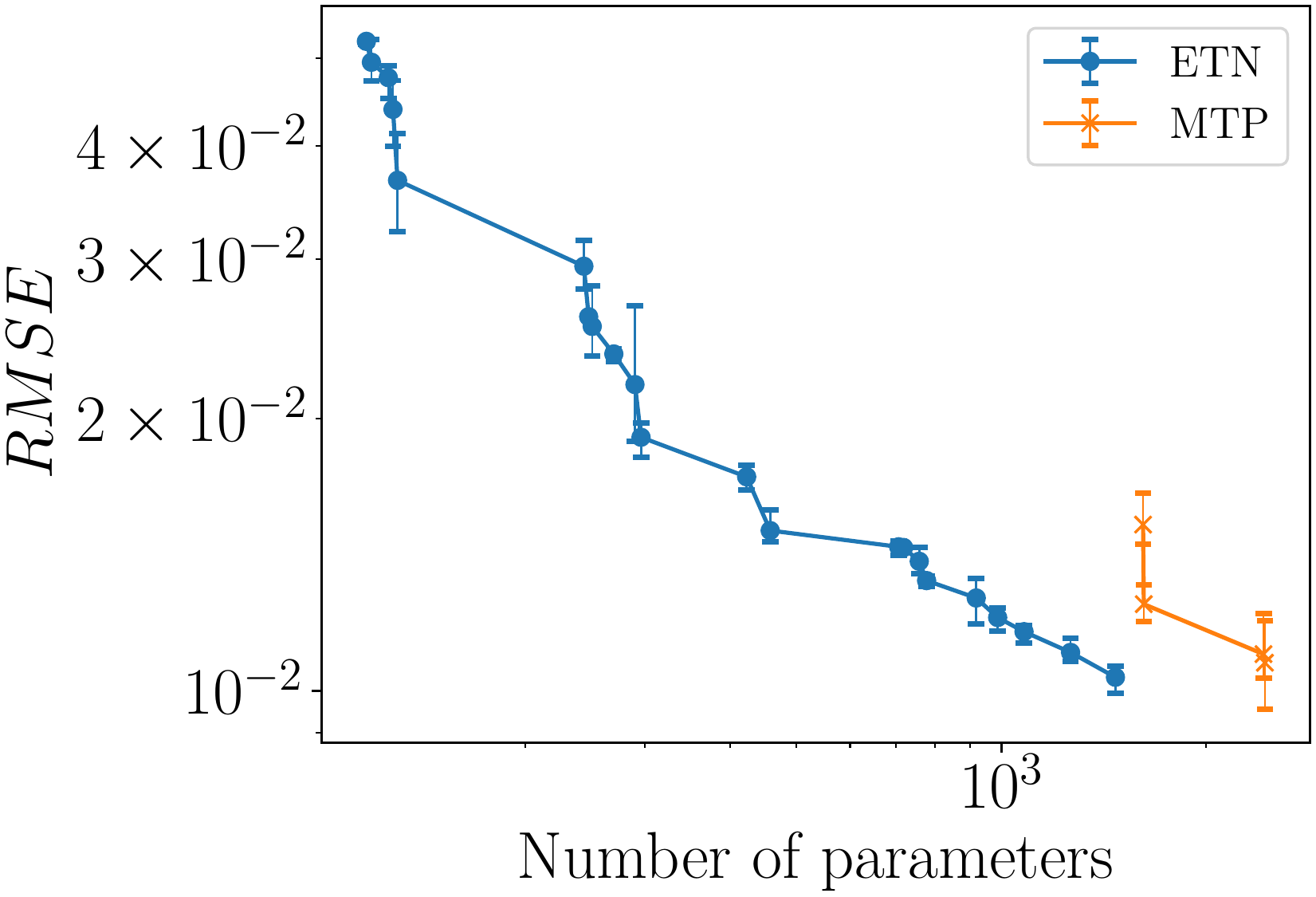}
    \end{minipage}
    \caption{(a) Mean validation loss, and (b) $RMSE$, as a function of the number of parameters for BA10.
    The horizontal bars correspond to the minimum and maximum values}
    \label{fig:results_BA10}
\end{figure}

As for the previous alloy datasets MoNbTaW-MEA and MoNbTaVW-HEA, ETN potentials generally require less coefficienst to reach the accuracy of MTPs, as shown in Figure \ref{fig:results_BA10}, although the difference is less pronounced ($\sim$\,1.5--1.8, as compared to $>$\,2 for MoNbTaW-MEA and MoNbTaVW-HEA).
Since BA10 only contains equilibrium structures, it also appears not too difficult to fit an MTP to it.
We yet remark that, for an ETN potential with 10 atomic species, the structure of the ETN \eqref{eq:tnp} may not be optimal since the size of the tensor $\uuA_\ell$ is already $100 \times N_{\rm spec}^{\rm rank}(\ell)$.
For example, for the ETN potential with $L = 5$ and $N_{\rm spec}^{\rm rank} = 3$ (cf., Figure \ref{fig:hyperparam-evolution}) the size of all $\uuA_\ell$'s is 800.
We think that an ETN of type \eqref{eq:tnp2} that also splits the species features could substantially reduce this amount.
We leave this for exploration in future work.

\subsubsection{Discussion}

\paragraph*{Evolution of the hyperparameters}

\begin{figure}[t!]
    \begin{minipage}{0.5\textwidth}
        \centering
        QM7
    \end{minipage}\hfill
    \begin{minipage}{0.5\textwidth}
        \centering
        QM9
    \end{minipage}\\[0.2em]
    \begin{minipage}{0.5\textwidth}
        \centering
        \includegraphics[width=0.8\textwidth]{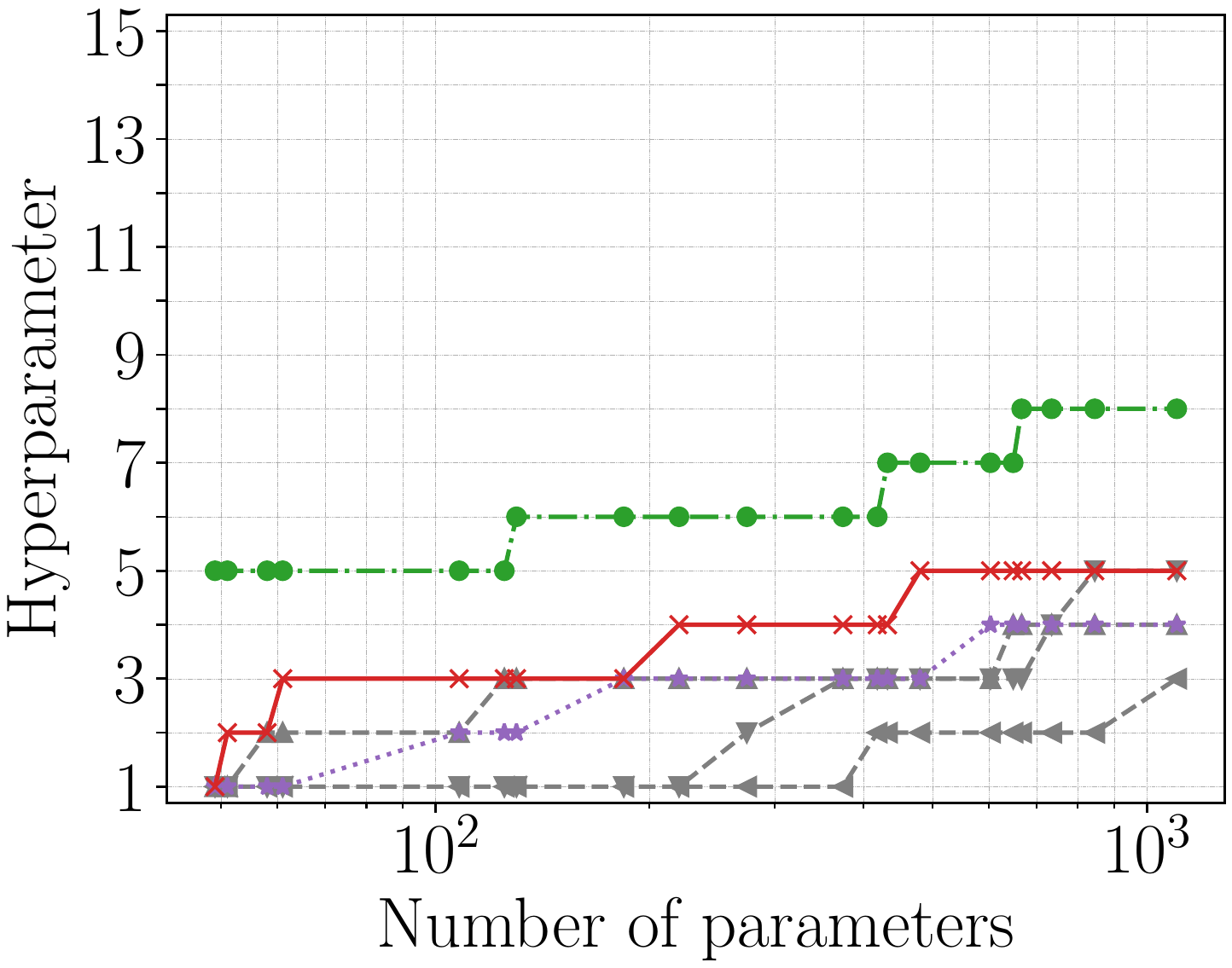}
    \end{minipage}\hfill
    \begin{minipage}{0.5\textwidth}
        \centering
        \includegraphics[width=0.8\textwidth]{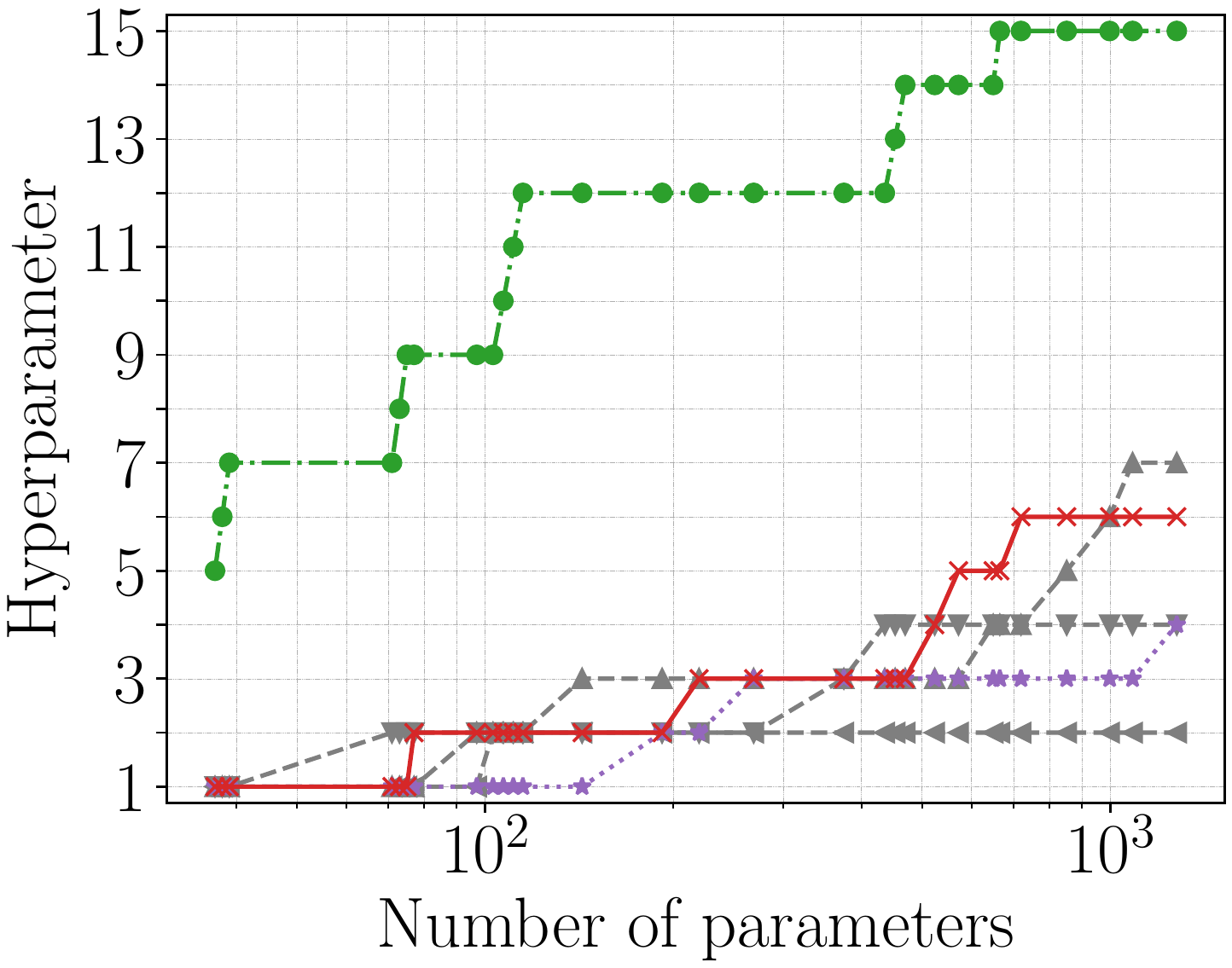}
    \end{minipage}\\[0.3em]
    \begin{minipage}{0.5\textwidth}
        \centering
        MoNbTaW-MEA
    \end{minipage}\hfill
    \begin{minipage}{0.5\textwidth}
        \centering
        MoNbTaVW-HEA
    \end{minipage}\\[0.2em]
    \begin{minipage}{0.5\textwidth}
        \centering
        \includegraphics[width=0.8\textwidth]{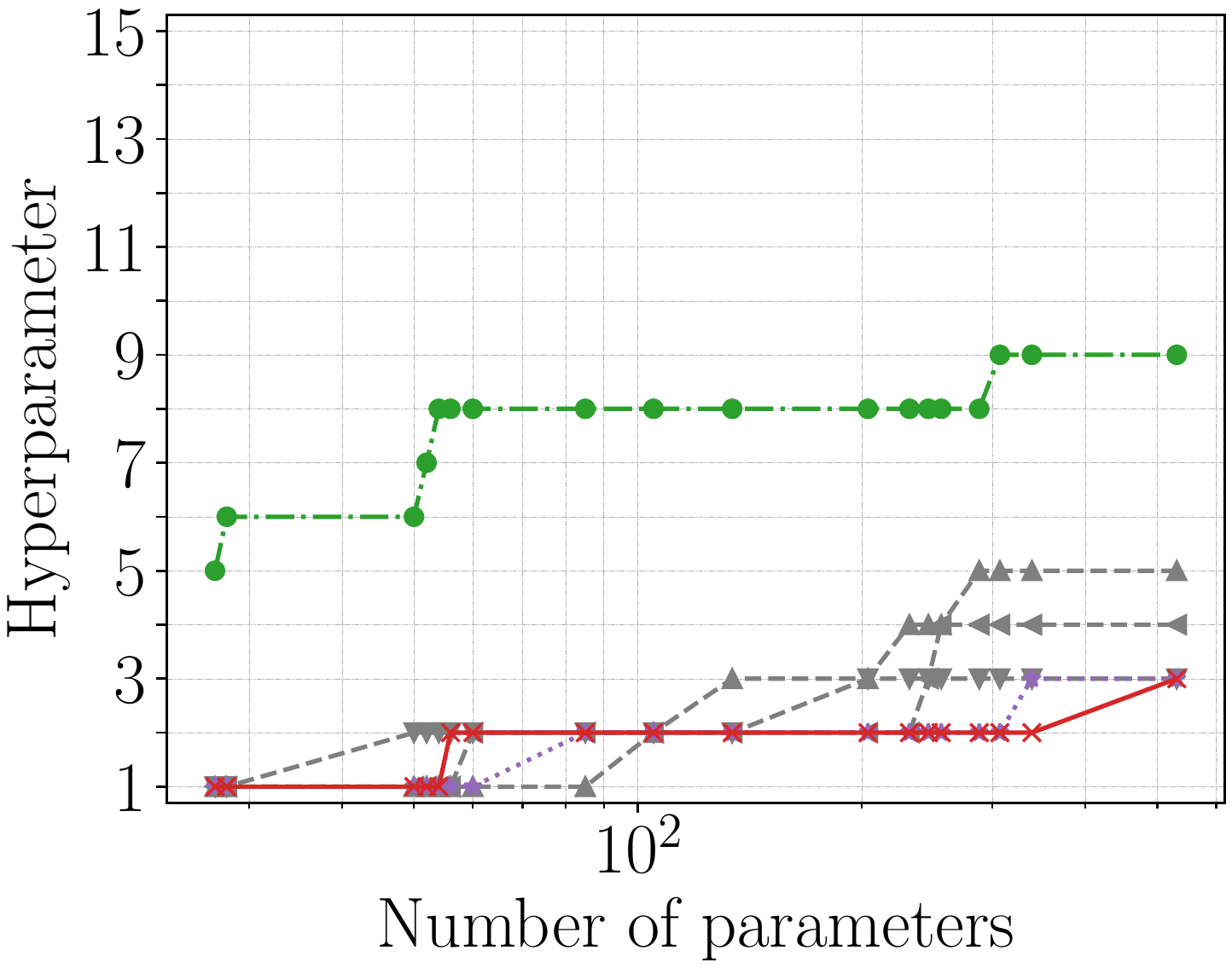}
    \end{minipage}\hfill
    \begin{minipage}{0.5\textwidth}
        \centering
        \includegraphics[width=0.8\textwidth]{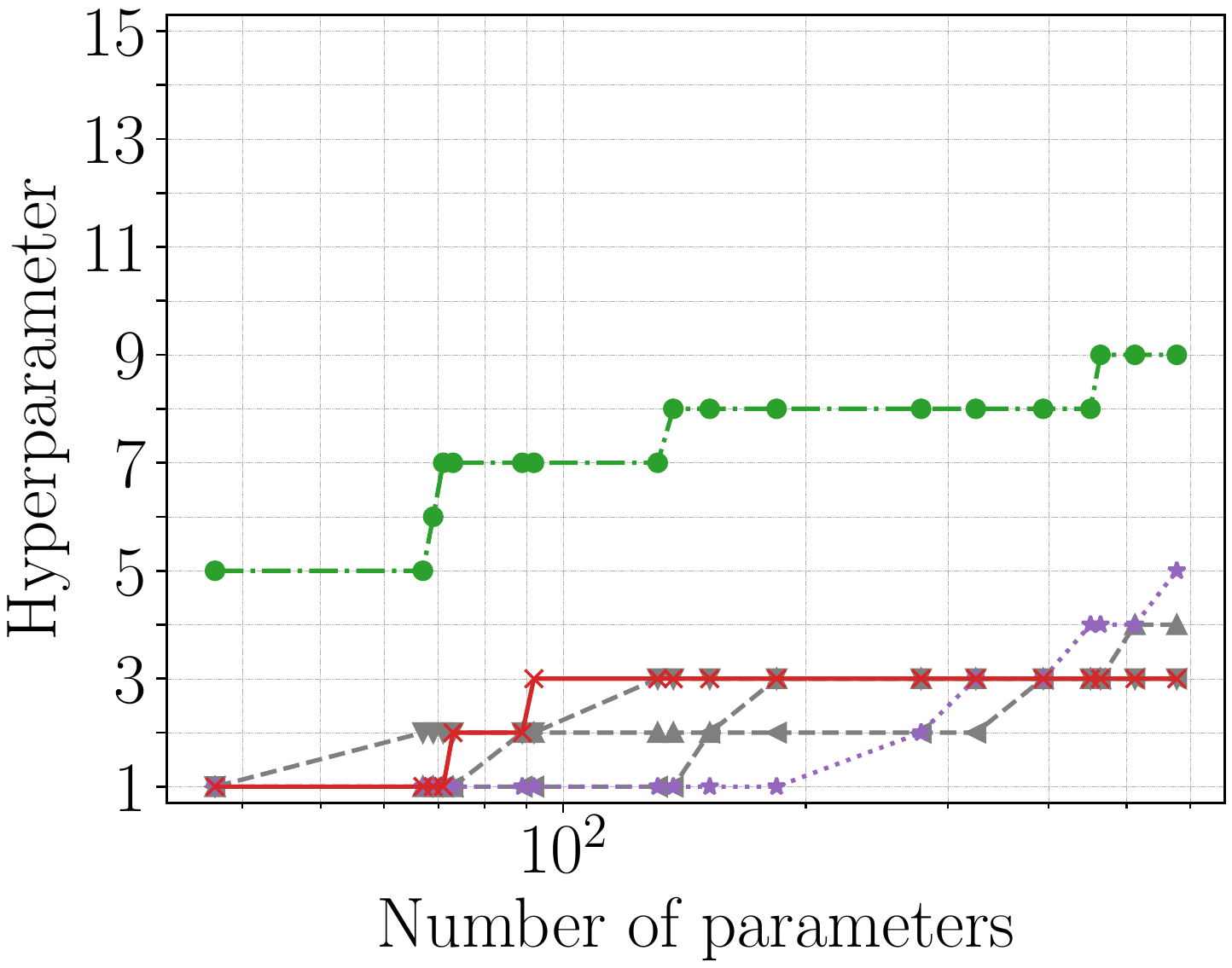}
    \end{minipage}\\[0.3em]
    \begin{minipage}{0.5\textwidth}
        \centering
        BA10
    \end{minipage}\hfill
    \begin{minipage}{0.5\textwidth}
    \end{minipage}\\[0.2em]
    \begin{minipage}{0.5\textwidth}
        \centering
        \includegraphics[width=0.8\textwidth]{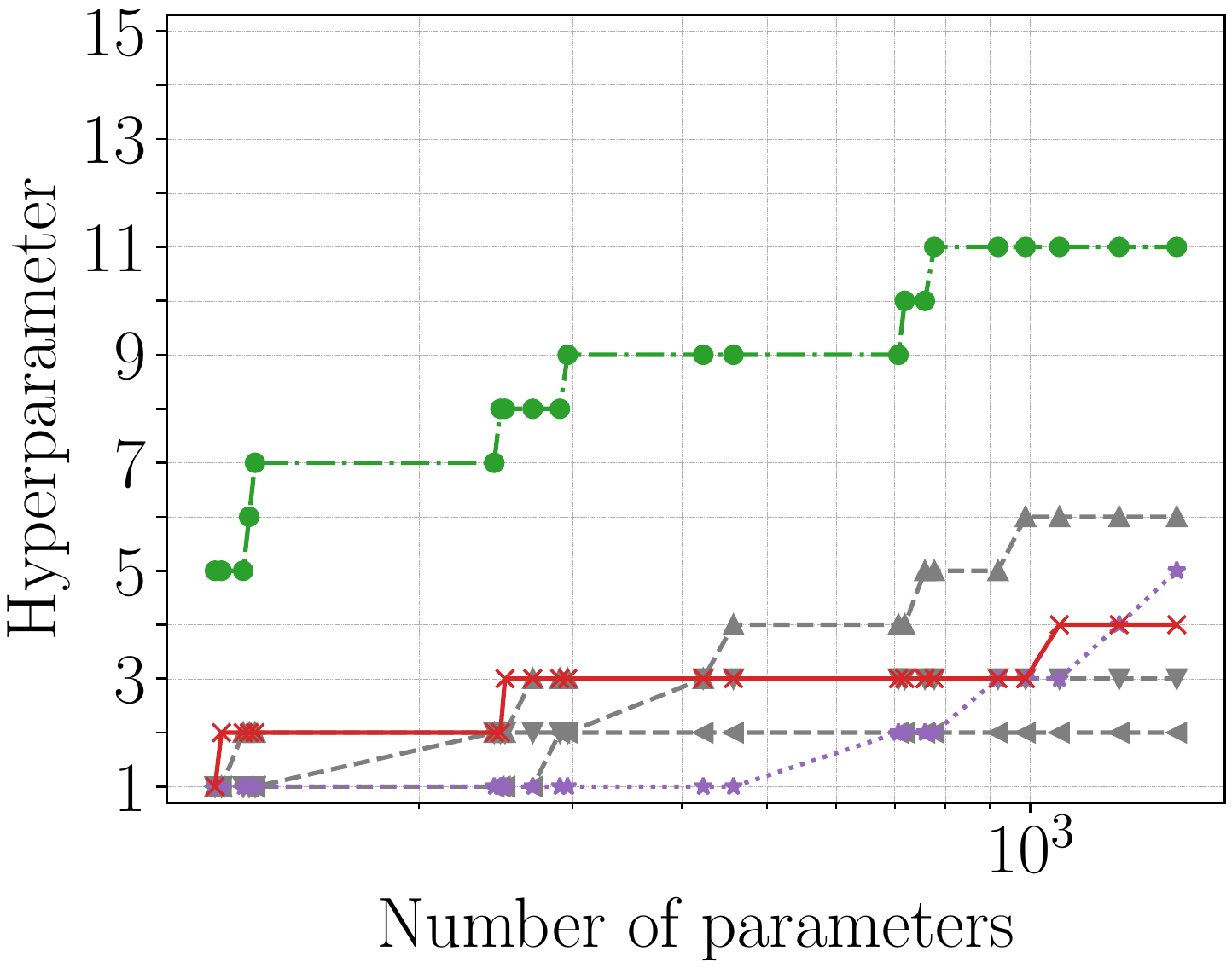}
    \end{minipage}\hfill
    \begin{minipage}{0.5\textwidth}
        \centering
        \includegraphics[width=0.35\textwidth]{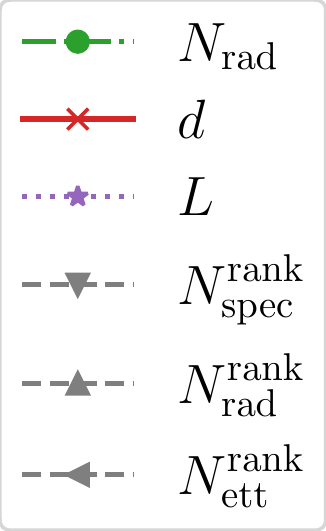}
    \end{minipage}
    \caption{Evolution of the ETN hyperparameters corresponding to the each iteration of the optimization algorithm for the five datasets from Figure \ref{fig:results_QM7}--\ref{fig:results_BA10}}
    \label{fig:hyperparam-evolution}
\end{figure}

We now analyze how the hyperparameters evolve during the optimization procedure.
For each dataset, the ETN hyperparameters $\scH$ are shown in Figure \ref{fig:hyperparam-evolution} as a function of the number of coefficients/iteration corresponding to Figure \ref{fig:results_QM7}--\ref{fig:results_BA10}.

The number radial basis functions $N_{\rm rad}$ grows approximately similar for all datasets, up to 8--11, except for QM9, where $N_{\rm rad}$ grows up to 15.

For the ETT order $d$ and the maximum angular momentum $L$, we observe an interesting difference between the molecule datasets and the alloy datasets:
while $d$ dominates over $L$ for the molecule datasets throughout the optimization procedure, $L$ starts dominating over $d$ for the alloy datasets with an increasing complexity of the potential.
This indicates that the body-order is more important for molecules than for alloys, and, vice-versa, the polynomial degree is more important for alloys than for molecules.
In addition, we observe that the order does not increase beyond 3 for the alloy datasets for interatomic potentials, i.e., for MoNbTaW-MEA and MoNbTaVW-HEA.
Order-3 ETN potentials are more sparse since we can neglect the imaginary parts of the equivariant tensors (cf., Section \ref{sec:O(3)-invariance}).
This implies that we can construct highly efficient ETN potentials for many-component alloys.

For the ranks, we observe that the species rank $N_{\rm spec}^{\rm rank}$ increases up to 5 and 4, respectively, for the molecule datasets, and up to 3 for the alloy datasets.
This is not surprising since QM7 and QM9 contain atoms from different groups (nonmetals and halogens), while the alloy datasets only contain atoms from the group of transition metals, except BA10, but BA10 does not contain all possible pair-wise combinations of atomic species.
Moreover, we observe that the radial rank $N_{\rm rad}^{\rm rank}$ always settles at some value that is $\sim$\,2--3 times lower than $N_{\rm rad}$.
This indicates that our assumption of letting the ranks decay with increasing angular momentum $\ell$ appears to be a reasonable choice.
The ETT rank $N_{\rm ett}^{\rm rank}$ is generally lower than $N_{\rm rad}^{\rm rank}$, which is is expected since a $N_{\rm ett}^{\rm rank} > N_{\rm rad}^{\rm rank}$ may ``overparameterize'' the potential.
The observed $N_{\rm rad}^{\rm rank} > N_{\rm ett}^{\rm rank}$ thus further confirms that our setup of constructing ETN potentials is indeed suitable and validates the correct functioning of the optimization algorithm.

\paragraph*{Influences due to commonalities and differences across the training sets}

In the future, it would be interesting to better understand the origin of the observed compression rates.
While this would require much more extensive testing with more data, some conclusions can already be drawn.
Transition metals have very similar physical and material properties.
Since the developers of MoNbTaW-MEA and MoNbTaVW-HEA were mainly concerned with those properties, it is not surprising that adding V to MoNbTaW did not alter the species ranks.
This picture might change however when considering chemical reactive properties with oxygen, e.g., for catalysts.
On the other hand, QM7 and QM9 consider a wider class of molecular structures that have different bonding properties and it thus perhaps not surprising that the species ranks are higher compared to the training sets composed of transition metals.
This may also explain why those datasets need a higher body-order because angles between bonds become more important.

To design more efficient networks for universal potentials, one strategy could be to first group similar species, contract them, and then take the contraction between different groups.
Another interesting direction could be establishing metrics that correlate element-specific or bond parameters like average number of d-valence electrons, or bandwidths, with the compression rates.
Such correlations could then be used to guide setting up a good initial ETN structure that would not require too much optimization.

\subsection{Comparison with non-polynomial MLIPs}
\label{sec:comparison}

From the two other popular classes of MLIPs, neural network potentials, and Gaussian process regression-based potentials, neural networks achieve state-of-the-art performance for molecule datasets.
However, we have found in Section \ref{sec:QM9} that converging ETN potentials on molecules with the python BFGS solver is difficult, in particular for ETN potentials with higher body-order.
While we have shown in Section \ref{sec:algebra} that alternating least-squares solvers are in principle stable for ETNs, they still need to be implemented and we thus postpone a comparison for molecules to future work.
We remark, however, that neural networks have already been frequently benchmarked against MTPs.
On unary systems (Li,Mo,Cu,Ni,Si,Ge), MTPs have been shown to reach higher accuracies for a comparable number of parameters than neural networks \citep{zuo_performance_2020}.
For the molecule database QM9, one of the most accurate implementations, HIP-NN \citep{lubbers_hierarchical_2018}, requires approximately an order of magnitude more parameters to reach the same mean absolute error of about 18\,meV than MTPs.
The according to \emph{Papers with Code} presently best neural network model for QM9, TensorNet \citep{simeon_tensornet_2023}, has reached a mean absolute error of about 4\,meV.
However, the network architecture in \citep{simeon_tensornet_2023} requires a number of parameters of the order of one million---this clearly emphasizes the need for more efficient representations like ETNs.

In the case of alloys, the---in our view---presently most interesting dataset is MoNbTaVW-HEA because it contains configurations with lattice defects and liquid configurations that are all relevant for predicting mechanical properties.
For this dataset, several Gaussian approximation potentials (GAPs) have been proposed in recent works by \citet{darby_compressing_2022,darby_tensorreduced_2023}, which also introduce compression schemes comparable to our tensor network approach.
The GAP in \citep{darby_compressing_2022} (in the following cGAP) uses a compression scheme that exploits symmetries in the power spectrum; the GAP in \citep{darby_tensorreduced_2023} (in the following trGAP) introduces a tensor-reduced parameter tensor using the canonical polyadic decomposition.

In order to compare those GAPs with ETN potentials, we have picked the three ETN potentials shown in Table \ref{tab:three_ETNs} that were found during the optimization (Figure \ref{fig:results_MoNbTaVW-HEA}).
All potentials use four-body interactions ($d=3$), but a different number of radial basis functions, angular momenta, and ranks.
As in \citep{darby_compressing_2022,darby_tensorreduced_2023}, we have taken the reduced subset of 2329 configurations from MoNbTaVW-HEA for the training, and benchmark the potentials on the test set from \citep{byggmastar_modeling_2021}.
Moreover, we now use 5000 BFGS iterations; all other parameters remain the same.

We then compare these three ETN potentials with the best (non-compressed) GAP, cGAP, and trGAP, which use three-body interactions.
The GAP and cGAP use 12 radial basis functions and a maximum angular momenta of 9.
The trGAP uses 8 radial basis functions and and a maximum angular momentum of 4, as for the ETN3 potential.
Hence, the complexity of contracting radial basis functions and spherical harmonics is similar or lower for the ETN potentials.

Interestingly, an ETN potential with a maximum angular momentum of only 1 (ETN2) achieves a comparable accuracy to that of the three GAPs (Table \ref{tab:hea_etn-gap-comparison}).
The energy error for the best ETN potential (ETN3) is the same as for the best GAP (trGAP), while the force error is lower, which might come from the higher body-order.
However, we remark that increasing the body-order in GAPs is computationally expensive, while ETNs can achieve a linear scaling (at least theoretically).
In addition, a GAP with higher body-order likely requires more sparse points/training data to achieve the same accuracy.

\begin{table}[t!]
    \centering
    \begin{tabular}{|c|c|c|}
        \hline
        Name & Hyperparameters & Coefficients \\ \hline\hline
        ETN1 & $\{7,3,1,2,2,1\}$ & 92  \\ \hline\hline
        ETN2 & $\{8,3,2,3,3,2\}$ & 278 \\ \hline\hline
        ETN3 & $\{9,3,5,3,4,3\}$ & 577 \\ \hline
    \end{tabular}
    \caption{Hyperparameters $\left\{ N_{\rm rad}, d, L, N_{\rm spec}^{\rm rank}, N_{\rm rad}^{\rm rank}, N_{\rm ett}^{\rm rank} \right\}$ and number of coefficients of the three ETN potentials used for the comparison in Section \ref{sec:comparison}, where $N_{\rm rad}$ is the maximum number of radial basis functions, $d$ is the order of the ETT, $L$ is the maximum number of angular momenta, and $N_{\rm spec}^{\rm rank}$, $N_{\rm rad}^{\rm rank}$, and $N_{\rm ett}^{\rm rank}$, are the ranks of $\uuuA$, $\uuuB$, and $\uuuT$, respectively}
    \label{tab:three_ETNs}
\end{table}

\begin{table}[t!]
    \centering
    \begin{tabular}{|c|c|c|c|c|c|c|c|}
        \hline
        Error measure & ETN1 & ETN2 & ETN3 & GAP & cGAP2 & trGAP \\ \hline\hline
        $RMSE$ [eV/atom] & 0.084 & 0.025 & 0.018 & 0.032 & 0.019 & 0.17 \\ \hline
        $RMSF$ [eV/\AA]  & 0.56  & 0.3   & 0.24 & 0.41  & 0.33  & 0.33 \\ \hline
    \end{tabular}
    \caption{MoNbTaVW-HEA test errors for the three ETN potentials from Table \ref{tab:three_ETNs} and the best GAPs reported in \citep{darby_compressing_2022,darby_tensorreduced_2023} (the values are deduced from Figure 8 in \citep{darby_compressing_2022}, and from Figure 3 \& 8 in \citep{darby_tensorreduced_2023}). The ETN test errors are the average errors over five training runs with different parameter initializations}
    \label{tab:hea_etn-gap-comparison}
\end{table}

Our findings and those of Darby et al. thus indicate that, with a suitable compression of the feature space, MLIPs with higher body-order may require less training data than without any compression.
Our optimization algorithm provides an efficient means for finding a suitable compression, something that would have been difficult to achieve with expensive grid searches.

\section{Generalizations to Other Tensor Networks}\label{sec:generalizations}

We think of the above construction as the simplest construction incorporating our main contribution---a way of incorporating equivariance in tensor networks.
As hinted in Section \ref{sec:algebra}, many ideas developed for the conventional tensor networks are transferable to the equivariant tensor networks. 
Below we list some of the ideas that we find can be immediately applied to the equivariant tensor networks.

\subsection{Arbitrary topology of tensor network cores}

The method of reshaping of two adjacent order-3 tensors, as illustrated Figure \ref{fig:dmrg}, as well as noting that it would also trivially work with one or two order-2 tensors, can be applied to construct an arbitrary topology involving order-3 tensors. For instance the hierarchical Tucker format or PEPs, as sketched in Section \ref{sec:tensor-intro:other} can readily be realized with our construction.

To implement PEPs or other tensor formats with tensors with order higher than three, one either has to use higher-order 3-j (i.e., ``4-j'', ``5-j'', etc.) tensors along with learnable coefficients or split the large order-5 tensor (as features in the two-dimensional PEPs) into three order-3 tensors which can then be symmetrized to avoid bias in the $x$ vs. $y$ coordinate.
A possible way of doing it for the two-dimensional PEPs is
\[
\begin{aligned}
   \includegraphics[scale=0.8]{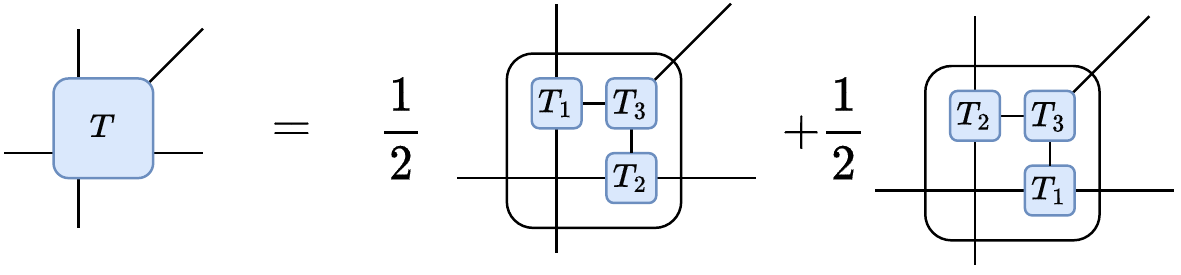}
\end{aligned}.
\]

\subsection{Weight sharing}

The ETNs as introduced in Section \ref{sec:algebra} depend linearly on the learnable weights in each of the tensors, however, already in our numerical experiments in Section \ref{sec:numerical} we used weight sharing for the tensors that compress the input features.
This resulted in a more parameter-efficient model, however, some of the (multi-)linear algebra algorithms (like DMRG or ALS) would not apply.

\subsection{Richer atomic features}

The flexible nature of equivariant tensor networks allows for equivariantly incorporating more atomic features like magnetic moments, charge, dipole moments, etc. \citep{drautz_atomic_2020,novikov_magnetic_2022}, and at the same time compress them in an efficient and \emph{physically interpretable} manner.

For instance, two possible way of treating magnetic moments ($m_i$ on the central atom and $m_j$ on the neighboring atom) are
\[
v^{(1)}_{ij} =
\begin{aligned}
   \includegraphics[scale=0.8]{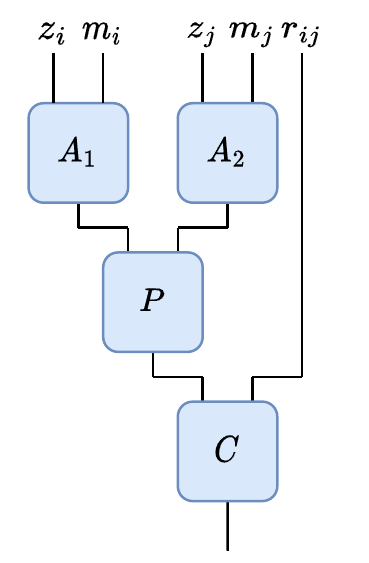}
\end{aligned},
\quad\quad
v^{(2)}_{ij} =
\begin{aligned}
   \includegraphics[scale=0.8]{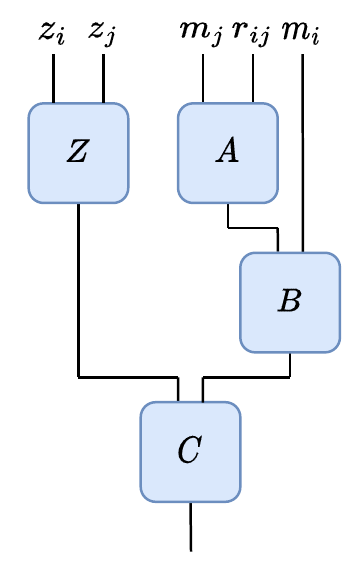}
\end{aligned}.
\]
Here in $v^{(1)}_{ij}$ we mix atomic species with magnetic moments into combined features with tensors $A_1$ and $A_2$.
Physically, we would say that we construct features that depend on the number ($z$) and magnetic state ($m$) of the core electrons of the $i$-th and $j$-th atoms.
Also, from the physical point of view it may be reasonable for them to share weights, i.e., to fix $A_1 = A_2$.
Then, these combined features are mixed together to produce information how the $(z,m)$ pair for the central atom should interact with another $(z,m)$ of the neighboring atom, which is then contracted with the relative positions $r_{ij}$ of these atoms to form the final feature vector of interaction of the $i$-th and $j$-th atoms.
A possible computational advantage of this method is that combining the $m$ and $z$ features with the $A_2$ vector can be precomputed and reused while processing different neighborhoods.

In $v^{(2)}$, instead, $z_i$ and $z_j$ are contracted using the learnable $Z$ tensor---this operation is essentially a look-up for the $(z_i, z_j)$-index feature vector which is then contracted with a similarly formed feature vector obtained from $m_i$, $m_j$, and $r_{ij}$.
Which of these two (or many other possible) variants is better in terms of accuracy or efficiency is yet to be tested.

Another example is adding non-atomic features like electronic temperature $T_{\rm el}$ to the potential \cite{zhang2020-DeepMD-temp,lopanitsyna2021-temp,ellis2021-temp}.
A way to do it could be adding an atomic feature to all the feature vectors $F$ or, instead, make it a separate model input like
\[
\begin{aligned}
   \includegraphics[scale=0.8]{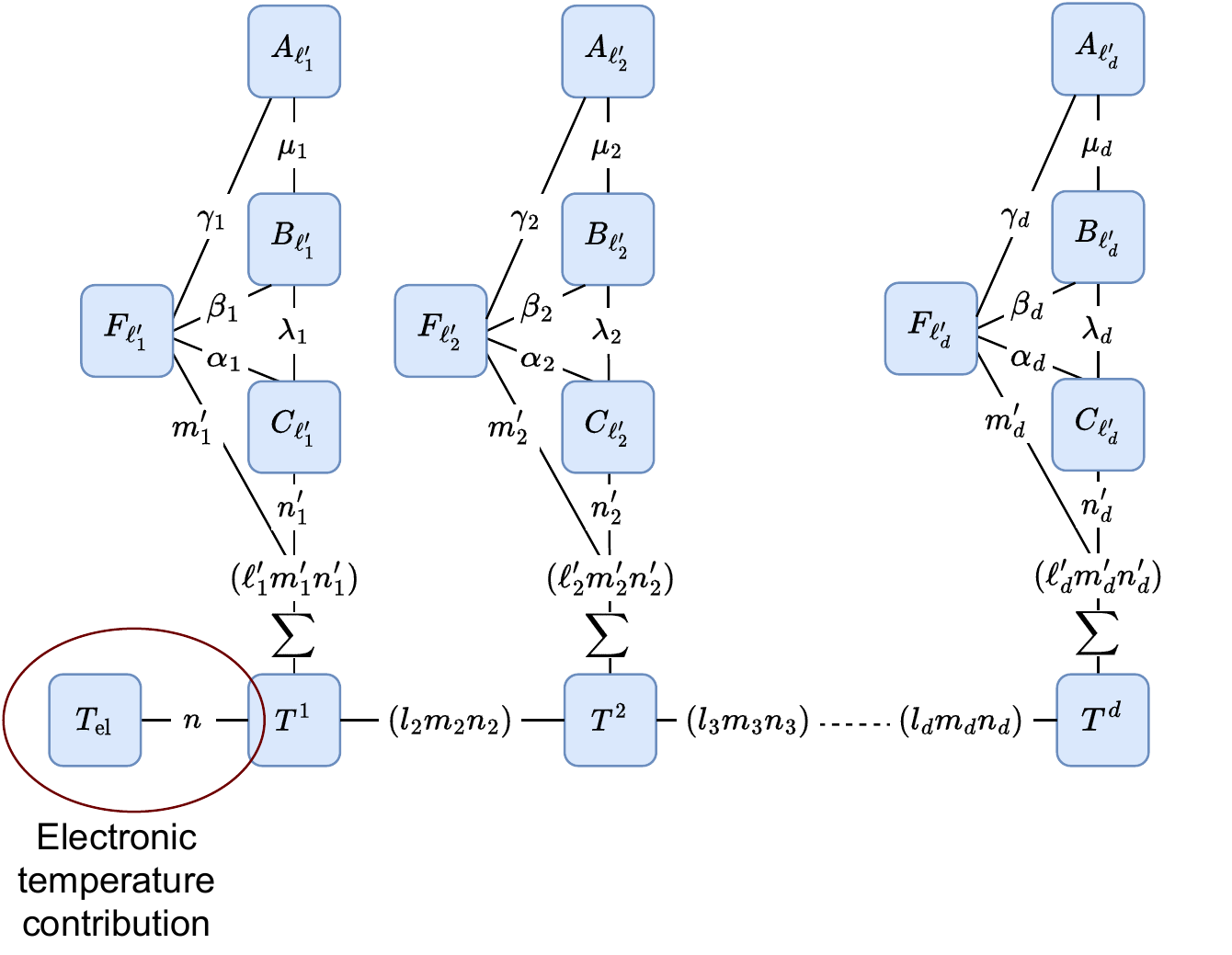}
\end{aligned}.
\]

\subsection{Message passing/graph convolutions, and pooling operations}

To introduce the message passing (also known as graph convolution operations) \citep{schutt_quantumchemical_2017,batzner_equivariant_2022,batatia_mace_2022} we recall that each of the input features $F_{i, (\ell m n)}$ also depend on the particular atom $i$ which forms another tensor dimension.
Instead of just summing all the features over the neighborhood of the $i$-th atom, we explicitly introduce the convolution operator $K$ as
\[
(K)_{ij}
=
\begin{aligned}
   \includegraphics[scale=0.8]{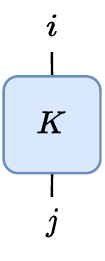}
\end{aligned}
= k(r_{ij}),
\]
where $k$ is some smooth cut-off function that serves as the convolution kernel.
The operation itself acts on features
\[
F_{i,(\ell m n)}
=
\begin{aligned}
   \includegraphics[scale=0.8]{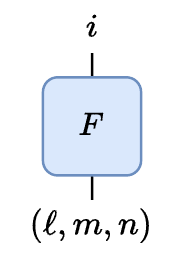}
\end{aligned}
,
\]
and distributes them over to the neighboring atoms:
\[
G_{i,(\ell mn)} = \sum_j (K)_{ij} F_{j,(\ell mn)},
\quad\text{or}\quad
\begin{aligned}
   \includegraphics[scale=0.8]{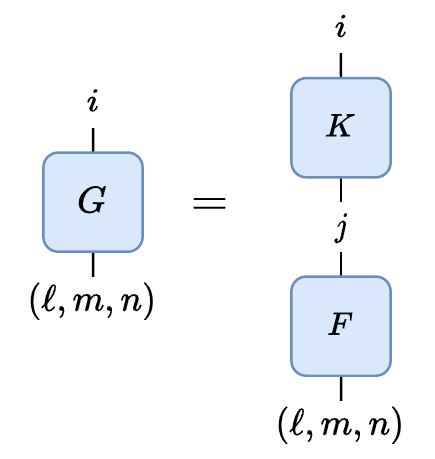}
\end{aligned}.
\]

Then, for example, a model in which features $F$ are passed to the neighbors, convolved with $F$, the result of which is passed to the neighbors and again convolved with $F$, is given by
$\uuB \big(K\big(\uuuA(KF)F\big) F\big)$ and diagrammatically expressed as
\[
\uuB \big(K\big(\uuuA(KF)F\big) F\big) ~=~
\begin{aligned}
   \includegraphics[scale=0.8]{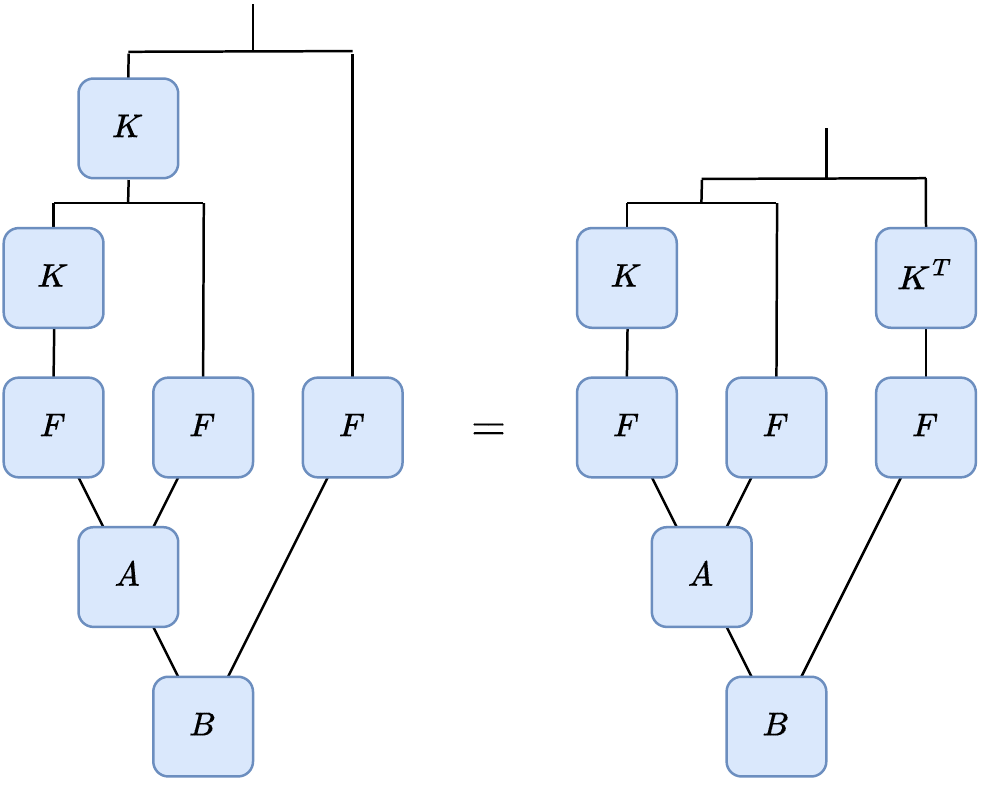}
\end{aligned}
\]
Again, operations on adjacent kernels, like $\uuuA$ and $\uuB$ in this example, are possible as they are independent of the convolution operations $K$.
The upper ``hanging'' link indicates that the result is the per-atom energy that should then be summed over this dimension (i.e., over $i$).

Pooling operations require simply defining an averaging operator over the atoms $i$.
This is generally considered not useful for interatomic potentials because they destroy the locality (short-sightedness) principle, but may be good for cheminformatics models like the prediction of the band gap.
For example, a model that takes atomic features $F$, passes it over to the neighbors, contracts with the original features $F$ using the tensor $A$, averages, and contracts these averaged features with the averaging over the original features $F$ would be
\[
\begin{aligned}
   \includegraphics[scale=0.8]{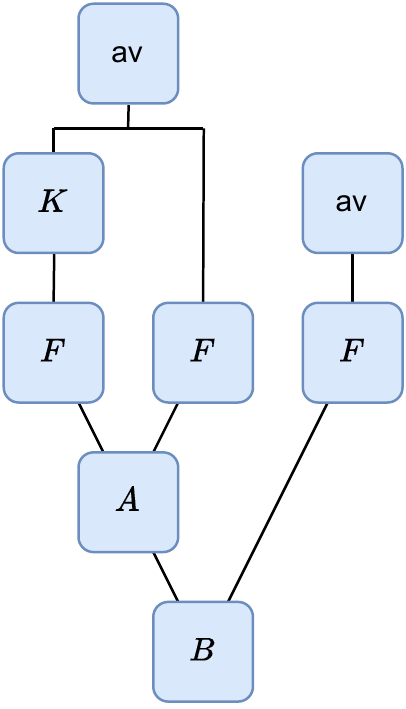}
\end{aligned}
\]
(here ${\rm av}$ is averaging over the atoms).
If one wants to use this block to inform a regular interatomic-potential-like model (leading to something resembling the attention mechanism), one could design something like
\[
\begin{aligned}
   \includegraphics[scale=0.8]{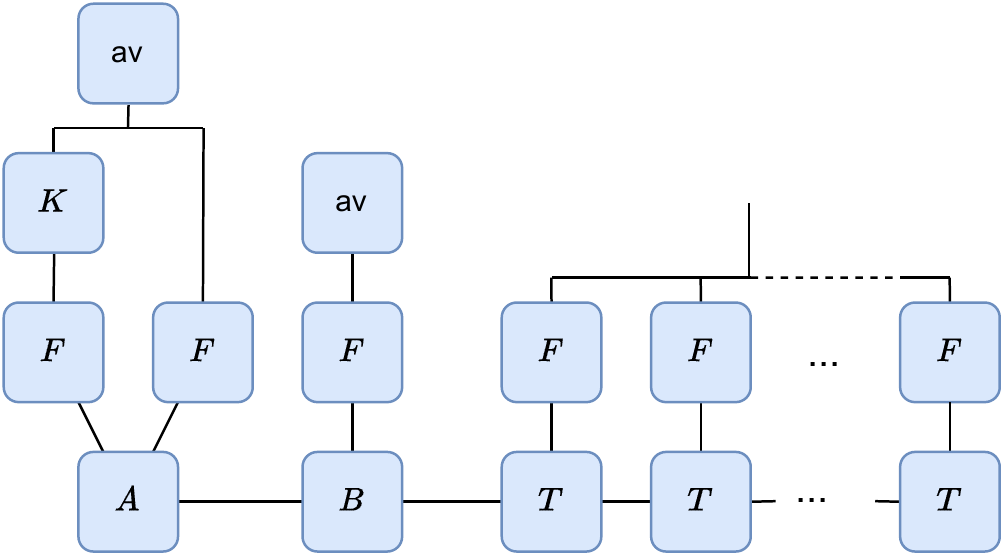}
\end{aligned},
\]
where in the diagram for conciseness we did not illustrate the part with compressing radial, species, etc., information into a single feature vector like it is done with tensors $A$, $B$, and $C$ in \eqref{eq:tnp2}.

\section{Concluding Remarks}
\label{sec:conclusions}

In this work, we have developed a formalism for constructing equivariant tensor networks (ETNs), i.e., tensor networks that remain invariant under group actions of SO(3).
Our main findings is that---analogous to conventional tensor networks---we only require up-to-order-3 equivariant tensors in order to construct arbitrary ETNs.
In our view, most importantly, our formalism of constructing ETNs can therefore lead to a \textbf{drastic simplification of constructing machine-learning interatomic potentials} of arbitrary complexity, which is our main target of application.
Our confidence grounds on the fact that many algorithms developed for conventional tensor networks translate almost verbatim to the equivariant case, e.g., automatic differentiation, or tensor orthogonalization (cf. Section \ref{sec:algebra}).
So, once these algorithms are implemented, the only thing that would be left is to code the tensor network diagram (and this might be even possible using visual programming environments at some point)---all the rest would literally ``fall from the tree''.

Representing interatomic interaction energies using ETNs can be considered as a high-performance representation of polynomial machine-learning interatomic potentials that naturally allows for basis- and feature-compression.
The numerical results presented in Section \ref{sec:ETNP-performance} indicates that ETNs can reach excellent performance for complex systems.
This ETN potential outperformed moment tensor potentials on several databases for multicomponent systems in terms of the required number of parameters by factor of about 2 to 3.
Interestingly, we have observed that the optimal hyperparameters do not evolve comparably for different species; for example, molecules benefit from a higher body-order, while metallic alloys benefit from a higher polynomial degree.
Future research may thus be devoted to developing improved algorithms that find the best ETN hyperparameters for a given problem.
The fact that we could significantly compress the feature space for an ETN potential with a still rather low number of features (positional and species) gives us hope that we can then reach even much better compression rates when adding new features, e.g., magnetic moments.
We are planning to explore this in future work, alongside with extending ETNs in many other directions some of which we have outlined in Section \ref{sec:generalizations}.

Also, unlike linearly parameterized potentials like MTP or ACE in which the construction of different basis functions requires different operations, the ETN potentials are based on tensorial contractions, which have been easier to parallelize on massively parallel architectures such as GPUs---this could also become another advantage of the ETN models.

Finally, we would also like to emphasize that ETNs are not limited to interatomic potentials but can generally be used to represent symmetry-preserving multivariate functions.
As such, they could also be applied to continuous problems like differential equations where they would share the same advantages as conventional tensor networks of avoiding a large sampling space (e.g., \citep{hong_functional_2022}), compared to neural networks.

\section{Acknowledgments}

M.H. gratefully acknowledges the financial support under the scope of the COMET program within the K2 Center “Integrated Computational Material, Process and Product Engineering (IC-MPPE)” (Project No 886385). This program is supported by the Austrian Federal Ministries for Climate Action, Environment, Energy, Mobility, Innovation and Technology (BMK) and for Labour and Economy (BMAW), represented by the Austrian Research Promotion Agency (FFG), and the federal states of Styria, Upper Austria and Tyrol.
A.S. was supported by the Russian Science Foundation (Grant No. 23-13-00332).

\section*{Appendix}

\begin{appendices}

\counterwithin*{equation}{section}
\renewcommand\theequation{\thesection\arabic{equation}}

\section{Other Tensor Networks}\label{sec:tensor-intro:other}

By utilizing the graphical representation of tensor contractions, let us give examples of other tensor networks.
One prominent example is the Hierarchical Tucker representation \citep{grasedyck_hierarchical_2010} which was proposed in the field of tensor representations at about the same time as tensor train.
On a diagram it looks like
\[
T_{\<4\>} = 
\begin{aligned}
    \includegraphics[width=0.43\textwidth]{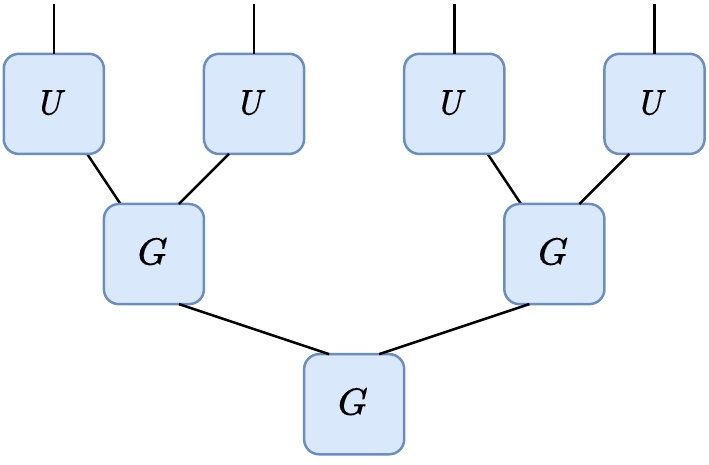}
\end{aligned}
.
\]

Another, more complex example is PEPS \citep{verstraete_renormalization_2004} which additionally features a two-dimensional arrangement of tensor indicies:
\[
T_{\<6\>} = 
\begin{aligned}
    \includegraphics[width=0.35\textwidth]{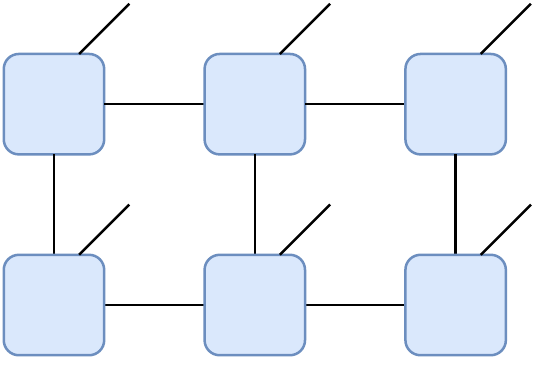}
\end{aligned}
.
\]

\section{Wigner 3-j Symbol for Real Spherical Harmonics}

\subsection{Selection Rules}
\label{appdx:w3j_rsph_selection_rules}

We first express the 3-j symbols for RSHs in terms of the usual 3-j symbols.
Therefore, we distinguish between between three cases
\begin{enumerate}[label=(\roman*)]
    \item $m_1 = m_2 = m_3 = 0$,
    \item $\exists! \, m = 0 \in \{ m_i \}$,
    \item $m_1,m_2,m_3 \neq 0$.
\end{enumerate}

Case (i) is trivial since
\begin{equation}\label{eq:w3j-rsph-expr1}
    \begin{Bmatrix}
        \ell_1 & \ell_2 & \ell_3 \\ 0 & 0 & 0
    \end{Bmatrix}
    =
    \begin{pmatrix}
        \ell_1 & \ell_2 & \ell_3 \\ 0 & 0 & 0
    \end{pmatrix}
    .
\end{equation}

For case (ii), we first exemplify our derivation for $m_3 = 0$; the other two cases follow analogously. From the conjugate transpose of $\uuU_\ell$ (eq. \eqref{eq:U_l})
\begin{equation}\label{eq:U_l_conj_transp}
    \uuU_\ell^{\sT\ast} =
    \frac{1}{\sqrt{2}}
    \begin{pmatrix}
        -\rmi &&&&&& 1 \\
        & -\rmi &&&& 1 & \\
        && \ddots && \reflectbox{$\ddots$} && \\
        &&& \sqrt{2} &&& \\
        && \reflectbox{$\ddots$} && \ddots && \\
        & (-1)^{\ell-1}\,\rmi &&&& (-1)^{\ell-1} & \\
        (-1)^\ell\,\rmi &&&&&& (-1)^\ell
    \end{pmatrix}
    ,
\end{equation}
we deduce
\begin{equation}
\begin{aligned}
    \begin{Bmatrix}
        \ell_1 & \ell_2 & \ell_3 \\ m_1 & m_2 & 0
    \end{Bmatrix}
    =&
    \begin{pmatrix}
        \ell_1 & \ell_2 & \ell_3 \\ m_1 & m_2 & 0
    \end{pmatrix}
    U^{\sT\ast}_{\ell_1 m_1m_1} U^{\sT\ast}_{\ell_2 m_2m_2} \\
    &+
    \begin{pmatrix}
        \ell_1 & \ell_2 & \ell_3 \\ -m_1 & -m_2 & 0
    \end{pmatrix}
    U^{\sT\ast}_{\ell_1 -m_1m_1} U^{\sT\ast}_{\ell_2 -m_2m_2} \\
    &+
    \begin{pmatrix}
        \ell_1 & \ell_2 & \ell_3 \\ m_1 & -m_2 & 0
    \end{pmatrix}
    U^{\sT\ast}_{\ell_1 m_1m_1} U^{\sT\ast}_{\ell_2 -m_2m_2} \\
    &+
    \begin{pmatrix}
        \ell_1 & \ell_2 & \ell_3 \\ -m_1 & m_2 & 0
    \end{pmatrix}
    U^{\sT\ast}_{\ell_1 -m_1m_1} U^{\sT\ast}_{\ell_2 m_2m_2}
    .
\end{aligned}
\end{equation}
To simplify the latter expression, we use the symmetry property of the Wigner 3-j symbol (cf., \citep{varshalovich_quantum_1988})
\begin{equation}\label{eq:3-j-sign-reversal}
    \begin{pmatrix}
        \ell_1 & \ell_2 & \ell_3 \\ m_1 & m_2 & m_3
    \end{pmatrix}
    =
    (-1)^{\ell_1 + \ell_2 + \ell_3}
    \begin{pmatrix}
        \ell_1 & \ell_2 & \ell_3 \\ -m_1 & -m_2 & -m_3
    \end{pmatrix}
\end{equation}
and the relations
\begin{align}\label{eq:UT*-relations}
    U^{\sT\ast}_{\ell_i -m_im_i} = \operatorname{sgn}(m_i)(-1)^{m_i}U^{\sT\ast}_{\ell_i m_im_i}
    &&&
    U^{\sT\ast}_{\ell_i m_im_i} = \operatorname{sgn}(m_i)(-1)^{m_i}U^{\sT\ast}_{\ell_i -m_im_i},
\end{align}
where
\begin{equation}
    \operatorname{sgn}(m_i) =
    \left\{
    \begin{aligned}
        \,  1 &&& m_i \ge 0, \\
        \, -1 &&& m_i < 0
    \end{aligned}
    \right.
\end{equation}
is the sign function.
Then,
\begin{equation}\label{eq:w3j-rsph-expr2.1}
\begin{aligned}
    \begin{Bmatrix}
        \ell_1 & \ell_2 & \ell_3 \\ m_1 & m_2 & 0
    \end{Bmatrix}
    =\,&
    U^{\sT\ast}_{\ell_1 m_1m_1} U^{\sT\ast}_{\ell_2 m_2m_2}
    \big( 1 \pm \operatorname{sgn}(m_1)\operatorname{sgn}(m_2) \big)
    \Bigg( \\
    &\,
    \begin{pmatrix}
        \ell_1 & \ell_2 & \ell_3 \\ m_1 & m_2 & 0
    \end{pmatrix}
    +
    \operatorname{sgn}(m_2)(-1)^{m_2}
    \begin{pmatrix}
        \ell_1 & \ell_2 & \ell_3 \\ m_1 & -m_2 & 0
    \end{pmatrix}
    \Bigg)
    ,
\end{aligned}
\end{equation}
where the $+$ and $-$ in the second parenthesis hold for even or odd $\ell_1 + \ell_2 + \ell_3$, respectively.
We omit the derivations for $m_1,m_2 = 0$ as they follow trivially from the previous one and just state the final results:
for $m_2 = 0$, we have
\begin{equation}\label{eq:w3j-rsph-expr2.2}
\begin{aligned}
    \begin{Bmatrix}
        \ell_1 & \ell_2 & \ell_3 \\ m_1 & 0 & m_3
    \end{Bmatrix}
    =\,&
    U^{\sT\ast}_{\ell_1 m_1m_1} U^{\sT\ast}_{\ell_3 m_3m_3}
    \big( 1 \pm \operatorname{sgn}(m_1)\operatorname{sgn}(m_3) \big)
    \Bigg( \\
    &\,
    \begin{pmatrix}
        \ell_1 & \ell_2 & \ell_3 \\ m_1 & 0 & m_3
    \end{pmatrix}
    +
    \operatorname{sgn}(m_3)(-1)^{m_3}
    \begin{pmatrix}
        \ell_1 & \ell_2 & \ell_3 \\ m_1 & 0 & -m_3
    \end{pmatrix}
    \Bigg)
    ,
\end{aligned}
\end{equation}
and for $m_1 = 0$
\begin{equation}\label{eq:w3j-rsph-expr2.3}
\begin{aligned}
    \begin{Bmatrix}
        \ell_1 & \ell_2 & \ell_3 \\ 0 & m_2 & m_3
    \end{Bmatrix}
    =\,&
    U^{\sT\ast}_{\ell_2 m_2m_2} U^{\sT\ast}_{\ell_3 m_3m_3}
    \big( 1 \pm \operatorname{sgn}(m_2)\operatorname{sgn}(m_3) \big)
    \Bigg( \\
    &\,
    \begin{pmatrix}
        \ell_1 & \ell_2 & \ell_3 \\ 0 & m_2 & m_3
    \end{pmatrix}
    +
    \operatorname{sgn}(m_3)(-1)^{m_3}
    \begin{pmatrix}
        \ell_1 & \ell_2 & \ell_3 \\ 0 & m_2 & -m_3
    \end{pmatrix}
    \Bigg)
    .
\end{aligned}
\end{equation}

For case (iii), by using \eqref{eq:U_l_conj_transp} in \eqref{eq:3-j-symbol-real-sph}, we have
\begin{equation}
\begin{aligned}
    \begin{Bmatrix}
        \ell_1 & \ell_2 & \ell_3 \\ m_1 & m_2 & m_3
    \end{Bmatrix}
    =&
    \begin{pmatrix}
        \ell_1 & \ell_2 & \ell_3 \\ m_1 & m_2 & m_3
    \end{pmatrix}
    U^{\sT\ast}_{\ell_1 m_1m_1} U^{\sT\ast}_{\ell_2 m_2m_2} U^{\sT\ast}_{\ell_3 m_3m_3} \\
    &+
    \begin{pmatrix}
        \ell_1 & \ell_2 & \ell_3 \\ -m_1 & -m_2 & -m_3
    \end{pmatrix}
    U^{\sT\ast}_{\ell_1 -m_1m_1} U^{\sT\ast}_{\ell_2 -m_2m_2} U^{\sT\ast}_{\ell_3 -m_3m_3} \\
    &+
    \begin{pmatrix}
        \ell_1 & \ell_2 & \ell_3 \\ m_1 & m_2 & -m_3
    \end{pmatrix}
    U^{\sT\ast}_{\ell_1 m_1m_1} U^{\sT\ast}_{\ell_2 m_2m_2} U^{\sT\ast}_{\ell_3 -m_3m_3} \\
    &+
    \begin{pmatrix}
        \ell_1 & \ell_2 & \ell_3 \\ -m_1 & -m_2 & m_3
    \end{pmatrix}
    U^{\sT\ast}_{\ell_1 -m_1m_1} U^{\sT\ast}_{\ell_2 -m_2m_2} U^{\sT\ast}_{\ell_3 m_3m_3} \\
    &+
    \begin{pmatrix}
        \ell_1 & \ell_2 & \ell_3 \\ m_1 & -m_2 & m_3
    \end{pmatrix}
    U^{\sT\ast}_{\ell_1 m_1m_1} U^{\sT\ast}_{\ell_2 -m_2m_2} U^{\sT\ast}_{\ell_3 m_3m_3} \\
    &+
    \begin{pmatrix}
        \ell_1 & \ell_2 & \ell_3 \\ -m_1 & m_2 & -m_3
    \end{pmatrix}
    U^{\sT\ast}_{\ell_1 -m_1m_1} U^{\sT\ast}_{\ell_2 m_2m_2} U^{\sT\ast}_{\ell_3 -m_3m_3} \\
    &+
    \begin{pmatrix}
        \ell_1 & \ell_2 & \ell_3 \\ -m_1 & m_2 & m_3
    \end{pmatrix}
    U^{\sT\ast}_{\ell_1 -m_1m_1} U^{\sT\ast}_{\ell_2 m_2m_2} U^{\sT\ast}_{\ell_3 m_3m_3} \\
    &+
    \begin{pmatrix}
        \ell_1 & \ell_2 & \ell_3 \\ m_1 & -m_2 & -m_3
    \end{pmatrix}
    U^{\sT\ast}_{\ell_1 m_1m_1} U^{\sT\ast}_{\ell_2 -m_2m_2} U^{\sT\ast}_{\ell_3 -m_3m_3}
    .
\end{aligned}
\end{equation}
Using \eqref{eq:3-j-sign-reversal} and \eqref{eq:UT*-relations}, we have
\begin{equation}\label{eq:w3j-rsph-expr3}
\begin{aligned}
    \begin{Bmatrix}
        \ell_1 & \ell_2 & \ell_3 \\ m_1 & m_2 & m_3
    \end{Bmatrix}
    =\,&
    U^{\sT\ast}_{\ell_1 m_1m_1} U^{\sT\ast}_{\ell_2 m_2m_2} U^{\sT\ast}_{\ell_3 m_3m_3}
    \big( 1 \pm \operatorname{sgn}(m_1)\operatorname{sgn}(m_2)\operatorname{sgn}(m_3) \big)
    \Bigg( \\
    &
    \begin{pmatrix}
        \ell_1 & \ell_2 & \ell_3 \\ m_1 & m_2 & m_3
    \end{pmatrix}
    +
    \operatorname{sgn}(m_3)(-1)^{m_3}
    \begin{pmatrix}
        \ell_1 & \ell_2 & \ell_3 \\ m_1 & m_2 & -m_3
    \end{pmatrix} \\
    &
    +
    \operatorname{sgn}(m_2)(-1)^{m_2}
    \begin{pmatrix}
        \ell_1 & \ell_2 & \ell_3 \\ m_1 & -m_2 & m_3
    \end{pmatrix} \\
    &
    +
    \operatorname{sgn}(m_1)(-1)^{m_1}
    \begin{pmatrix}
        \ell_1 & \ell_2 & \ell_3 \\ -m_1 & m_2 & m_3
    \end{pmatrix}
    \Bigg)
    ,
\end{aligned}
\end{equation}
where, again, the $+$ and $-$ in the second parenthesis hold for even or odd $\ell_1 + \ell_2 + \ell_3$, respectively.
From \eqref{eq:w3j-rsph-expr1}, \eqref{eq:w3j-rsph-expr2.1}--\eqref{eq:w3j-rsph-expr2.3}, and \eqref{eq:w3j-rsph-expr3}, it follows that the 3-j symbol for RSHs obeys the selection rules \hyperlink{SR3}{\textbf{[SR3]}} and \hyperlink{SR4}{\textbf{[SR4]}}, in addition to the triangular inequality \hyperlink{SR1}{\textbf{[SR1]}} for the three $\ell$'s.

Further, for the 3-j symbol for RSHs being a real number $\neq 0$, one of the $m_i$'s must be positive which follows from the definition of $U^{\sT\ast}_{\ell_i m_im_i}$ \eqref{eq:U_l_conj_transp}.
Vice versa, for being a complex number with pure imaginary part $\neq 0$, one of the $m_i$'s must be negative.
Therefore, the 3-j symbol for RSHs is always real whenever $\ell_1 + \ell_2 + \ell_3$ is even---and pure imaginary whenever $\ell_1 + \ell_2 + \ell_3$ is odd.

\subsection{Sparsity}
\label{appdx:w3j_rsph_sparsity}

\begin{figure}[hbt]
    \begin{minipage}{0.5\textwidth}
        \centering
        (a)
    \end{minipage}
    \begin{minipage}{0.5\textwidth}
        \centering
        (b)
    \end{minipage}\hfill
    \begin{minipage}{0.5\textwidth}
        \centering
        \includegraphics[width=0.9\textwidth]{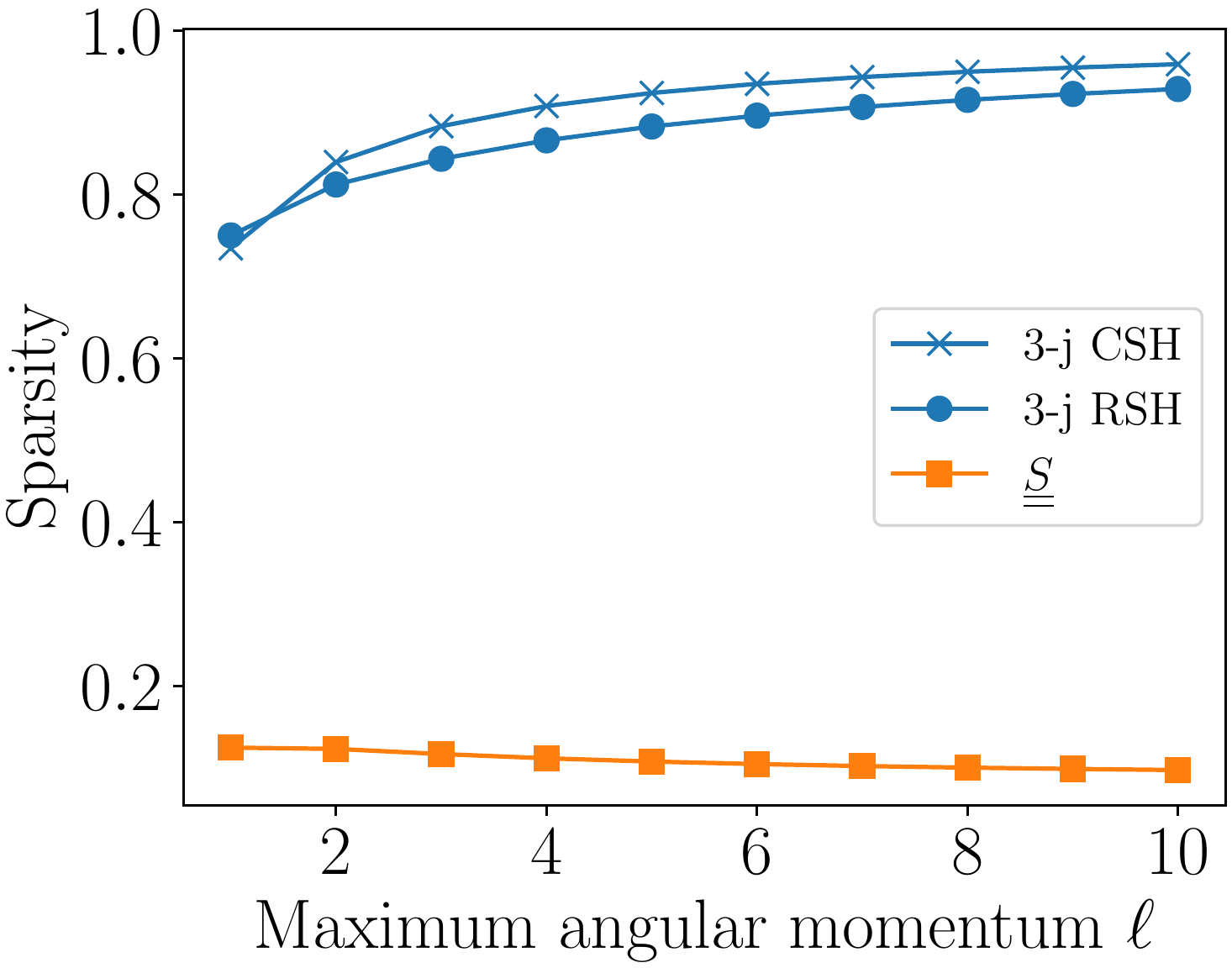}
    \end{minipage}
    \begin{minipage}{0.5\textwidth}
        \centering
        \includegraphics[width=0.9\textwidth]{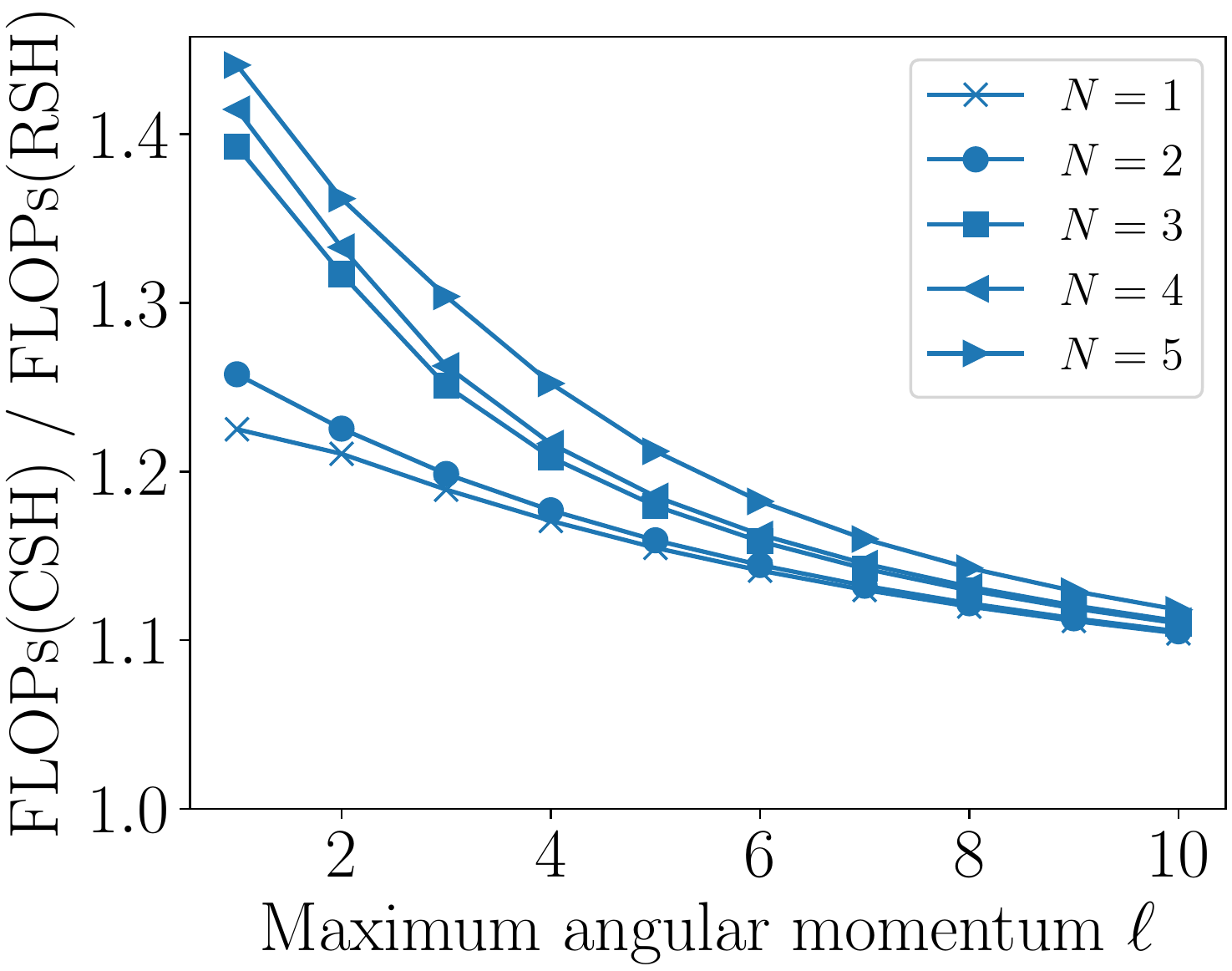}
    \end{minipage}
    \caption{(a) Sparsity of the 3-j symbols for real and complex spherical harmonics, and the order-2 tensor $\uuS$ from \eqref{eq:instructions-tensor-contraction}. (b) Ratio of required FLOPs for the contractions $u_i = T_{ijk} v_j w_k$ with a different maximum number of channels $N$ when using a complex and a real spherical harmonics basis, respectively}
    \label{fig:w3j_sparsity}
\end{figure}

To analyze the sparsity, i.e., the number of nonzero entries divided by the size of the tensor, we consider 3-j symbols with some fixed maximum angular momentum.
The sparsity of the 3-j symbols for real and complex spherical harmonics is shown in Figure \ref{fig:w3j_sparsity} (a).

Interestingly, even though the 3-j symbol for RSHs is slightly less sparse, using an RSH basis requires slightly less floating point operations (FLOPs) for tensor contractions than using a complex spherical harmonics basis.
To see this, let $\uuuT$ be an equivariant tensor of size $M\times M\times M$ and let its channels decay with increasing $\ell$ according to eq. \eqref{eq:decay-channels}.
The maximum number of channels is denoted by $N = N(\ell = 0)$.

We exemplify our analysis by means of the contraction operation $u_i = T_{ijk} v_j w_k$, that is most relevant for contracting our ETNs, where $u_i$ and $w_i$ are two covariant but otherwise arbitrary vectors, and $v_j$ is the input feature vector.
To that end, first suppose that we are using a complex spherical harmonics basis.
Then, $T_{ijk}$ is a real-valued tensor with $N_{\rm nz}^{\rm csh}$ nonzero entries, and the feature vector $v_j$ is complex-valued.
The contraction can thus be split into the following operations
\begin{equation}\label{eq:instructions-tensor-contraction}
    \begin{aligned}
        S_{ik}^{\rm re} &= T_{ijk} v_j^{\rm re} && \text{($N_{\rm nz}^{\rm csh}$ FLOPs)} \\
        S_{ik}^{\rm im} &= T_{ijk} v_j^{\rm im} && \text{($N_{\rm nz}^{\rm csh}$ FLOPs)} \\
        u_i^{\rm re}    &= S_{ik}^{\rm re} w_k^{\rm re} + S_{ik}^{\rm im} w_k^{\rm im} && \text{($2M^2$ FLOPs)} \\
        u_i^{\rm im}    &= S_{ik}^{\rm re} w_k^{\rm im} + S_{ik}^{\rm im} w_k^{\rm re} && \text{($2M^2$ FLOPs)}
    \end{aligned}
    ,
\end{equation}
with the corresponding FLOPs given in parenthesis;
we have assumed fused multiply-add instructions and that the intermediate tensor $S_{ik}$ is stored in dense format as it is not very sparse (cf., Figure \ref{fig:w3j_sparsity} (a)).

Now assume a real basis.
Then, $T_{ijk}$ is a complex-valued tensor with $N_{\rm nz}^{\rm rsh}$ nonzero entries, and the feature vector $v_j$ is real-valued.
The first two operations from \eqref{eq:instructions-tensor-contraction} therefore change to
\begin{equation}
    \begin{aligned}
        S_{ik}^{\rm re} &= T_{ijk}^{\rm re} v_j && \text{($N_{\rm nz}^{\rm rsh/re}$ FLOPs)} \\
        S_{ik}^{\rm im} &= T_{ijk}^{\rm im} v_j && \text{($N_{\rm nz}^{\rm rsh/im}$ FLOPs)}
    \end{aligned}
    ,
\end{equation}
where $N_{\rm nz}^{\rm rsh/re}$ and $N_{\rm nz}^{\rm rsh/im}$ are the numbers of nonzeros of the real and imaginary part of $\uuuT$, respectively.
Due to the fact that an entry of $\uuuT$ is either pure real or pure imaginary, we have $N_{\rm nz}^{\rm rsh} = N_{\rm nz}^{\rm rsh/re} + N_{\rm nz}^{\rm rsh/im}$.
Hence, using RSHs requires less FLOPs since $N_{\rm nz}^{\rm rsh} + 4M^2 < 2N_{\rm nz}^{\rm csh} + 4M^2$  for all tensor sizes shown in Figure \ref{fig:w3j_sparsity} (b).

\subsection{Contraction Properties}
\label{appdx:w3j_rsph_contraction_properties}

Generating a new covariant vector by contracting the usual 3-j symbol with two covariant vectors, $u_{\ell_1 m_1}$ and $u_{\ell_2 m_2}$, yields another covariant vector $(-1)^{m_3} w_{\ell_3 -m_3}$ with a phase shift.
To see this, consider the 3-j symbol in terms of the Clebsch-Gordan coefficients $C_{\ell_1m_1\ell_2m_2}^{\ell_3m_3}$
\begin{equation}
    \begin{pmatrix}
        \ell_1 & \ell_2 & \ell_3 \\ m_1 & m_2 & m_3
    \end{pmatrix}
    =
    \frac{(-1)^{\ell_1 - \ell_2 - m_3}}{\sqrt{2\ell_3 + 1}} \, C_{\ell_1m_1\ell_2m_2}^{\ell_3-m_3},
\end{equation}
and recall that the Clebsch-Gordan coefficients expand the basis $\vert \ell_3 m_3 \>$ in terms of $\vert \ell_1 m_1 \>$ and $\vert \ell_2 m_2 \>$.

This phase shift is naturally reversed when using the 3-j symbol for RSHs.
To understand why this is the case, consider the contraction
\begin{equation}
    \sum\limits_{\substack{\ell_3,m_3,\\m_3',m_3''}}
    \begin{pmatrix}
        \ell_1 & \ell_2 & \ell_3 \\ m_1 & m_2 & m_3'
    \end{pmatrix}
    U^{\sT\ast}_{\ell_3 m_3'm_3}
    \begin{pmatrix}
        \ell_3 & \ell_4 & \ell_5 \\ m_3'' & m_4 & m_5
    \end{pmatrix}
    U^{\sT\ast}_{\ell_3 m''_3m_3}.
\end{equation}
Since $U^{\sT\ast}_{\ell_3 m_3'm_3}U^{\sT\ast}_{\ell_3 m''_3m_3} = U^{\sT\ast}_{\ell_3 m_3'm_3}U^{\ast}_{\ell_3 m_3m''_3}$ is the anti-diagonal matrix
\begin{equation}
    \uuU^{\sT\ast}_\ell \uuU^\ast_\ell =
    \begin{pmatrix}
        &&&&&& (-1)^\ell \\
        &&&&& (-1)^{\ell-1} & \\
        &&&& \reflectbox{$\ddots$} && \\
        &&& 1 &&& \\
        && \reflectbox{$\ddots$} &&&& \\
        & (-1)^{\ell-1} &&&&& \\
        (-1)^\ell &&&&&&
    \end{pmatrix},
\end{equation}
we have $v_{\ell m_3'} U^{\sT\ast}_{\ell_3 m_3'm_3}U^{\sT\ast}_{\ell_3 m''_3m_3} = w_{\ell m_3''}$, with $w_{\ell m_3} = (-1)^{-m_3} v_{\ell -m_3}$.

\subsection{Symmetry Properties}
\label{appdx:w3j_symmetries}

The 3-j symbols for RSHs obey the same symmetry properties for even and odd permutations than the usual 3-j symbol, i.e.,
\begin{equation}
    \begin{Bmatrix}
        \ell_1 & \ell_2 & \ell_3 \\ m_1 & m_2 & m_3
    \end{Bmatrix}
    =
    \begin{Bmatrix}
        \ell_3 & \ell_1 & \ell_2 \\ m_3 & m_1 & m_2
    \end{Bmatrix}
    =
    \begin{Bmatrix}
        \ell_2 & \ell_3 & \ell_1 \\ m_2 & m_3 & m_1
    \end{Bmatrix}
    ,
\end{equation}
and
\begin{equation}
    (-1)^{\ell_1 + \ell_2 + \ell_3}
    \begin{Bmatrix}
        \ell_1 & \ell_2 & \ell_3 \\ m_1 & m_2 & m_3
    \end{Bmatrix}
    =
    \begin{Bmatrix}
        \ell_2 & \ell_1 & \ell_3 \\ m_2 & m_1 & m_3
    \end{Bmatrix}
    =
    \begin{Bmatrix}
        \ell_1 & \ell_3 & \ell_2 \\ m_1 & m_3 & m_2
    \end{Bmatrix}
    =
    \begin{Bmatrix}
        \ell_3 & \ell_2 & \ell_1 \\ m_3 & m_2 & m_1
    \end{Bmatrix}
    .
\end{equation}
On the other hand, changing the sign of the $m$'s gives, under the assumption that \hyperlink{SR3}{\textbf{[SR3]}} holds,
\begin{equation}
    \begin{Bmatrix}
        \ell_1 & \ell_2 & \ell_3 \\ -m_1 & -m_2 & -m_3
    \end{Bmatrix}
    =
    0,
\end{equation}
as opposed to
\begin{equation}
    \begin{pmatrix}
        \ell_1 & \ell_2 & \ell_3 \\ -m_1 & -m_2 & -m_3
    \end{pmatrix}
    =
    (-1)^{\ell_1 + \ell_2 + \ell_3}
    \begin{pmatrix}
        \ell_1 & \ell_2 & \ell_3 \\ m_1 & m_2 & m_3
    \end{pmatrix}
    .
\end{equation}

\subsection{Orthogonality Properties}
\label{appdx:w3j_orth}

Recall that the usual 3-j symbol has the following orthogonality property \citep{varshalovich_quantum_1988}
\begin{equation}\label{eq:w3j_orth}
    (2\ell + 1) \sum_{m_1,m_2}
    \begin{pmatrix}
        \ell_1 & \ell_2 & \ell \\ m_1 & m_2 & m
    \end{pmatrix}
    \begin{pmatrix}
        \ell_1 & \ell_2 & \ell' \\ m_1 & m_2 & m'
    \end{pmatrix}
    =
    \delta_{\ell\ell'}
    \delta_{mm'}
    .
\end{equation}
For the 3-j symbols for RSHs, we obtain almost the same property, namely,
\begin{equation}
    (-1)^{\ell_1 + \ell_2 + \ell}
    (2\ell + 1) \sum_{m_1,m_2}
    \begin{Bmatrix}
        \ell_1 & \ell_2 & \ell \\ m_1 & m_2 & m
    \end{Bmatrix}
    \begin{Bmatrix}
        \ell_1 & \ell_2 & \ell' \\ m_1 & m_2 & m'
    \end{Bmatrix}
    =
    \delta_{\ell\ell'}
    \delta_{mm'}
    ,
\end{equation}
which can be derived from the expressions of the 3-j symbols for RSHs \eqref{eq:w3j-rsph-expr1}, \eqref{eq:w3j-rsph-expr2.1}--\eqref{eq:w3j-rsph-expr2.3}, and \eqref{eq:w3j-rsph-expr3}, in terms of the usual 3-j symbol, and the orthogonality property \eqref{eq:w3j_orth}.
Since the 3-j symbol for RSHs is always real whenever $\ell_1 + \ell_2 + \ell_3$ is even, and complex whenever $\ell_1 + \ell_2 + \ell_3$ is odd, the latter is in fact the same as \eqref{eq:w3j_orth} in the sense of unitarity, i.e.,
\begin{equation}\label{eq:w3j-rsph_orth}
    (2\ell + 1) \sum_{m_1,m_2}
    \begin{Bmatrix}
        \ell_1 & \ell_2 & \ell \\ m_1 & m_2 & m
    \end{Bmatrix}
    \begin{Bmatrix}
        \ell_1 & \ell_2 & \ell' \\ m_1 & m_2 & m'
    \end{Bmatrix}^\ast
    =
    \delta_{\ell\ell'}
    \delta_{mm'}
    ,
\end{equation}
where $\{\bullet\}^\ast$ denotes the complex conjugate of the 3-j symbol for RSHs.

\section{Functional forms of other Machine-Learning Interatomic Potentials}
\label{appdx:TN-MLIPs}

\subsection{Moment Tensor Potentials}
\label{appdx:MTPs}

For MTPs, the per-atom energies are defined as a linear combination of parameters $\mtheta_\alpha$ and scalar basis functions $B_\alpha$
\begin{equation}\label{eq:per-atom-energy-mtp}
    \clE(\{ \ur_{ij} \}) =
    \sum_{\alpha} \mtheta_\alpha B_\alpha(\{ \ur_{ij} \}).
\end{equation}
The basis functions themselves are obtained from scalar contractions of the moment tensor descriptors
\begin{equation}
    M_{\< \nu \>}^\mu =
    \sum_j C_{\mu n\alpha\beta} \, Q_n(\| \ur_{ij} \|) z_\beta^i z_\gamma^j
    \, \Big( \underbrace{\ur_{ij} \otimes \ldots \otimes \ur_{ij}}_{\nu \; \textnormal{times}} \Big),
\end{equation}
where $C_{\mu n\alpha\beta}$ is another parameter tensor.
The numbers $\mu$ and $\nu$ define the level of $M_{\< \nu \>}^\mu$, $\operatorname{lev}M_{\mu,\nu} = 2 + 4\mu + \nu$.
The polynomial degree and body-order of MTPs are defined by the level $\operatorname{lev}_{\rm MTP}$.
Following \citet{gubaev_accelerating_2019}, for a given MTP level, all possible scalar contractions of $M_{\< \nu \>}^\mu$'s that satisfy $\operatorname{lev}_{\rm MTP} \ge \sum_i \operatorname{lev}M_{\< \nu_i \>}^{\mu_i}$ are included as basis functions.

In order to cast \eqref{eq:per-atom-energy-mtp} into a tensor network structure, we slightly re-define the formalism above.
To that end, we define the feature tensor
\begin{equation}
    F_{n\beta\gamma(\tilde{\ell}\tilde{m})} =
    Q_n(\| \ur_{ij} \|) z_\beta^i z_\gamma^j
    \, \Big( \underbrace{\ur_{ij} \otimes \ldots \otimes \ur_{ij}}_{\tilde{\ell} \; \textnormal{times}} \Big)_{(\tilde{\ell}\tilde{m})},
\end{equation}
which includes all $M_{\< \nu \>}^\mu$'s for a given MTP level, reshaped to a vector;
we have chosen its index to be $(\tilde{\ell}\tilde{m})$ to highlight the similarity with respect to the spherical harmonics basis.
The per-atom energy can then be written as
\begin{equation}
    \clE(\{ \ur_{ij} \}) =
    \sum_{\alpha} \mtheta_\alpha \tilde B_{\alpha (\mu_1 \tilde\ell_1 \tilde m_1) \ldots (\mu_d \tilde\ell_d \tilde m_d)}
    v_{(\mu_1 \tilde\ell_1 \tilde m_1)} \ldots v_{(\mu_d \tilde\ell_d \tilde m_d)}
    ,
\end{equation}
with
\begin{equation}
    v_{(\mu_i \tilde\ell_i \tilde m_i)} = \sum_j C_{\mu_i n_i\beta_i\gamma_i} F_{n_i\beta_i\gamma_i(\tilde\ell_i\tilde m_i)}
    .
\end{equation}
Here, $\tilde B_{\bullet}$ is an integer tensor that encodes all O(3)-invariant contractions of the features for a given level.
MTPs can then be written in the tensor network diagram notation as follows
\begin{equation}\label{eq:tn_mtp}
\clE^{\rm mtp}(F_{\< 4\>}) = \;
\begin{aligned}
    \includegraphics[scale=0.8]{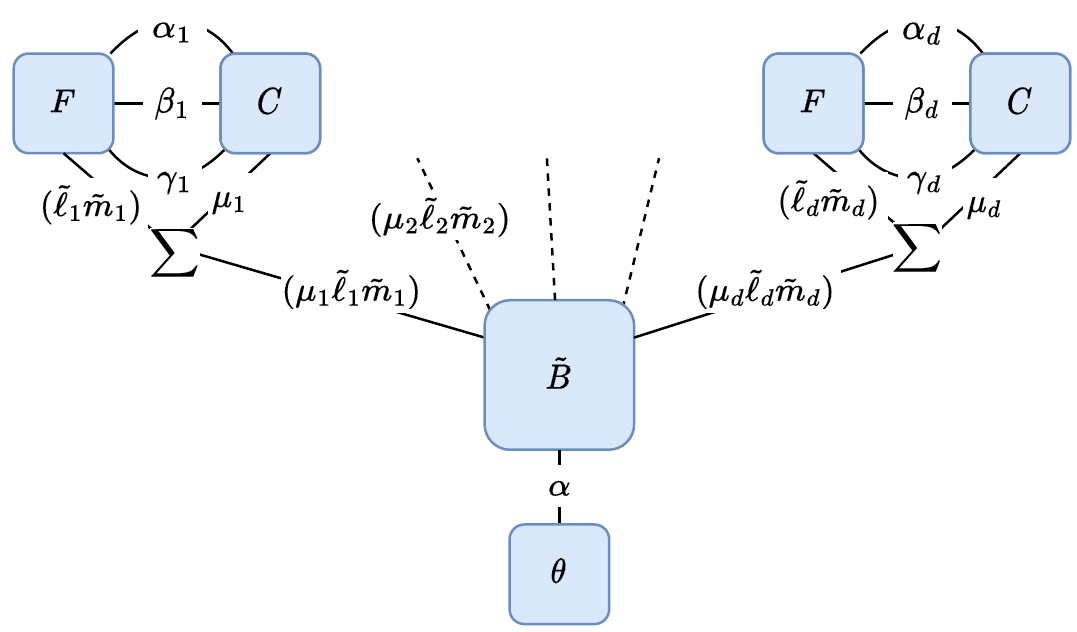}
\end{aligned}.
\end{equation}
This diagram should be understood that, unlike using the 3-j (or for that matter Clebsch-Gordan tensor) multiple times to get higher (than three) order polynomials, MTPs use an alternative way of contracting the moments to produce a complete set of basis functions.

\subsection{Spectral Neighbor Analysis and Atomic Cluster Expansion Potentials}
\label{appdx:ACEs}

The Spectral Neighbor Analysis (SNAP) \citep{thompson_spectral_2015} potentials are four-body potentials (i.e., that multiply the features $F$ with each other three times) and have the following diagram:
\begin{equation}\label{eq:tn_snap}
\clE^{\rm snap}(F) = \;
\begin{aligned}
   \includegraphics[scale=0.8]{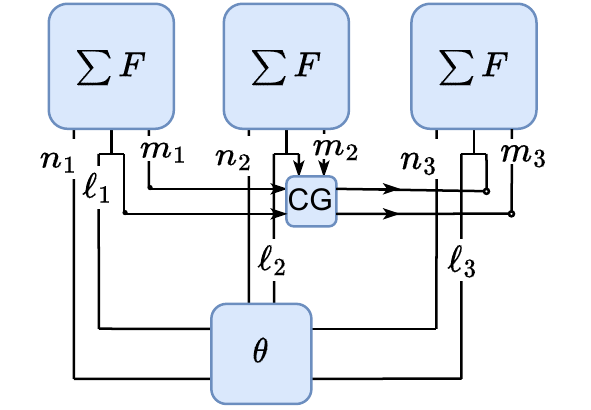}
\end{aligned}.
\end{equation}
Here we simplify the diagram by not explicitly distinguishing the summation-over-neighbors operations and not considering separately radial from chemical features (thus not distinguishing the original SNAP from explicit multielement SNAP \citep{cusentino_explicit_2020}).
These potentials use Clebsch-Gordan coefficients to ensure the rotational symmetry and have the order-3 $(\ell n)$-parametrized tensor $\mtheta$ with learnable coefficients.
If we replace the Clebsch-Gordan coefficients with the 3-j coefficients the way they are introduced in this paper, then SNAP can be considered as simply contracting the three feature vectors with a single equivariant order-3 tensor as introduced in Figure \ref{fig:order-3-algebra}(a).

The Atomic Cluster Expansion (ACE) potentials \citep{drautz_atomic_2019} replicate the construction \eqref{eq:tn_snap} to a chain of Clebsch-Gordan tensors and a large-order tensor of coefficients $\mtheta$, the example of which being
\begin{equation}\label{eq:tn_ace}
\clE^{\rm ace}(F) = \;
\begin{aligned}
   \includegraphics[scale=0.8]{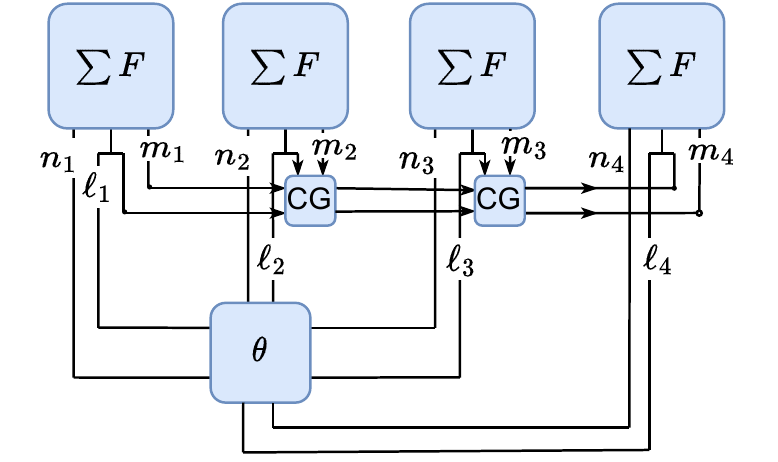}
\end{aligned}.
\end{equation}
Compared to MTP, ACE uses the procedure based on the Clebsch-Gordan coefficients to produce the set of basis functions (also complete) that is indexed by $(\ell_i, m_i)$ and whose coefficients form a tensor.
In a very recent work a canonical tensor decomposition was applied to ACE to compress the coefficient tensor \citep{darby_tensorreduced_2023}.

Thus, comparing ACE and ETN to SNAP (all three three potentials use spherical harmonics), one can say that ACE generalizes SNAP by chaining the Clebsch-Gordan tensors and forming a large coefficient tensor preserving linearity of the model, while ETN generalizes SNAP by chaining \emph{order-3 equivariant tensors}, moving from linear SNAP to a multilinear model.
\end{appendices}

\printbibliography[heading=bibintoc]

\end{document}